# Enhanced Light-Matter Interaction in Two-Dimensional Transition Metal Dichalcogenides

Lujun Huang[1#], Alex Krasnok[2#], Andrea Alú[2,3], Yiling Yu[4], Dragomir Neshev[5] and Andrey E Miroshnichenko[1*]

1. School of Engineering and Information Technologies, University of New South Wales, Canberra, ACT, 2600, Australia
2. Photonics Initiative, Advanced Science Research Center, City University of New York, New York, New York 10031, USA
3. Physics Program, Graduate Center, City University of New York, New York, NY 10016, USA
4. Center for Nanophase Materials Sciences, Oak Ridge National Laboratory, Oak Ridge, Tennessee 37831, USA
5. ARC Centre of Excellence for Transformative Meta-Optical Systems (TMOS), Department of Electronic Materials Engineering, Research School of Physics, The Australian National University, Canberra, ACT 2601, Australia

# These authors contributed equally to this work
To whom correspondence may be addressed: andrey.miroshnichenko@unsw.edu.au

Two dimensional (2D) transition metal dichalcogenide (TMDC) materials, such as $MoS_2$, $WS_2$, $MoSe_2$, and $WSe_2$, have received extensive attention in the past decade due to their extraordinary electronic, optical and thermal properties. 2D TMDC materials evolve from indirect bandgap semiconductors to direct bandgap semiconductors while their layer number is reduced from few layers to a monolayer limit. Consequently, there is strong photoluminescence in a monolayer (1L) TMDC due to the large quantum yield. Moreover, such monolayer semiconductors have two other exciting properties: large binding energy of excitons and valley polarization. These properties make them become ideal materials for various electronic, photonic and optoelectronic devices. However, their performance is limited by the relatively weak light-matter interactions due to their atomically thin form factor. Resonant nanophotonic structures provide a viable way to address this issue and enhance light-matter interactions in 2D TMDCs. Here, we provide an overview of this research area, showcasing relevant applications, including exotic light emission, absorption and scattering features. We start by overviewing the concept of excitons in 1L-TMDC and the fundamental theory of cavity-enhanced emission, followed by a discussion on the recent progress of enhanced light emission, strong coupling and valleytronics. The atomically thin nature of 1L-TMDC enables a broad range of ways to tune its electric and optical properties. Thus, we continue by reviewing advances in TMDC-based tunable photonic devices. Next, we survey the recent progress in enhanced light absorption over narrow and broad bandwidths using 1L or few-layer TMDCs, and their applications for photovoltaics and photodetectors. We also review recent efforts of engineering light scattering, e.g., inducing Fano resonances, wavefront engineering in 1L or few-layer TMDCs by either integrating resonant structures, such as plasmonic/Mie resonant metasurfaces, or directly patterning monolayer/few layers TMDCs. We then overview the intriguing physical properties of different types of van der Waals heterostructures, and their applications in optoelectronic and photonic devices. Finally, we draw our opinion on potential opportunities and challenges in this rapidly developing field of research.

## 1. Introduction

Light-matter interactions, which include light absorption, scattering and emission, constitute an important branch of physics. They lay the foundation of modern optics and optoelectronic research. The demands for low power consumption, fast speeds, compact footprints, and low cost require the miniaturization of photonic devices and high device densities. Researchers and scientists have devoted tremendous efforts to pushing the dimensions of photonics and optoelectronic devices to a nanoscale. Typical sizes in the vertical dimension are a few tens of nanometers. Reduction to several or even subnanometer scale is quite challenging since the electronic and optical properties may no longer be maintained. It poses the challenge for developing atomically thin photonic or optoelectronic devices. In 2004, Novoselov et al. reported the findings of the two-dimensional (2D) phase of graphite-graphene, representing a single layer of carbon atoms arranged in a honeycomb lattice[1]. This work has signified the advent of the era of 2D materials[2]. Unlike conventional materials, 2D materials are strongly bonded within the layer but weakly bonded by van der Waals (vdW) forces between layers, making monolayer (1L) and few layers readily accessible by mechanical exfoliation. Although graphene's unique electronic and optical properties induced by linear dispersion relation have brought up many intriguing applications, its zero bandgap limits applications where semiconductors with a direct bandgap are required. Therefore, researchers and scientists have strived to expand the family of two-dimensional layered materials with a wide array of electronic, optical, mechanical, and thermal properties in the past decades[3–5]. These two-dimensional materials include semi-metallic graphene with zero bandgap, semiconducting transition metal dichalcogenide (TMDC) and black phosphorene monolayers, ferromagnetic chromium triiodide ($CrI_3$) and $Cr_2Ge_2Te_6$, insulator hexagonal boron nitride (hBN), etc. Their atomically smooth surface free of dangling bonds allows for building van der Waals heterostructures by stacking one on another without the need of lattice matching, providing a wealth of space in exploring new materials[2]. The formation of van der Waals heterostructures combines the advantages of two different materials and enables rich physical properties.

In this review, we focus on two-dimensional (2D) TMDC materials, which include $MoS_2$, $WS_2$, $MoSe_2$, $WSe_2$, $MoTe_2$, $WTe_2$. In contrast to gapless graphene, TMDC materials evolve from indirect bandgap to direct bandgap semiconductors when thinned down to the monolayer limit[6,7]. This unique property makes them promising candidates for electronic, photonic, and optoelectronic devices. On the one hand, Kis et al. successfully fabricated the first atomically thin field-effect transistor with a high on/off ratio by exploring the direct bandgap of 1L-TMDC[8]. The field-effect transistor is the fundamental component for modern integrated circuits. Thus, 1L-TMDC opens a new door towards building atomically thin integrated circuits. On the other hand, the transition from indirect to direct bandgap results in a significant improvement of quantum yield in a monolayer semiconductor, making the photoluminescence (PL) in the monolayer several orders larger than that of few-layer materials. The optical bandgap of 1L-TMDCs varies between 1.08 eV and 2.0 eV, spanning from visible to the near-infrared spectral range. The exciton binding energy in these monolayer semiconductors is few hundreds meV at room temperature due to quantum confinement and reduced dielectric screening[9]. Such large binding energy makes the observation of exciton-polaritons at room temperature possible via strong coupling of 1L-TMDC with a photonic cavity[10]. Apart from their intrinsic nature of direct-bandgap semiconductors, the broken inversion symmetry in monolayer structures causes the split of valence bands by spin-orbital interaction and gives rise to valley contrast at K and K' points in the band diagram, at which each valley can be selectively excited by specific circular polarization incidence [11]. Given so many salient properties from 1L-TMDC, these structures serve as an excellent platform for developing atomically thin photonic and optoelectronic devices with high performance, such as optical modulators, lasers, light-emitting diodes, solar cells, photodetector, etc.

However, their atomically thin nature limits the coupling strength of light-matter interaction. For example, the absorption of 1L-$MoS_2$ is <10% in the spectral range from 450 nm to 700 nm, limiting the performance of 1L-TMDC-based devices. Resonant photonic structures provide an alternative to fully exploiting the potential of TMDC materials in photonics and optoelectronics. As a rapidly growing field, light-matter interactions of two-dimensional materials have been reviewed in many surveys, from fundamentals to applications[11–20]. This work provides our perspective on enhanced light-matter interactions in TMDCs by covering a wide range of topics, from light emission to absorption and scattering.

The structure of this review paper is organized as follows. Section 2 discusses the basic properties of excitons in 1L-TMDC and theoretical approaches for cavity-enhanced light-matter interactions. Then, in Section 3 we review enhanced light emission (i.e., lasing, light-emitting diode, second harmonic generation, etc.) from monolayer semiconductors by either improving the material quality via chemical and physical treatment or embedding them into photonic cavities. In Section 4, we discuss recent efforts in realizing strong coupling of 1L-TMDC with different types of optical cavities. Next, we survey the recent progress in valleytronics of 1L-TMDC in Section 5 and discuss advances in tunable photonic devices based on 1L-TMDC in Section 6. In Section 7, we overview narrowband and broadband absorption enabled by 1L or few-layer TMDCs and their application in photovoltaics and photodetectors. Section 8 presents recent progress on light scattering and wavefront engineering with 1L-TMDC or multilayer structures. In Section 9, we review the light-matter interactions of van der Waals heterostructures based on 1L-TMDCs and their photonic and optoelectronic applications. Finally, we present our concluding remarks and outlook of future research directions in this area.

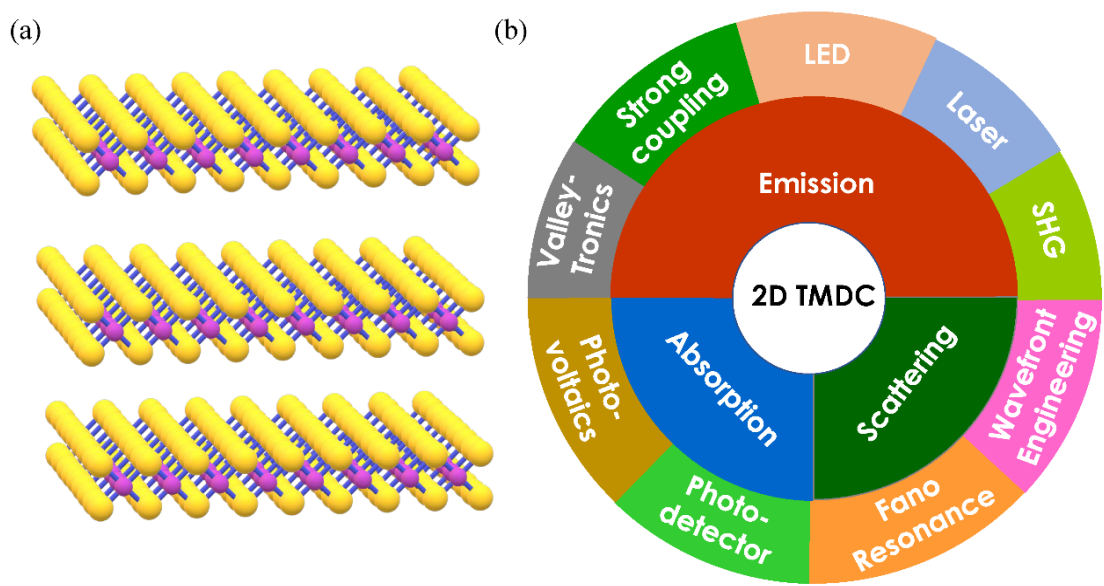

Figure 1. (a) Schematic illustration of TMDC materials. (b) Light-matter interaction of 2D TMDC, which includes light emission, absorption, and scattering. Light emission constitutes of five parts: LED, Laser, SHG, strong coupling, valleytronics. Light absorption mainly involves narrowband and broadband absorption and their applications in photovoltaics and photodetectors. Light scattering includes Fano resonance and wavefront engineering.

## *2. Theoretical Background*

**2.1. Excitons in 2D TMDCs**

Transition metal chalcogenides are layered structures in their bulk form with weak interlayer van der Waals interactions[21]. Here, we focus on the trigonal prismatic semiconducting group VI TMDCs described by the formula $MX_2$ (M = Mo, W; X = S, Se, Te). Such TMDCs have been attracting significant interest for many years[21–23]. In particular, it was discovered that in their bulk condition, TMDCs possess an indirect bandgap corresponding to the transition between the valence band (VB) maximum at the Γ-point and the conduction band (CB) minimum situated about halfway along the Γ-K direction[24,25]. However, the 2D material revolution caused by the emergence of graphene[1,26] and the development of mechanical exfoliation[1,27,28] along with chemical vapor deposition[29–31] fabrication techniques led to the emergence of a 2D phase of TMDCs in the early 2010s. In what follows, we discuss the main changes in the properties of TMDCs upon transition from bulk to the 2D phase, along with the significant features of 2D TMDCs that make these materials unique for advanced photonics and optoelectronics[15].

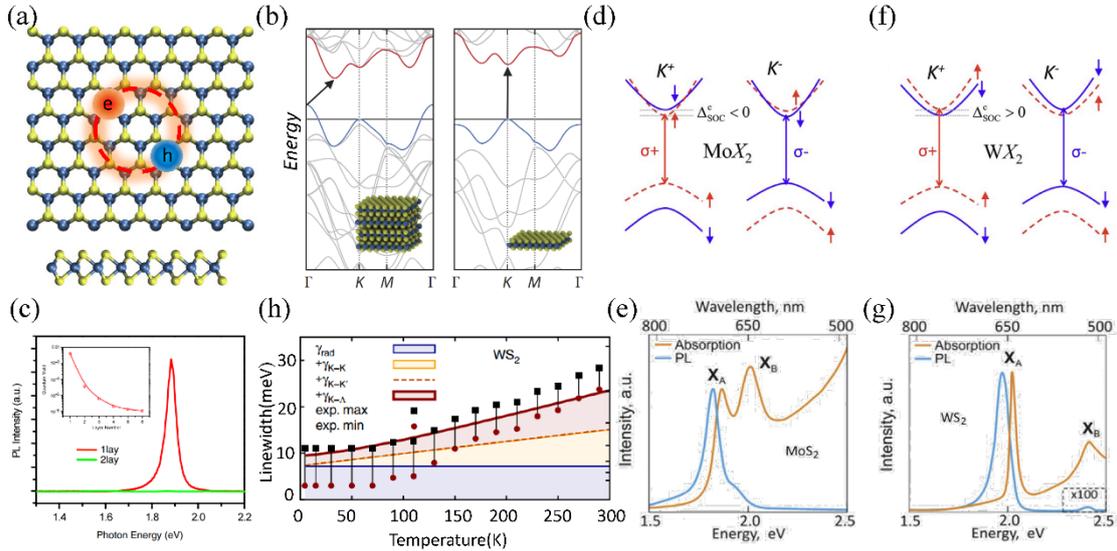

Figure 2. (a) Crystal structure of monolayer transition metal dichalcogenide. The transition metal (chalcogen) atoms appear in blue (yellow). (b) Indirect-direct bandgap transition (first-principles calculation) in $MoS_2$. (c) Enhanced layer-dependent PL intensity for $MoS_2$ monolayers confirming the transition to a direct-gap semiconductor. (d), (f) Schematic illustrations in a single-particle picture show that the order of the CBs is opposite in $MoX_2$ (d) and $WX_2$ (f) monolayers. (e),(g) Absorption and fluorescence spectra of 1L-$WS_2$ and 1L-$MoS_2$ at room temperature, respectively. (h) Dependence of the linewidth and lifetime of the A exciton in $WS_2$ on temperature: black and red dots are the upper and lower experimental values for the linewidths; the thick red line is the total theoretically predicted linewidth. The shaded color regions and corresponding labels contribute to the radiative decay along with intra- and intervalley electron-phonon scattering terms. ((b) from Ref. [32], (c) from Ref. [7], (d)-(g) from Ref. [12], (h) from Ref. [33])

**2.1.1 Transition to 2D Phase**

We begin by discussing the main changes in electro-optical properties occurring during the transition of TMDCs from bulk to the 2D phase. Bulk TMDCs consist of multilayers that are weakly bonded by vdW interlayer interactions between each other. As a result, these single layers can be exfoliated relatively easily, preserving their in-plane crystal structure. Fig.2a shows the crystal structure of

1L-TMDC in-plane and out-of-plane. The layers of TMDCs have a trigonal prismatic crystal structure, with hexagonally arranged transition metals (blue atoms) sandwiched between two chalcogen layers (yellow atoms). These crystals typically have the 2H phase, which restores inversion symmetry in even-numbered layers but breaks it in a single layer giving rise to many intriguing phenomena, including enhanced second harmonic generation and valley polarization, as discussed below.

The indirect bulk TMDCs undergo a transition to direct gap semiconductors at the monolayer level[6,7,34], as shown in Fig.2b for MoS$_2$[32]. The simulations show that when two TMDC layers are in close proximity, the chalcogen orbitals interact and raise the valence band's energy at the Γ point. A similar effect occurs in the CB along the path between the Γ and K points, leading to an indirect bandgap away from the K points. As the layers are separated, these interactions are reduced, leaving a direct bandgap at the K points. This transition was first predicted by density functional theory (DFT) simulations[34] in 2007 and soon after observed experimentally as a significant enhancement of photoluminescence (PL) quantum yield in monolayer phase[6,7]. For example, the fluorescence of 1L-MoS2 is found to be four orders of magnitude greater than that of bulk MoS$_2$[7], as shown in Fig.2c. Later DFT calculations[35] revealed that the direct bandgap occurs at the high-symmetry extrema located at the $K^+$ and $K^-$ points of the hexagonal Brillouin zone and gives rise to interband transitions in the visible to the near-infrared spectral range. As a result, the indirect gap energy corresponding to the separation between the Γ and the midpoint along Γ-K increases in the few-layer limit and eventually makes the TMDCs direct in the 2D phase. This behavior has been confirmed experimentally via angle-resolved PL spectroscopy on 2D, few-layer, and bulk MoS$_2$ and MoSe$_2$[36,37].

Except for PL spectroscopy, Raman spectroscopy is another powerful characterization tool to identify the layer number of TMDC materials because it witnesses significant change with the reduced layer number[38,39]. For example, it was found that there are two Raman modes for monolayer and multilayer MoS$_2$ within the range from 360cm$^{-1}$ to 430cm$^{-1}$, which are assigned as E$_{2g}^1$ and A$_{1g}$ mode. E$_{2g}^1$ mode corresponds to the lattice vibration within the plane and is very sensitive to the strain, while A$_{1g}$ is out of plane vibration mode. With the increasing layer number, E$_{2g}^1$ (A$_{1g}$) mode becomes soften (stiffen) and their intensities increase monotonically. Thus, the difference between these two Raman modes and intensity can be used to evaluate the layer number of TMDCs (e.g. MoS$_2$.).

The indirect-to-direct bandgap transition is also accompanied by a significant enhancement in the oscillator strength of the corresponding transitions resulting in strong absorption (1 - 15%) of incoming light at the lowest resonances in the 2D phase[6,7,40] that is about one order of magnitude larger than absorption in thin semiconductors such as GaAs and Si[41]. The absorbance ($A$) of a monolayer TMDC at the resonance is controlled by the ratio of the radiative $\gamma_{ex}^r$ and the non-radiative $\gamma_{ex}^{nr}$ decay rate of the excitons,

$$A = \frac{2\gamma_{ex}^r \gamma_{ex}^{nr}}{(\gamma_{ex}^r + \gamma_{ex}^{nr})^2}. \tag{1}$$

Due to the mismatch of the radiative and nonradiative rates, the absorption in pristine 1L-TMDCs never reaches the maximum value of 50% at the critical coupling condition ($\gamma_{ex}^r = \gamma_{ex}^{nr}$). The enormously high reflectivity of 2D TMDCs caused by the large exciton-photon coupling strength has also been demonstrated recently[42,43]. The high reflectivity of 2D TMDCs has been recently explored to create atomically thin optical lenses and other ultimately-thin devices[44–47]. These are discussed in detail in section 8.

Thus, in contrast to graphene that is a gap-less semimetal 2D material, 1L-TMDCs are direct bandgap semiconductors with strong optical transitions and strong absorption. These properties make them attractive for applications in phototransistors[48], sensors[49], logic circuits[50,51], and various light-producing and harvesting devices[52–54].

**2.1.2 Excitons in 2D TMDCs**

2D TMDCs, like any other semiconducting materials, support not only conducting electrons in the CB and holes in the VB but also excitons caused by their Coulomb attraction. These excitations can respond to the external optical field coherently and collectively in the form of collective motions of electron-hole pairs giving rise to exciton resonances that can be observed experimentally. The formation of excitons requires the system to spend some binding energy ($E_b$) required to bind the electron and hole together. As a result, these excitonic states' energy lies below the CB minimum, and excitons manifest themselves in experiments as narrow scattering or emission lines. Remarkably, due to the 2D confinement of the charges in 2D TMDC, their Coulomb interaction (and the binding energy) is about 1 to 2 orders of magnitude higher than in traditional quasi-2D systems such as GaAs, GaN, ZnSe quantum wells[55–57]. The large values of the bounding energy allow for the observation of not only *neutral (X) excitons* (binding energies of 0.2 to 0.8 eV)[58–62] occurring at the $K^+$ and $K^-$ points in the Brillouin zone, also divided into *A* and *B excitons* caused by the lifting of spin degeneracy due to strong spin-orbit coupling, see below, but also other species of excitons, including *charged (T) excitons* (trions)[63–68], *excitonic molecules* (biexcitons)[69], localized excitons associated with various defects (impurity-induced potentials) in TMDCs, dark intervalley ($K-K'$ or $K-\Gamma$) excitons, dark higher angular momentum states, interlayer excitons (IX)[70–73], and other varieties of bound electron-hole states[9,32] that we discuss in what follows. Such considerable bounding energies of excitons well exceeding the thermal energy at room temperature (~30 meV) make them dominate the optical properties of 2D TMDCs[74].

The exciton states' appearance below the free-particle bandgap ($E_g$) gives rise to the formation of the so-called *optical band gap* ($E_0$) defined as the lowest energy excitonic state in absorption. The free-particle bandgap of 2D materials can be measured using scanning tunneling spectroscopy. In its turn, the optical band gap is defined via PL measurements of excitons. The difference $E_b = E_g - E_0$ yields the exciton binding energy, determined to be, e.g., about 0.55 eV in 2D MoSe$_2$ on bilayer graphene[62].

$E_b$ in TMDCs and in other semiconductors depends on the dimensionality, dielectric screening, and effective mass. It can be estimated in the framework of the 2D hydrogen-like model as $E_b \approx 4\text{Ry}\mu/(m_0\varepsilon_{\text{eff}}^2)$, where Ry=13.6 eV is the Rydberg constant, $\varepsilon_{\text{eff}}$ stands for the effective dielectric constant of the system, roughly averaged from the contributions of the TMDC and the surroundings, $m_0$ is the free electron mass, and $\mu = m_e m_h/(m_e + m_h)$ is the reduced mass. In 2D TMDCs, due to the reduced dielectric screening effect (small $\varepsilon_{\text{eff}}$), the binding energy gets increased compared to bulk materials. For realistic parameters of $\mu = 0.25 m_0$ and $\varepsilon_{eff} = 5$, this approach gives 0.5 eV in agreement with the experiment.

Excitons, particularly the X excitons, are similar to a hydrogen atom and exhibit the *Rydberg series* in their energy spectrum below the free-particle bandgap[60]. These states in the Rydberg series of the TMDCs excitons can be numbered by the principal quantum number $n$, so that the ground state is $n = 1$ and the excited states are $n > 1$. The energies of excitons' excited states $n > 1$ can be obtained directly from linear absorption or reflectance spectroscopy[75]. These states manifest themselves as resonances with decreasing oscillator strength $f_n = f_{n=1}/(2n-1)^3$ [9,76]. Moreover, because of the considerable binding energy of excitons, they are hydrogen atom-like formations of a small size characterized by the Bohr radius of $a_B \simeq 1$ nm. Consequently, this small size of excitons results in their delocalization in the k-space[77,78] and place them between Wannier and Frenkel type excitons[79].

Finally, the reduced dielectric screening effect in 2D TMDCs results in tightly bounded X excitons and other excitonic states. It makes them very sensitive to the variations of the surrounding dielectric environment and feasible to tune the exciton binding energy. As a result, the excitonic states themselves drastically via the dielectric environment engineering[80–83].

**2.1.3 Transition Selection Rules**

Here, we discuss the energy, spin, and momentum conservation laws when a photon with a certain momentum and spin 1 excites an electron from the VB to the CB, leaving an empty electron state, which can be described by a hole state with a net positive charge in the valence band. If $\hbar\omega$ is the photon's energy, then only states for which the *energy conservation law* is satisfied can be excited, i.e. $\hbar\omega = E_2 - E_1$. For the interband transitions, $E_1$ and $E_2$ are the energies of the corresponding states in different bands. Next, if $\mathbf{q}_\parallel$ is the projection of the momentum of a photon on the TMDC's plane, then according to the *momentum conservation law*: $\mathbf{q}_\parallel = \mathbf{k}_h + \mathbf{k}_e$, where $\mathbf{k}_h$ and $\mathbf{k}_e$ are momenta of the hole in the VB and the momenta of the electron in the CB, respectively. The electron and hole are fermions, and their spin can take either +1/2 or -1/2. According to the *spin conservation law*, the photon can cause transitions between different electronic states with *the same spin projection* since the spin of a hole is opposite to the spin of the corresponding empty electron state (see Fig.2d,f). These transitions are called bright, as they are allowed by the selection rules.

These selection rules are typical for any semiconductor, but there is another set of rules inherent specifically to TMDC–*valley selection rules*. The valley selection rules arise from the different quantum numbers $m$ associated with the bands at the two valleys, K and K'. Namely, at the K valley, the VBs are assigned $m = -1/2$ and $m = +1/2$ for the spin-down and spin-up, respectively, while the CBs are assigned $m = -3/2$ and $m = -1/2$ for the spin-down and spin-up, respectively. Interband transitions from the VB to CB can occur through absorption of left- ($\sigma^-$) or right- ($\sigma^+$) circularly polarized photon that carries an angular momentum $\Delta m = -1$ or $\Delta m = +1$, respectively.

**2.1.4 Strong Spin-Orbit Coupling-A and B Excitons.**

Another attractive property of 2D TMDCs is their strong spin-orbit coupling (SOC) caused by the heavy elements. Remarkably, the spin-orbit interaction in TMDCs is much stronger than in ordinary semiconductors, quantum wells, and graphene. This strong spin-orbit coupling lifts the spin degeneracy of the conductance and valence bands resulting in the spin splitting at the K point ($\Delta_{soc}^{c}$) in the VB of ~0.2 eV (Mo based) and ~0.4 eV (W based). Also, depending on the metal atom (Mo or W), the CB spin splitting has a different sign: $\Delta_{soc}^{c} < 0$ for MoX$_2$ and $\Delta_{soc}^{c} > 0$ for WX$_2$. This splitting leads to the formation of A and B excitons[35,37,84,85] (Fig.2e,g). X$_A$ (X$_B$) corresponds to the lower (higher) energy exciton. Note that, while both A and B excitons give rise to the excitation and absorption spectra, the high energy exciton X$_B$ is typically much weaker in the PL spectra.

**2.1.5 Dark Excitons in 2D TMDCs.**

If an excitonic transition does not satisfy the selection rules, such an excitonic state is dark. Even though dark excitons are forbidden, they can be excited via interactions with another particle or an external DC electric or magnetic fields[61,77,86]. Dark excitons are decoupled from the radiative channels and can have longer lifetimes, thus having great potential for Bose-Einstein condensation, optoelectronic devices, and sensing[33,87,88]. Dark excitons can interact strongly with bright excitons and alter their properties and, as such, various optical and optoelectronic phenomena[61,86–96], e.g., including light emission quenching at reduced temperatures[86], nonlocal effects[97], and alternation of the excitonic linewidths and lifetimes[33].

We distinguish between momentum-forbidden and spin-forbidden dark excitons. In *momentum-forbidden* dark excitons, electrons and holes are located at different valleys in the momentum space. They cannot be accessed by light due to the lack of required momentum transfer. Intervalley momentum-forbidden dark excitons present important scattering channels for exciton-phonon processes. In particular, the dark K-Λ excitons with the hole located at the K and the electron at the Λ valley are of great interest in tungsten-based TMDCs, where they are believed to be the energetically lowest states (Λ valley is about 80 meV below the K valley in 1L-WSe$_2$[98]). *Spin-forbidden* dark excitons are composed of holes and electrons with an *opposite spin* and hence require spin-flip to be excited. The spin-forbidden dark excitons in 2D TMDCs can be brightened in PL experiments with applied magnetic fields that mix up the spin-split bands and softens the spin-selection rule[95,96,99]. In these experiments, dark excitons manifest themselves as resonant features about 50 meV below the bright exciton resonance in tungsten-based TMDCs.

The CB splitting caused by spin-orbit coupling leads to an energy separation between bright (spin-allowed) and dark (spin-forbidden) excitons. As a consequence, the lowest-energy transition in 2D MoX$_2$ is the bright exciton[100,101]. In contrast, the lowest-energy transitions in 2D WX$_2$ materials are dark excitons[86,102–105] For example, spin-allowed bright excitons in MoS$_2$ and MoSe$_2$ have the lowest energy, whereas, in WS$_2$ and WSe$_2$, the spin-forbidden states consisting of holes and electrons with opposite spin are the energetically lowest[86].

The splitting in CB gives rise to two intravalley singlet-like bright excitons (A$_B$) and two intravalley triplet-like spin-forbidden dark excitons (A$_D$), one of each at the K and K′ valleys[86,106]. The existence of the dark exciton states in WSe$_2$ was first experimentally verified with temperature-dependent steady-state and time-resolved PL measurements[86]. If $N_{A_B}$ and $N_{A_D}$ are the populations of bright (A$_B$) and dark (A$_D$) excitons, they must obey the Boltzmann distribution in the thermal equilibrium state.

$$\frac{N_{A_B}}{N_{A_D}} = e^{-\Delta^c_{soc}/k_b T}, \tag{2}$$

where $k_b$ is Boltzmann's constant and T is the temperature. If the dark state lies below the bright one in the energy spectrum, the bright state's excitation leads to the population of the dark state, where the excitons can live longer. On the other hand, the increase of temperature can populate the bright states taking the excitons from the dark state, leading to the PL enhancement with the temperature increase. For example, the PL intensity of WSe$_2$ increases as temperature increases, which is attributed to the lowest exciton levels being dark states. On the contrary, the PL intensity for MoS$_2$ decreases as temperature increases, implying that the lowest exciton level is a bright state. The dark excitonic states' existence significantly affects the exciton dynamics and transport in 2D TMDCs, as discussed below.

Since the dark exciton level has a much longer radiative lifetime, it can serve as a reservoir for the exciton population for W-based TMDCs at room temperature[90].

### 2.1.6 Localized Excitons

TMDCs are always imperfect and contain various defects, strains, and impurities. The concentration of defects significantly depends on the fabrication technique. For example, the defect density in CVD WS$_2$ has been defined to be ($\sim 3 \times 10^{13}$ cm$^{-2}$)[107], which is four orders of magnitude higher than in exfoliated WS$_2$ ($\sim 2 \times 10^{9}$ cm$^{-2}$)[108]. Due to defects, electrons, and holes in TMDCs samples can be trapped in resulting potentials giving rise to *localized excitons*. These localized exciton states manifest themselves as a series of spectrally narrow resonances below the bright excitons at low temperatures and vanishing at higher temperatures when the thermal energy is sufficient to overcome the trapping potential[109]. These localized excitonic states can have excellent single-photon emission properties, coherence [second-order correlation function $g_2(\tau = 0) < 0.5$ ][110–114], and the spectral width of emission below 120 μeV in free-standing monolayers[110]. The localized exciton emitters, also known as TMDC quantum dots (TMDC QDS), often appear at TMDC samples' edges. Remarkably, such localized excitons can be created at will by profiling a substrate under a TMDC flake (or placing TMDCs on a metasurface)[115–117], making wrinkles[118], or inducing a strain gradient on a 1L-TMDC by placing it atop of nanostructures and nanoantennas[119–121]. In this case, the nanostructures induce strains in the 2D materials in a controllable way generating quantum emitters near the high-field region and giving rise to the *deterministic activation of quantum emitters*.

The localized excitons and other exciton species can be observed at low temperature and can be distinguished by their different slope ($\alpha$) of PL and by their dependence on the excitation laser intensity[122,123]. A more detailed discussion is presented in Section 6.

### 2.1.7 Valley Polarization

The broken inversion symmetry (P) of hexagonal crystalline structure and the strong spin-orbit coupling in 2D TMDC lead to inequivalent K and K' valleys with antisymmetric spin states. In the experiment, the fact that 2D TMDCs have inequivalent valleys manifests itself in the different behavior under excitation by the circularly polarized light. For example, the light of clockwise polarization $\sigma^+$ (counter-clockwise polarization $\sigma^-$) causes transitions from the VB to the CB only in the K (K') valley, giving rise to the so-called *valley polarization*[124]. This fascinating property gives rise to various effects, including valley-selective photoexcitation[124–128], valley Hall effect[129,130], valley-tunable magnetic moment[131], and valley-selective optical Stark effect[132,133]. We discuss the valley polarization in detail below in Section 5.

**2.1.8 Exciton Dynamics and Relaxation.**

The theory of dynamics of excitons, their transport, and relaxation is less developed, and some results are still under debate. Nevertheless, here we present the settled and well-established results related to the exciton dynamics in 2D TMDCs. Since recently, the exciton dynamics in 2D TMDCs attract a great deal of attention[9,99]. One can utilize the time-resolved photoluminescence (PL) spectroscopy or transient absorption (TA) spectroscopy to investigate the exciton dynamics experimentally. The results of these experiments show that after the light hits the TMDC, the photons cause the optical excitation of the material if allowed by the selection rules, that is, the formation of the free electrons and holes with the subsequent formation of the excitons (ultrafast process with a characteristic time of < 1 ps). Then, the excitons undergo thermalization and condense to the lowest allowed states. If the lowest states are bright, then these excitons can decay radiatively with photons' emission after annihilation (recombination of the electron and hole). However, if the lowest states are dark, these excitons first have to be converted to the bright excitons before annihilation to make them emit effectively. The excitons do not rest after excitation but rather move from the excitation spot and propagate several 100 nm up to micrometers before they decay.

After their excitation, excitons preserve the phase and polarization of the photon (coherent excitons)[91], which can scatter light resonantly and coherently and manifest themselves as resonant enhancement in the 2D TMDCs refractive index[40,134,135]. The phonon-assisted effects lead to the formation of incoherent excitons[136].

A simplified two-level quantum approach can describe the dynamics of exciton excitation and decay. In this approach, two fundamental times are attributed to the excitonic state: the population decay time $T_1$, which defines the state population relaxation rate $\gamma_{ex}$ ($T_1 = 1/\gamma_{ex}$), and coherence time $T_2$, which establishes the dephasing rate $\Gamma$ ($T_2 = 1/\Gamma$). The state population relaxation rate can also be divided into radiative $\gamma_{ex}^r$ and non-radiative $\gamma_{ex}^{nr}$ parts so that $\gamma_{ex} = \gamma_{ex}^r + \gamma_{ex}^{nr}$. The radiative $\gamma_{ex}^r$ decay rate is the rate at which the exciton energy converts into the energy of light emission, whereas the nonradiative rate $\gamma_{ex}^{nr}$ corresponds to the nonradiative decay channels. The smaller $\gamma_{ex}^{nr}$ compared to $\gamma_{ex}^r$, the more efficient the system. The efficiency of the quantum system emission is characterized by the *quantum yield*

$$QY = \frac{\gamma_{ex}^r}{\gamma_{ex}^r + \gamma_{ex}^{nr}}. \tag{3}$$

The experimentally defined radiative lifetimes at low temperature (4 K) {and room temperature} for MoS$_2$ and WSe$_2$ are ~5 ps[137,138] and ~4 ps[139] {0.85 ns[140] and 4 ns[141]}, respectively and coincide well with the results of rigorous numerical simulations[142]. Thus, the exciton radiative lifetime at room temperature is about three orders of magnitude larger than at low temperatures. On the other hand, the nonradiative decays in 2D TMDCs are mainly attributed to very fast exciton trapping (1-10 ps). This mismatch in $\gamma_{ex}^{nr}$ and $\gamma_{ex}^r$ especially at room temperature leads to small quantum yield values (~0.1-1 %), which grow significantly with lowering the temperature and passivation of defects. For example, it has been demonstrated that these low $QY$ can be substantially improved (up to ~95%) by chemical treatment by organic superacids[143,144]. The $QY$ of 2D TMDCs can also be improved via coupling with wisely tailored optically resonant nanocavities and nanoantennas[145–153]. A more detailed discussion can be found in Section 3.

In turn, the coherence time defines the coherent superposition of the ground state (maximum of the valence band) and the exciton state and gives rise to the homogeneous linewidth of an exciton resonance. The state population relaxation also contributes to the coherence relaxation rate so that we can write $\Gamma = \gamma/2 + \Gamma^*$, where $\Gamma^* = 1/T_2^*$ is the rate of pure dephasing processes, including elastic exciton-exciton and exciton–phonon scattering. The coherence time is hence bounded by the value of $1/\gamma$ and is typically equal to 0.1-1 ps at low temperatures[154].

For example, Fig.2h shows the dependence of the linewidth and lifetime of the A exciton in WS$_2$ on temperature: black and red dots are the upper and lower experimental values for the linewidths; the thick red line is the total theoretically predicted linewidth. The shaded color regions and corresponding labels indicate the contributions from the radiative decay along with intra- and intervalley electron-phonon scattering terms[33].

The presence of dark excitons significantly affects the exciton dynamics in 2D TMDCs. We define $N_{A_B}$ and $N_{A_D}$ to be the populations of bright (A$_B$) and dark (A$_D$) excitons in a TMDC. Then the dynamics of the A$_B$ and A$_D$ exciton states can be described by the following coupled rate equations:

$$\frac{dN_{A_B}}{dt} = -(k_B + k_1)N_{A_B} + k_{-1}N_{A_D}, \tag{4}$$

$$\frac{dN_{A_D}}{dt} = -(k_D + k_{-1})N_{A_D} + k_1 N_{A_B}, \tag{5}$$

where $k_B$ and $k_D$ stand for relaxation rates of the bright and dark states. The quantities $k_1$ and $k_{-1}$ are the rates for bright-dark and dark-bright converting respectfully and related as $k_{-1}/k_1 = N_{A_B}/N_{A_D} = e^{-\Delta_c/k_b T}$ at equilibrium[101].

The exciton dynamics are altered at the high excitation rate when the excitons start interacting. In this case, the exciton-exciton annihilation effect appears and gives rise to faster exciton decay. In Ref.[155], exciton-exciton annihilation in 1L-, 2L-, 3L-WS$_2$ was investigated using time-resolved PL spectroscopy. It was found that at low exciton density, PL dynamics of 1L-WS$_2$ exhibit single-exponential decay, whereas at the increased pumping, the exciton-exciton annihilation emerges, giving rise to the faster processes in excitonic dynamics, which can be described by the following rate equation:

$$\frac{dn}{dt} = -\gamma n - \gamma_{ee} n^2, \tag{6}$$

where $n = N_{A_B} + N_{A_D}$ is the total exciton population, $\gamma$ is the exciton lifetime without annihilation, and $\gamma_{ee}$ is the annihilation rate constant. The fitting of the experimental data by Eq. (6) is used to obtain exciton-exciton annihilation rate $\gamma_{ee}$, which is found to be 0.41 ± 0.02, (6.0 ± 1.1) × 10$^{-3}$ and (1.88 ± 0.47) × 10$^{-3}$ cm$^2$ s$^{-1}$ for 1L-, 2L-, and 3L-WS$_2$, respectively[101].

### 2.1.9 Excitonic Tuning

The exciton states of 2D TMDCs can be tuned in various ways, including chemical doping[156], electrical doping[75], laser intensity[157,158], mechanical strain[159], substrate[160], surrounding dielectric environment[80], and interaction with optical cavities[66,161,162]. This intriguing property is discussed in detail in Sections 3 and 6.

**2.2 Excitons in Multilayer van der Waals Heterostructures**

Vertically stacking two 1L-TMDCs with weak interlayer vdW bounding forces gives rise to the formation of vdW hetero-bilayers (HBLs) with the reach physics of interlayer excitons (IX) when an electron and hole are located in different layers[163,164]. The spatial separation of electrons and holes in different layers is attributed to the type II band alignment in vdW HBLs, which has been theoretically predicted by first principle calculation[165] and directly measured with sub-micrometer angle-resolved photoemission spectroscopy (μ-ARPES)[166] and microbeam X-ray photoelectron spectroscopy (μ-XPS)[167]. Besides, the spatial separation results in IX's long lifetime, which causes a great interest for optoelectronic and photonic applications[168,169]. If the TMDC layers have close lattice matching and if the twist angle is small enough, the resulting HBL can support a long-period *moiré pattern*[170–173]. A more detailed discussion on vdW HBLs is given in Section 9.

**2.3. Optical response of optical resonators and nanocavities**

**2.3.1 Resonators for light-matter interaction**.
A resonator is a system that can store the electromagnetic field energy in its reactive form for at least several periods of the field oscillations. The larger number of oscillations the stored field can do before it dissipates and/or leaks out, the larger quality (Q) factor of the resonator. The intrinsic Q-factor of a resonator (not loaded) is defined as

$$Q = \frac{\omega_0}{2\gamma_{res}}, \qquad (7)$$

where $\omega_0$ is the resonant frequency and $\gamma_{res}$ stands for the total decay rate, $\gamma_{res} = \gamma_{res}^{r} + \gamma_{res}^{nr}$, and $\gamma_{res}^{r}$ ($\gamma_{res}^{nr}$) is the radiative (nonradiative or dissipative) decay rate. Note that in contrast to the excitonic states discussed above, the classical resonant system is characterized by only one decay rate because there is no analog to the population decay rate in classical physics.

The Q-factor's importance is explained by the fact that it gives the upper-limit of stored energy in the cavity mode in the steady-state: the larger the Q-factor, the greater the stored energy [174,175] distributed over the mode volume. Dielectric cavities that provide very large Q-factors (whispering gallery mode resonators, Bragg resonators, photonic crystals) are extended structures much larger than the resonant wavelength. As such, their high-Q modes are very delocalized in space, providing modest or even weak values of local field enhancement. To further increase the electric field of the mode per one photon, it is needed to reduce the effective mode volume ($V_{eff}$), which can be defined in the low loss approach as

$$V_{eff} = \int \frac{\varepsilon(r)|E(r)|^2 \, dV}{\left(\varepsilon(r)|E(r)|^2\right)_{max}}, \qquad (8)$$

where the integration runs over the mode field $E(r)$, and $\varepsilon(r)$ is the permittivity of the resonator and surrounding area[176–180]. Usually, the effective mode volume is less than the physical volume of the resonator. For example, plasmonic nanocavities demonstrate unprecedented effective mode shrinking up to values of the order of $\lambda^3/10^4$ and smaller[179], whereas the dielectric resonators are bounded by the diffraction limit. However, Eq.(8) is not always applicable to nanocavities (especially to plasmonic ones) because they often demonstrate a high dissipative rate[181]. This issue has been addressed in the literature[178–182].

The major effect of a cavity or a resonator on an excitonic subsystem such as 2D TMDCs is twofold. Firstly, the resonator provides an *enhanced excitation rate*, which is essential for excitonic systems that purely interact with light. Even though pristine high-quality 2D TMDCs can have relatively high excitation rates, their further enhancement is still beneficial, especially for the dark excitons and interlayer excitons in vdW heterostructures having much weaker oscillator strength and hence low excitation rates. The enhancement of the excitation rate ($\kappa_{\mathrm{exc}}$) is consistent with the increasing of the absorption cross-section of the exciton transition placed in the enhanced E-field $\mathbf{E}$ at the vicinity of the resonator. The excitation enhancement can be characterized quantitatively in the dipole approximation by the factor

$$\frac{\kappa_{\mathrm{exc}}}{\kappa_{\mathrm{exc},0}} = \frac{|\mathbf{n}_d \cdot \mathbf{E}(\mathbf{r}_{ex})|^2}{|\mathbf{n}_d \cdot \mathbf{E}_0(\mathbf{r}_{ex})|^2}, \qquad (9)$$

where $\mathbf{E}_0$ is the free-space E-field strength, $\kappa_{\mathrm{exc},0}$ is the intrinsic excitation rate, $\mathbf{n}_d$ is the orientation of dipole transition[183]. A resonator (cavity, nanoantenna etc.) enhances $\mathbf{E}$ via storing energy, mode squeeze, resulting in maximization of the excitation rate.

Secondly, the interaction of excitons with light in the form of coherent energy exchange can be significantly enhanced in a resonator's vicinity. The excitons in 2D TMDC can interact with a resonator in two drastically different regimes, weak and strong coupling[10,12,184,185]. The interaction regime depends on the *coupling figure of merit* (cFOM) defined below via the Rabi frequency and the total decay rate. If cFOM<1, the system is in the weak coupling. While for cFOM>1, it is in the strong coupling (see below)[186,187].

In the *weak coupling regime* (cFOM<1), the excitons decay exponentially in time, with the rate $\gamma_{\mathrm{ex}} = \gamma_{\mathrm{ex}}^{\mathrm{r}} + \gamma_{\mathrm{ex}}^{\mathrm{nr}}$, where $\gamma_{\mathrm{ex}}^{\mathrm{r}}$ and $\gamma_{\mathrm{ex}}^{\mathrm{nr}}$ are radiative and nonradiative decay rates. This total decay rate is changed due to coupling to the resonator with respect to the intrinsic rate ($\gamma_{\mathrm{ex}}^0$) that can be calculated as $\gamma_{\mathrm{ex}}^0 = nd^2\omega^3 / (3\pi\hbar\varepsilon_0 c^3)$ with $d$ being the dipole moment, $n$ is the refractive index of hosting dielectric, $\varepsilon_0$ is the dielectric constant. The total Purcell factor ($F$) quantitatively quantifies the Purcell effect and is defined as $F \equiv \gamma_{\mathrm{ex}} / \gamma_{\mathrm{ex}}^0$ [188,189]. In the single-mode approximation for the resonator and the dipole approximation for the excitons, the maximal value of the Purcell factor for an exciton with a polarization aligned with the mode E-field and placed in the mode maxima can be calculated as[177,190,191]:

$$F_p = \frac{3}{4\pi^2}\left(\frac{\lambda}{n}\right)^3 \frac{Q}{V_{\mathrm{eff}}}, \qquad (10)$$

with $V_{\mathrm{eff}}$ being the effective mode volume defined above, $\lambda$ is the radiation wavelength ($\lambda = 2\pi c/\omega$), $\omega$ is the angular frequency. Note that in this approximation, the Purcell factor is determined by the characteristics of a resonator only[192].

Under continuous wave (CW) excitation in the steady-state regime, the Purcell factor is equal to the enhancement of total power emitted by an exciton, which is either radiated out to the far-field ($P_{\mathrm{rad}}$) or dissipated ($P_{\mathrm{nrad}}$) in the resonator's material, $F = (P_{\mathrm{rad}} + P_{\mathrm{nrad}})/P^0$, with $P^0$ being the radiation power without a resonator. The coupling of 2D TMDCs with various resonators in the weak coupling regime is essential as it enables enhancement of the quantum yield of exciton emission[151,193–199] reaching ~65%[195], enhanced single-photon emission[195], brightening up dark excitons[148,195,200], exciton light emission tailoring and shaping[145–147,201], and low-threshold lasing[202,203].

The *strong coupling* between TMDC excitons and photonic cavity modes is achieved when cFOM>1 and manifests itself as a hybridization of excitonic and photonic modes with the formation of hybrid light-matter modes. In systems containing 2D TMDCs, the strong coupling regime has been reported since 2015[10]. More discussion on the strong coupling is presented in Section 4.

Finally, the optical resonators (nanoantennas, cavities, etc.) enable excitation and control of valley polarization in 2D TMDCs and TMDCs vdW heterostructures and facilitate valleytronic devices even at room temperature via resonantly enhanced *spin-orbit interaction* (SOI) effects that effectively lock the valley polarization and direction of emission/propagation and photonic Rashba effect[127,201,204–211].

Thus, various characteristics of 2D TMDCs, including quantum yield, PL intensity, valley polarization, etc., can be improved by coupling them to resonant optical structures (cavities or resonators). These optical resonators can be divided into *plasmonic* and *dielectric*, according to their physical mechanisms. We discuss the characteristics of these resonators below.

**2.3.2 Plasmonic nanoresonators**

Single metal NPs with sizes of tens nanometers possess so-called *plasmonic resonances* (plasmons), i.e., hybrid states of strongly coupled light and electronic oscillations. In contrast to dielectric resonators, plasmons in metallic structures are characterized by wavelengths much shorter than the photons' wavelength in free space, making it possible to overcome the diffraction limit[212].

The interaction of light with such plasmonic particles can be modelled as follows. When light impinges on the surface of a metal, different plasma oscillations can occur. First, *bulk plasma* can sustain longitudinal oscillations with the resonant frequency $\omega_p$, resulting from the restoring force caused by nonequilibrium charge distribution. In the Drude model, this plasma frequency is defined as $\omega_p = \sqrt{ne^2/(\varepsilon_0 m)}$, where $n$ stands for the density of electrons, $e$ and $m$ are their charge and mass, $\varepsilon_0$ is the permittivity of vacuum. These bulk plasma oscillations lead to the incident field's screening when the incident waves get reflected at a frequency below the plasma frequency. Above the plasma frequency, the plasma cannot respond quickly enough to screen out the incident field, and the metal becomes transparent. These properties are well described by the Drude model, where the frequency-dependent relative permittivity of a metal, $\varepsilon_m$, is given by $\varepsilon_m = \varepsilon_\infty - \omega_p^2/(\omega^2 + i\gamma_{pl}\omega)$, with $\varepsilon_\infty$ being the permittivity at high frequency, $\gamma_{pl}$ is the relaxation frequency associated with the characteristic time interval between scattering events damping the plasma collective motion. If the permittivity at a high frequency $\varepsilon_\infty = 1$, then the bulk plasma resonance denotes the frequency at which the $\varepsilon_m$ turns to zero.

Planar metallic surfaces support another type of plasmon modes, known as surface plasmon-polaritons (SPPs). In contrast to the bulk plasma resonance, the SPP modes associated with planar interfaces can be excited by light if the momentum matching condition is fulfilled. These polaritons can propagate along the surface, typically over a few wavelengths of light. The dispersion relation of these plasmons is $k_{SPP} = k_0\sqrt{\varepsilon_m\varepsilon_d/(\varepsilon_m + \varepsilon_d)}$, where $k_{SPP}$ is the wavevector of the surface plasmon, $\varepsilon_d$ is the relative permittivity of the surrounding dielectric, $k_0$ is the free-space wavevector ($k_0 = 2\pi/\lambda_0$), with $\lambda_0$ being the wavelength of light in free space. Hence, the SPP mode wavevector is greater (wavelength is shorter) than that of light of the same frequency.

The third plasmon mode is the so-called localized surface plasmon resonance (LSPR) in metallic NPs. When light hits the particle, the oscillating electric field of the light produces a force on the electrons, which leads to an excitation of NP's polarization. When the particle is small compared to the incident wavelength, one can use the electrostatic approximation for this scattering scenario, and Mie scattering theory gives the polarizability for spherical particles

$$\chi^e = 4\pi R^3 \frac{\varepsilon_m - \varepsilon_d}{\varepsilon_m + 2\varepsilon_d}, \tag{11}$$

where R is the NP radius. The particle's strongest response is associated with the polarizability's pole at $\varepsilon_m \approx -2\varepsilon_d$, when the denominator in Eq. (11) is closest to zero, giving rise to LSPR.

In the LSPR regime, the electromagnetic fields are confined to volumes much smaller than a cubic wavelength. This mode volume shrinking effect is a crucial property of plasmonic NPs, making them very useful for nanophotonics. Higher-order multipoles (quadrupole, octupole, etc.) in larger particles can also be excited due to the field retardation effect on the particle's size, leading to the dependence of the spectral position of the LSPR on the particle size. The shape is another critical parameter determining the properties of the spectral position of LSPR. Different particles have been investigated, including nonspherical core/shell structures[213], nanocages[214], star-shaped[215], nanocrescents[216] etc. Some of them show very sharp resonances, e.g., the star-shaped and cube-shaped[217] particles.

The small mode volumes ($V_{eff}$) of LSPRs are beneficial for sensing, light emission enhancement, and strong coupling effect at the single-molecule level. However, in the applications associated with wave propagation (wavefront engineering, polarization conversion, lensing), metallic NPs are not functional due to their dissipative losses that significantly reduce light manipulation efficiency. Hence, loss control in plasmonic materials is one of the most critical challenges in nanoplasmonics. Various approaches have been proposed to alleviate material losses, including embedding optical gain materials into metallic structures, core-shell geometries[218], and new materials like graphene or wide-bandgap semiconductor oxides[219–222]. Among them, the latter is the most attractive. Graphene and wide-bandgap semiconductor oxides might have superior plasmonic properties at terahertz or infrared frequencies than noble metals (i.e., Au, Ag, Al)[222]. Materials like graphene[221,223] may further provide an intriguing possibility of controlling and dynamically tailoring the material dispersions, which is not allowed in noble metals. Nonetheless, there is no universal recipe and material for metamaterials and plasmonic systems.

A variety of new optical effects stem from 2D TMDC coupled to plasmonic nanoscale objects[12]. It has been demonstrated that the coupling of 2D TMDCs with plasmonic nanoantennas can lead to quantum yield enhancement of emission via Purcell effect[12,148,151,201], the formation of exciton-polaritons in strong coupling regime[12,66,82,224], pronounced Fano resonance and optically induced transparency[82], and valley-selective response at room temperature[205]. However, all these effects suffer from high dissipative losses and Joule heating inherent for metals in the optical range[219,222,225]. Since the excitonic system's behavior is modified dramatically with temperature, as we discussed above, the heat generation by plasmonic structures is detrimental for 2D TMDCs. This circumstance causes more and more researchers to refocus their investigations on the coupling of 2D TMDCs with high-index dielectric NPs and nanostructures, free of thermal heating.

**2.3.3 High-index dielectrics**

Mie resonances of high-index dielectric nanoparticles (DNPs) pave an alternative route toward developing nanostructures with unique optical properties. The Mie theory predicts the resonant behavior of subwavelength dielectric particles[226], which was demonstrated recently in silicon colloids[227,228], single NPs, and their ordered structures. The interest in such dielectric structures has been gained by developments in modern nanofabrication methods[229–233]. It has been shown that such subwavelength particles have negligible dissipative losses in visible and IR ranges and support both electric and magnetic resonances. These properties make them an excellent counterpart to plasmonic materials.

In a case of a subwavelength dielectric particle ($R < \lambda$, $R$ is the particle's radius), the scattering pattern in the far-field at the magnetic or electric resonances resembles that of magnetic or electric point-like dipoles. It is thereby possible to introduce magnetic $\chi^m$ and electric $\chi^e$ polarizabilities of such a particle[234–236] with the Mie scattering coefficients $b_1$ and $a_1$ as follows

$$\chi^m = i\frac{6\pi}{k_d^3}b_1, \quad \chi^e = i\frac{6\pi\varepsilon_0\varepsilon_h}{k_d^3}a_1. \tag{12}$$

These equations give $\chi^m \approx 0$ and $\chi^e \approx 4\pi R^3(\varepsilon_m - \varepsilon_d)/(\varepsilon_m + 2\varepsilon_d)$, in the quasistatic approximation ($R \ll \lambda$) that coincides with Eq. (11) for metal NPs. However, in this case $\mathrm{Re}(\varepsilon_m) \geq 0$ and the denominator has no zeros. The first resonance (fundamental) of a dielectric NP is of magnetic dipole (MD) nature, with electric dipole (ED) resonance and high-order resonances lying at shorter wavelengths.

Dielectric resonators do not exhibit significant mode volume shrinking and are limited by the diffraction limit. Although DNPs have small dissipative losses, the radiative losses are large, giving rise to a modest Q-factor of fundamental modes, which grows with the mode order[237]. In some circumstances, the broadband response may be useful, e.g., for ultrafast applications like fs-laser tuning and switching. For other applications relying on the narrowband response (i.e. sensing), the radiative losses may be suppressed with anapole states[238–243] or (quasi) embedded eigenmodes[244–247].

Because of their optical properties briefly discussed above, the high-index dielectric NPs are very attractive for various applications, including the formation of desired radiation and absorption patterns[248], enhanced spontaneous emission (Purcell effect)[249–254], tailoring of scattering[255–260], nonlinear action[261–268], strong coupling[83,269], sensing[188], optical interconnections, etc. Nowadays, DNPs are often used in metamaterials[270–273] and metasurfaces[265,274–280].

DNPs enable significant enhancement of exciton PL from 2D TMDCs because of relatively high Q-factors, small dissipative losses, magnetic dipole resonance that is well-matched with an exciton polarized in the plane. Also, DNPs provide a rich platform for tailoring emission patterns and maximizing the collection efficiency important in experiments[197]. It also enables the realization of strong coupling of exciton in 1L-TMDC and Mie resonator[83,269].

## 2.4. Coupling of Excitons to Optical Cavities

The coupling of excitons in 2D TMDCs with resonators can be described in the *Jaynes-Cummings* formalism[281–283], one of the most fundamental models of how light and matter interact. The essential ingredient of this formalism is the coupling Hamiltonian $H_{int} = -\mathbf{d} \cdot \mathbf{E}_v(\mathbf{r}_{ex})$, where $\mathbf{d} = e\langle e | \hat{\mathbf{r}} | g \rangle$ is the transition dipole between the excited $|e\rangle$ and ground $|g\rangle$ states of the exciton, $e$ and $\hbar$ are the elementary charge and reduced Planck constant, and $\hat{\mathbf{r}}$ is the radius-vector operator. The quantity $\mathbf{E}_v(\mathbf{r}_{ex})$ is the mode *vacuum* E-field or the E-field *per one photon* at an exciton position. This description can be further simplified by introducing the *interaction constant* ($g$) assuming that $\hbar g = -\mathbf{d} \cdot \mathbf{E}_v(\mathbf{r}_{ex})$. In the Jaynes-Cummings formalism, the polariton and exciton modes are described by the Hamiltonian of an oscillator, $H_{pol} = \hbar \omega_{pol} \hat{a}^\dagger \hat{a}$ and $H_{ex} = \hbar \omega_{ex} \hat{\sigma}^\dagger \hat{\sigma}$. Here, $\hat{\sigma}$ and $\hat{\sigma}^\dagger$ ($\hat{a}$ and $\hat{a}^\dagger$) are the annihilation and creation operators of the exciton (photon in the mode), $\omega_{pol}$ and $\omega_{ex}$ are eigenfrequencies of polariton (polarization mode of the resonator) and exciton. The electric field operator at the emitter position and the exciton dipole moment operator can be written as $\mathbf{E}_v(\mathbf{r}_{ex}) = E(\hat{a}^\dagger + \hat{a})$ and $\mathbf{d} = \mathbf{d}_{eg}(\hat{\sigma} + \hat{\sigma}^\dagger)$, respectively.

In the rotating wave approximation (RWA), i.e., in the regime in which we can neglect the counter-rotating terms $\hat{a}\hat{\sigma}$ and $\hat{a}^\dagger \hat{\sigma}^\dagger$, the Hamiltonian of the Jaynes-Cummings formalism is then written as

$$H_{JC} = \hbar \omega_{ex} \hat{\sigma}^\dagger \hat{\sigma} + \hbar \omega_{pol} \hat{a}^\dagger \hat{a} + \hbar g(\hat{\sigma} \hat{a}^\dagger + \hat{\sigma}^\dagger \hat{a}), \tag{13}$$

In the case of no coupling, $g = 0$, the system's eigenstates are the direct product of the atomic eigenstates $|g\rangle$, $|e\rangle$, and the cavity Fock states $|n\rangle$, $n \geq 0$. However, in the presence of coupling $g \neq 0$, the eigenspectrum of this Hamiltonian is (assuming $\omega_{pol} = \omega_{ex}$)

$$E_n^\pm = n\hbar \omega_{pol} \pm g\sqrt{n}, \tag{14}$$

with the corresponding eigenstates, $|n,\pm\rangle = 1/\sqrt{2}(|g,n\rangle \pm |e,n-1\rangle)$ corresponding to the superposition of two states: (1) exciton in the ground state and *n* photons in the mode and (2) exciton in the excited state and *n*-1 photons in the mode [284,285]. Thus, the coupling between the exciton and photonic mode leads to *hybridization of two states* with a gap between the new eigenstates (*dressed states*): $\Omega = E_n^+ - E_n^- = 2g\sqrt{n}$, also known as Rabi splitting, and the quantity $\Omega$ being the Rabi frequency ($\hbar\Omega$ is the Rabi energy). We see that for a Fock state with one photon in the cavity (single photon regime), the Rabi frequency is $\Omega = 2g$. This Rabi frequency - coupling strength correspondence is often used in experiments since the coupling strength $g$ can be deduced from the fitting of experimental results (see below).

The Jaynes-Cummings formalism allows deducing several peculiarities of the exciton-photon coupling. However, this picture is not complete as it does not consider the dissipation in excitonic and photonic systems. A more rigorous analysis of such a coupled system can be conducted in the Lindblad master equation formalism[286,287]. This approach is, however, less illustrative, motivating the use of phenomenological approaches.

The simplest phenomenological approach that allows considering losses in the excitonic and photonic subsystems is based on the following coupling Hamiltonian

$$H_{int} = \begin{pmatrix} \omega_{ex} - i\gamma_{ex} & g \\ g & \omega_{pol} - i\gamma_{pol} \end{pmatrix}, \quad (15)$$

which represents two oscillators with complex eigenfrequencies $\omega_{ex} - i\gamma_{ex}$ and $\omega_{pol} - i\gamma_{pol}$, coupled to each other with strength $g$. The diagonalization of the Hamiltonian yields the following solution for the system dressed eigenstates:

$$E^{\pm} = \left(\frac{\omega_{pol} + \omega_{ex}}{2}\right) - i\left(\frac{\gamma_{pol} + \gamma_{ex}}{2}\right) \pm \sqrt{g^2 + \frac{1}{4}[(\omega_{pol} - \omega_{ex}) - i(\gamma_{pol} - \gamma_{ex})]^2}. \quad (16)$$

Thus, in this picture, the newly dressed states not only split, but their imaginary part also becomes renormalized. The splitting between the dressed states is again the Rabi frequency $\Omega = E^+ - E^- = \sqrt{4g^2 + [(\omega_{pol} - \omega_{ex}) - i(\gamma_{pol} - \gamma_{ex})]^2}$. In the resonant case with no frequency detuning, this formula gives $\Omega = \sqrt{4g^2 - (\gamma_{pol} - \gamma_{ex})^2}$. Thus, if $2g > |\gamma_{pol} - \gamma_{ex}|$ ($2g < |\gamma_{pol} - \gamma_{ex}|$) then the Rabi frequency is real-valued [imaginary-valued], giving rise to the strong [weak] coupling regime. However, we note that in experiments, satisfying this criterion is not enough to observe the strong coupling regime because the splitting has to be also large than the total spectral widths of both resonances, that is $\Omega > (\gamma_{pol} + \gamma_{ex})$. Thus, it is illustrative to introduce the coupling figure of merit (cFOM) defined as

$$\text{cFOM} = \frac{\Omega}{\gamma_{pol} + \gamma_{ex}}. \quad (17)$$

If cFOM>1, the system is in the strong coupling regime, and this can be observed in the experiment as a sufficient splitting of the resonances into two dressed states[185,288]. Otherwise, if cFOM<1, the system is in the weak coupling regime.

It is instructive to consider now the phenomenological coupled-oscillator model (COM), where the polaritonic mode and excitonic resonance are described by an oscillator with corresponding eigenfrequency and decay rate[289,290]:

$$\ddot{x}_{pol}(t) + \gamma_{pol}\dot{x}_{pol}(t) + E_{pol}^2 x_{pol}(t) + g\dot{x}_{ex}(t) = F_{pol}(t), \quad (18)$$

$$\ddot{x}_{ex}(t) + \gamma_{ex}\dot{x}_{ex}(t) + E_{ex}^2 x_{ex}(t) + g\dot{x}_{pl}(t) = F_{ex}(t), \quad (19)$$

where $x_{pol}$, $\gamma_{pol}$, $E_{pol} = \hbar\omega_{pol}$, and $F_{pol}$ ($x_{ex}$, $\gamma_{ex}$, $E_{ex} = \hbar\omega_{ex}$, and $F_{ex}$) are the oscillation "coordinate", linewidth, energy, and force in the external field driving the oscillations of the polariton (exciton), and $g$ again accounts for the exciton-polariton coupling strength. The coupling is assumed to be through the near field, so the coupling strength is the real-valued positive quantity. This picture can be further simplified, making a realistic assumption that $F_{pol} \gg F_{ex}$ we assume that $F_{ex} \approx 0$ in Eqs. (18,19), which allows us to derive the following important formula for scattering cross-section:

$$\sigma_{\text{scat}}(E) \propto E^4 \left| \frac{(E_{\text{ex}}^2 - E^2 - i\gamma_{\text{ex}}E)}{(E^2 - E_{\text{pol}}^2 + i\gamma_{\text{pol}}E)(E^2 - E_{\text{ex}}^2 + i\gamma_{\text{ex}}E) - E^2 g} \right|^2, \quad (20)$$

which is often used to fit experimental data (dark-field scattering) and judge whether the structure is in the strong or weak coupling regime (see below)[66,82,188,291]. This model can be easily generalized to include more excitonic or polaritonic resonances[66,186].

## 3. Enhanced Light Emission of a TMDC Monolayer

Although 1L-TMDCs are direct band-gap semiconductors, their quantum yields (QYs) are still relatively low due to the defects (i.e., vacancy) and intrinsic doping from either supporting substrate or ambient environment. In the steady-state under CW excitation, the fluorescence QY (see Eq.(3)) can also be defined as the ratio of the number of emitted photons to the number of absorbed photons. It usually serves as the quality indicator for 1L-TMDCs because QY is very sensitive to the defects and sub-bandgap states. Defects can act as non-radiative recombination centers, while the intrinsic doping may lead to the formation of the positive or negative charged exciton (trion) that will redshift and broaden the photoluminescence (PL) spectrum. Both play essential roles in suppressing the QY, thus quenching their PL efficiency, which significantly hampers the development of 1L-TMDCs' applications on high-performance photonic and optoelectronic devices, such as light-emitting diodes and lasers.

Currently, there are two main strategies to promote the PL efficiency of 1L-TMDC. One is to modify the intrinsic properties of 1L-TMDC through either chemical and physical treatments or substrate and environmental engineering. The other is to integrate luminescent material-1L-TMDC onto various resonant cavities (i.e., Fabry-Perot cavity, plasmonic nanocavity, dielectric nanocavity, photonic crystal cavity, whisper-gallery-mode cavity, and others), where the operating principles are detailed discussed in section 2. In the following, we give a comprehensive discussion of the mechanism behind these methods.

**3.1 Chemical and Physical Treatment for Enhanced Light Emission**

The PL efficiency can be significantly enhanced by conducting chemical treatments with different species on the 1L-TMDCs[156]. In 2015, Amani et al. experimentally demonstrated that the QY of as exfoliated 1L-$MoS_2$ increases from 0.6% to near-unity after chemical treating with an air-stable, solution-based organic superacid-bis(trifluoromethane) sulfonimide (TFSI)[143], which are viewed as strong protonating agents with Hammett acidity function lower than that of pure sulfuric acid. Fig.3a shows that the PL intensity increases by 190 folds after TFSI treatment compared to that of as exfoliated case, while the line-shape shows almost no change. Moreover, as shown in Fig.3b-c, spatial uniformity of PL enhancement was demonstrated in PL-mapping on the whole monolayer flake before and after TFSI treatment, indicating the validity of such chemical treatments. Calibrated PL measurements were conducted to measure the QY of monolayer flake across a wide range of power ($10^{-4}$ to $10^2$ A/$m^2$). Before TFSI treatment, the QY of as-exfoliated flake is less than 1%. Surprisingly, the QY of TFSI treated sample approaches to almost unity for excitation power below $10^{-2}$ A/$m^2$ and showed a significant drop off at high pumping power. That also means that the exciton radiative recombination plays a dominant role in emission at low pumping power while the nonradiative biexciton recombination rate gradually increases with the increasing power. Time-resolved PL measurements indicate that the luminescence lifetime increases from ~0.3 ns to 10.8 ± 0.6 ns before and after TFSI treatment at extremely low pumping fluences.

The giant enhancement of QY is ascribed to the passivation of surface defects by TFSI, confirmed by a series of materials characterization (i.e., AFM, Raman, XPS). It is interesting to note that TFSI treatment also improves QY of as exfoliated 1L-WS$_2$ but has a negligible effect on exfoliated 1L-MoSe$_2$ and 1L-WSe$_2$. A followed work from the same group experimentally demonstrated that the PL QY of as exfoliated 1L-WS$_2$ increases from almost 20% to >95% after chemical treatment. At the same time, PL emission shows one order of magnitude enhancements[144], which is less than the emission enhancement factor (more than two orders) in 1L-MoS$_2$. This is understandable because the quality of an exfoliated 1L-WS$_2$ is usually higher than that of MoS$_2$, which is also manifested by the fact that QY for WS$_2$ is almost 20 times as MoS$_2$. By contrast, for 1L-MoSe$_2$ and 1L-WSe$_2$, the magnitude of PL decreases by half and two-third after such treatments.

The improvement (or degradation) of QY induced by TFSI treatment is also evidenced by the longer (or shorter) radiative lifetime. The combined results suggest that TFSI can only repair/passivate the defects in sulfur-based 2D materials but does not work for selenium-based systems. This is confirmed by scanning tunnelling microscopy measurements, which show many structural defects and acceptor impurities for sulfurized surfaces while donor impurities dominate the selenides surfaces.

More recently, the fundamental mechanism of TFSI-treated MoS$_2$ with almost unity QY was revealed by Lien et al.[292]. They found that the TFSI treatment is not the only way to promote PL efficiency. QY can be made to approach unity by simply electrostatic doping for both as-process MoS$_2$ and WS$_2$. The negative back gating voltage $V_g$ can inject holes into n-type 1L-MoS$_2$ and 1L-WS$_2$ to make them become intrinsic semiconductor so that PL emission is dominated by radiative recombination of the neutral exciton. Moreover, similar results were found between TFSI treated and untreated samples at $V_g$=-20V from PL QY and time resolve PL measurements. Also, a threshold voltage shift was found for the transistor treated by TFSI. These results indicate that hole doping plays a vital role in giant PL (or QY) enhancement after TFSI treatment. However, it is interesting to note that both the PL spectrum profile and Raman spectrum almost remains the same before and after treating monolayer with TFSI. Usually, hole injection results in a shorter emission wavelength, narrowing the PL linewidth and simultaneously blueshift of $A_{1g}$ peaks of the Raman spectrum[293]. Therefore, further experiments may need to be conducted to clarify the exact mechanisms of PL enhancement.

Note that all the above studies are based on exfoliated 1L-TMDC. TFSI treatment's effect becomes more complicated for the chemical vapor deposition (CVD) as-grown or transferred CVD grown sample to other substrates compared to as-exfoliated ones. To clarify the role of chemical treatment and reveal the underlying mechanism of emission enhancement, Yu et al. treated both as-grown and transferred MoS$_2$, WS$_2$, MoSe$_2$ and WSe$_2$ onto another substrate with acids (i.e., sulfuric acid and TFSI) and Lithium salt (Li-TFSI)[294]. For transferred monolayer, as shown in Fig.3d-g, PL intensity shows one order of magnitude enhancement for WS$_2$ after treatment while it shows a remarkable decrease for 1L-WSe$_2$. Similar enhancement of PL intensity can also be found for transferred 1L-MoS$_2$ and 1L-MoSe$_2$. This is different from the treated as exfoliated samples, for which TFSI treatment only works for MoS$_2$ and WS$_2$ but fails in MoSe$_2$ and WSe$_2$. The opposite trend between transferred WS$_2$ (or MoS$_2$, MoSe$_2$) and WSe$_2$ was attributed to the fact that 1L-WS$_2$ and 1L-WSe$_2$ are, respectively, n-type and p-type semiconductors. Acid-like sulfuric acid and TFSI could induce p-doping to WS$_2$ so that the radiative recombination is dominated by neutral exciton while it contributed to the formation of positive trion (hole bounded to exciton). Except for the enhancement of

PL intensity, the p doping effect was also evidenced by the blueshift of PL peak and linewidth narrowing of PL spectrum as well as stiffen (blueshift) of $A_{1g}$ peak of Raman spectrum[293]. As grown samples, however, show entirely different phenomena. Both $WS_2$ and $WSe_2$ show substantial PL enhancements after TFSI (or Li-TFSI) treatment, which suggests that the underlying enhancement mechanism should not be the same as that of transferred ones.

Further experimental results indicate the cation (i.e., H+ and Li+) intercalation decouples the interaction between as-grown monolayer and substrate or introduces the hole injection into as transferred monolayer. The effect of cation intercalation between the substrate and 2D materials is verified by the increases of height for as-grown monolayer as shown in Fig.3h-i. Almost no change of the PL was observed for suspended samples before and after treatment in Fig.3j-k. Weaker interaction between as-grown 1L-TMDC and substrate was also confirmed by the time resolve transition reflection measurements, which indicates that the exciton lifetime becomes longer after treatments. It has been reported that giant PL enhancements were realized for as grown 1L-$WSe_2$ by mitigating the substrate effects. The PL QY of synthetic $WSe_2$ is improved to 60%, one order higher than the exfoliated one. Such high QY is the synergic effects of substrate decoupling via solvent (e.g., acetone) evaporation and optimization of growth condition. It is worth pointing out that the effect of TFSI treatment on the transferred CVD grown 1L-$MoSe_2$ and 1L-$WSe_2$ onto other substrates is not consistent with the results of as-exfoliated cases. This might be caused by the intrinsic property of selenium-based 2D materials. Further research should be conducted to explore the fundamental physics behind these differences.

Given the superior functionality of TFSI treatment on QY improvement of both 1L-$MoS_2$ and 1L-$WS_2$, Amani et al. found that CVD grown samples must be transferred to secondary substrates to obtain high QY with TFSI treatments because it can release tensile strain that may induce the transition of 1L-TMDC from direct bandgap to indirect bandgap semiconductors[295]. Also, sulfur precursor temperature during the synthesis of 1L-$MoS_2$ is crucial in ensuring TFSI treatment's effectiveness, whereas the PL QY can be improved from 0.1% to almost 30% at 200° for sulfur precursor temperature. Except for the superacid TFSI, other acid treatments are also demonstrated effectively to enhance PL QY. For example, hydrohalic acid (HBr) treatment results in 30 times PL enhancement in highly defective 1L-$MoSe_2$ because it can promote the neural exciton and trion emission through p-doping as well as the removal of the structural defects[296]. Another example is oleic acid treatment, which can not only significantly improve the PL QY of 1L-$WS_2$ but also induce an increase in mobilities in field-effect transistors[297].

As pointed out previously, chemical doping can effectively modify the carrier density and modulate the optical properties of 1L-TMDC because charged exciton (or trion) can be formed as a result of the interplay between exciton and charge carrier. Such charge excitons can reduce the PL QY and broaden the linewidth of the PL spectrum. Researchers from Kyoto University demonstrated that PL intensities of 1L-$MoS_2$ could be tuned by p-dopant or n dopant solution-based chemicals[298]. Both 2,3,5,6-tetrafluoro-7,7,8,8-tetracyanoquinodimethane ($F_4$TCNQ) and 7,7,8,8-tetracyanoquinodimethane (TCNQ) are served as p-type chemical dopants, while nicotinamide adenine dinucleotide (NADH) is used as n-type dopants. From Fig.3l, it can be observed that the PL peak intensity increases by more than 20 times after treating 1L-$MoS_2$ with $F_4$TCNQ and TCNQ. At the same time, the PL peak shows an obvious blueshift. Effective p doping to n-type $MoS_2$ makes the radiative recombination of the neutral exciton become dominant. On the contrary, NADH treatment reduces PL intensity, which can be explained by n-doping to 1L $MoS_2$. Similar treatment has also been applied to improve the PL of 1L-$WS_2$[299]. Other chemicals like $H_2O$ molecules and chloride molecules were also demonstrated to be able to introduce dopings[300,301]. Functional self-assembled monolayers (SAMs) of organic molecules are another convenient way to introduce carrier doping by dipole moments[302,303]. Three commonly used functional SAM are andtrichloro-(1H,1H,2H,2H-perfluorooctyl), silane (FOTS, $CF_3$-SAM), octyltrichlorosilane(OTS, $CH_3$-SAM), and 3-(trimethoxysilyl)-1-propanamine(APTMS, $NH_2$-SAM). The functional process is schematically illustrated in Fig. 3m. APTMS exhibits electron-doping behavior because it has long pair electrons. Nevertheless, FOTS ($CF_3$-SAM) act as hole-donors due to the large electron- negativity of F atoms. The slight dipole moment of $CH_3$-SAM (OTS) has a weak p-doping effect. For transferred 1L-$MoS_2$ on functional SAM modified substrates, n-doping and p doping behaviors were validated by the soften (redshifts) and stiffen (blueshifts) of $A_{1g}$ peak in the Raman spectrum. Due to the effective p-doping or n-doping, the PL enhancement or quenching is observed for $MoS_2$, $WS_2$ and $WSe_2$. Fig.3n shows the PL spectra of $WS_2$ on $SiO_2$/Si substrate with functional SAM sandwiched between $WS_2$ and substrate[304]. The sample with FOTS has the strongest PL intensity and its PL peak is located at around 612 nm, indicating it is dominated by neutral exciton emission due to effective p-doping. OTS has a similar effect on 1L-$WS_2$ but with weaker p-doping, evidenced by the similar spectra shape and 20 times reduction of PL intensity compared to that of FOTS. Unlike OTS and FOTS, the APTMS redshifts the PL peak to 650 nm, and the intensity is two orders lower than that of FOTS, suggesting the n-doping effect.

Strong PL enhancement can also be achieved through defect engineering and oxygen bonding[305]. The defects can be created by annealing and oxygen plasma radiation. Micro-PL measurements indicate that the PL intensity improves by more than three orders at the defect sites after considering the laser spot size. Such a huge PL enhancement can be correlated to two factors: (1) Defect sites can chemically adsorb oxygen with strong bonding energy introduce heavy p doping, converting trion emission into neutral exciton emission. (2) The nonradiative recombination is suppressed at defect sites. First principle calculations also confirm that more effective charge transfer happens between $O_2$ and sulfur vacancy than between $O_2$ and ideal $MoS_2$.

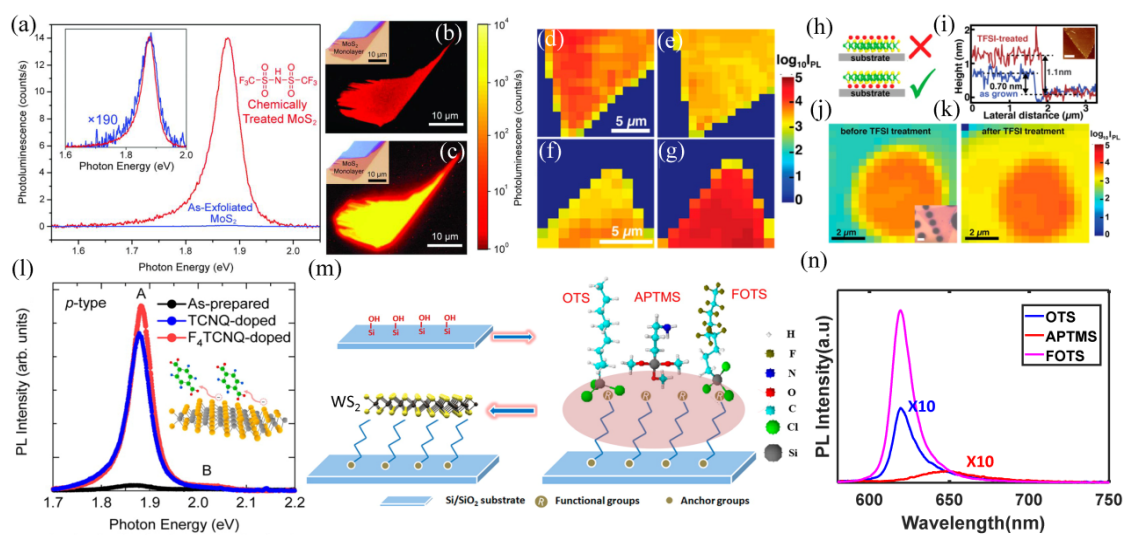

Figure 3. Enhanced PL emission by chemical and physical treatment. (a) PL spectrum for as exfoliated 1L-$MoS_2$ before and after TFSI treatment. (b-c) PL mapping of 1L-$MoS_2$ before (b) and after treatment (c). (d-e) PL mapping of as transferred 1L-$WSe_2$ on a sapphire substrate before (d) and after TFSI treatment (e). (f-g) PL mapping of as transferred 1L-$WS_2$ on a sapphire substrate before (f) and after TFSI treatment (g). (h) Schematic illustration of cation intercalation between monolayer and substrates. (i) The AFM measured the height of as grown 1L-$WS_2$ before and after TFSI treatment. (j-k) PL mapping of suspended 1L-TMDC

before (j) and after TFSI treatment (k). (l) PL spectra of 1L-MoS$_2$ before and after treatment with TCNQ and F4TCNQ (p-type molecules). (m) Schematic illustration of WS$_2$ transferred onto SiO$_2$ substrates with functional self-assembled monolayers. (n) PL spectra of 1L-WS$_2$ treated by OTS, APTMS and FOTS. ((a)-(c) from Ref. [143], (d)-(k) from Ref. [294], (l) from Ref. [298], (m) from Ref. [302], (n) from Ref. [304])

### 3.2 Environment and Substrate Engineering

A substrate should be introduced to support 1L-TMDC when electronic, photonic, or optoelectronic devices are fabricated. Therefore, it is necessary to explore how the substrate and environment affect the optical properties of monolayer thin film. The linewidth of exciton emission for 1L-TMDC on SiO$_2$/Si substrate is a few tens of meV. Such broad linewidths can be attributed to the homogeneous broadening due to the interaction between exciton and phonons as well as inhomogenous broadening effects via the formation of trion (positive or negative charged exciton), such as defects and substrate-induced doping. When 1L-TMDC is sandwiched between the hBN[109,306,307] (see Fig.4a-b), the full width at half maximum (FWHM) for both PL and reflectivity spectrum is reduced to 2~5 meV at 4K due to the suppression of inhomogeneous contributions, which is almost one order lower than FWHM at room temperature. Fig.4c shows the typical PL emission spectrum and differential reflectivity of 1L-MoS$_2$ encapsulated in hBN that are measured at 4K while the PL emission spectrum for 1L MoS$_2$ on SiO$_2$/Si is also plotted as a reference. PL spectrum for capped samples is dominated by neutral exciton emission with FWHM only 4.5 meV. However, the situation is more complicated for uncapped samples directly sitting on SiO$_2$/Si substrate. PL emission spectrum is together contributed by broad defect assisted trion emission (FWHM=38 meV) and neutral exciton emission (FWHM=16 meV). Asymmetric lineshape of Fano resonance for reflectivity spectrum is induced by the interference between sharp exciton resonance of MoS$_2$ and direct reflection from the substrate.

The narrowing effect of linewidth by encapsulation was also demonstrated in other materials (MoSe$_2$, WS$_2$ and WSe$_2$), as shown in Fig.4d. It is worth pointing out that such linewidth narrowing effects work for both as-exfoliated and CVD-grown monolayers. Due to the smooth surface of 2D materials, top hBN protects the 1L-TMDC from direct exposure to the air, which may induce physio and chemisorption. Simultanesouly, the bottom hBN layer perfectly separates the 1L-TMDC from the substrate, blocking doping channels and suppressing the defect-assisted nonradiative recombination originating from substrates as well as dramatically increasing electron mobility. This also suggests that one can obtain high-quality materials by directly growing 1L-TMDC (i.e., MoS$_2$ and WS$_2$) on hBN substrate[308], as demonstrated by Cong et al. The sharp exciton resonance from 1L-TMDC allows for investigating their intrinsic properties, such as the valley polarization and coherence and Valley-Zeeman splitting. It is also widely used in developing photonics and optoelectronic devices with high performance, such as light-emitting diode and atomically thin mirror, which are discussed thoroughly in the following sections. Except for HBN capping, a novel material-ultrathin Ga$_2$O$_3$ glass has been found as an ideal large-scale coating materials for simultaneously enhancing monolayers' optical performance and protecting them from material deposition[309].

PL features can also be tuned by the surrounding dielectric environments. Lin et al. studied the intrinsic influence of dielectric constants of environments on the exciton dynamics in single layer MoS$_2$[310]. When the dielectric constant increases from 2 to 33, both exciton and trion peaks blueshifts by 40 meV. Besides, the PL intensity is enhanced by a factor of 10 and 4 for exciton and trion, respectively. The intensity ratio of exciton and trion emission is also varied. The dielectric screening of

Columb interaction well explains these changes. Also, Kim et al. combined amorphous fluoropolymer encapsulation and TFSI treatment to realize highly stable near unity PL QY of 1L-MoS$_2$ and 1L-WS$_2$[311]. The encapsulation of monolayer into fluoropolymer makes treatments still effective after various postprocessing. This would bear significance in fabricating optoelectronic devices with high performances.

The substrate can affect the optical properties of MoS$_2$ by changing the doping level and modifying the radiative decay rate of exciton and trions[312–314]. Additionally, water trapped at the interface of MoS$_2$ and substrates plays a vital role in modulating the PL emission spectrum. Vargehese et al. found that the water monolayer sandwiched between MoS$_2$ and mica substrates can reduce PL intensity by a factor of 30-50 compared to water-free region[313], and simultaneously redshifts the peak position by 20 meV. Both indicate the formation of trion. Besides, out of plane $A_{1g}$ mode shows blueshift 1cm$^{-1}$ for 1L MoS$_2$ on mica in the absence of water. Also, Kevin probe force microscope measurements indicated that the water adlayer shifts the Fermi level of 1L-MoS$_2$ by 75 meV.

All the above pieces of evidence suggest that trapped water acts as an effective n-dopant. The n-doping behavior of water moisture is also verified by the high PL intensity of transferred WSe$_2$ on SiO$_2$/Si substrates because the intrinsic WSe$_2$ is a p-type semiconductor[314]. Therefore, the substrates can be divided into two categories: hydrophobic ones and hydrophilic ones. PL efficiency can be dramatically enhanced by mitigating the doping effects from either water moisture and substrate itself. The substrate effect can be eliminated by suspending the monolayers, as shown in Fig.4e. By doing so, giant PL enhancement was found for suspended monolayers compared to as-grown ones on substrates (see Fig.4f-g). When 1L-MoS$_2$ and 1L-WS$_2$ are transferred onto hydrophilic substrates (i.e., SiO$_2$/Si, ITO, GaN, etc.), trion emission at longer wavelength dominates with low and broad PL spectrum because water moisture does always exists at the interface of substrates and monolayers. The results are shown in Fig.4h. When monolayers are transferred onto hydrophobic substrates like Teflon and OTS, PL is dramatically improved due to water absence. The weaker n-doping effect was also corroborated by the larger $A_{1g}$ peak of the Raman spectrum in Fig.4i. Another interesting phenomenon is that high PL was also experimentally observed when monolayers are transferred onto 2D substrates, such as mica and hBN. This is because less water moisture can be trapped at the interface due to a small number of defects. The case for WSe$_2$ is the opposite. High PL was found for WSe$_2$ on n-doping substrates like polystyrene. Except for the role of the doping effect, substrates also shorten the exciton lifetime and facilitate the defect-assisted nonradiative recombinations.

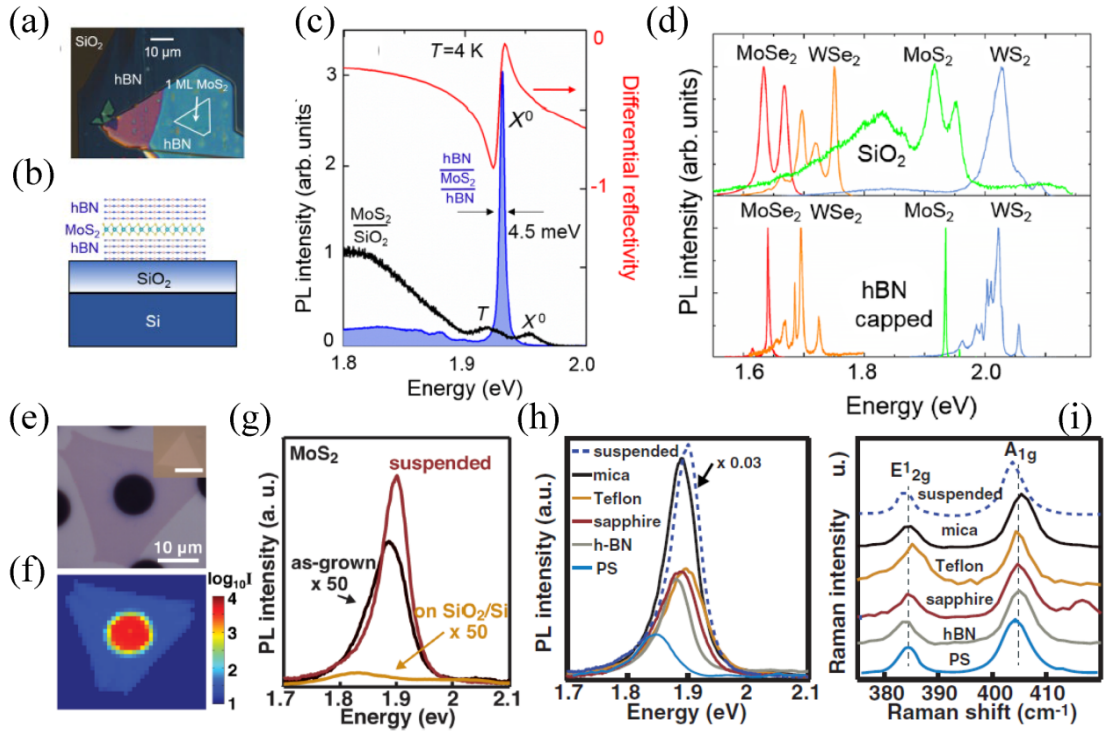

Figure 4. PL emission of 1L-TMDC controlled by substrate engineering. (a) Optical microscope image of van der Waals heterostructure hBN/1L-MoS$_2$/hBN on SiO$_2$/Si substrate. (b) Schematic drawing of the multilayer structure. (c) PL and reflection spectrum at T=4K for capped and uncapped 1L-MoS$_2$. (d) PL spectra at T=4K for 1L MoS$_2$, MoSe$_2$, WS$_2$, WSe$_2$ without (top) and with hBN capping (bottom). (e) Optical image of the suspended monolayer. Inset is an optical image of the as-grown monolayer. (f) PL mapping for suspended monolayer. (g) PL spectra of suspended 1L MoS$_2$, as-grown 1L MoS$_2$ on sapphire substrates, and as transferred 1L MoS$_2$ onto SiO$_2$/Si substrates. (h) PL spectra of transferred MoS$_2$ on different substrates. PL spectra of as-grown 1L MoS$_2$ on sapphire substrate and suspended 1L MoS$_2$ are given as references. (i) Raman spectra of suspended 1L MoS$_2$ and 1L MoS$_2$ on different substrates. ((a)-(d) from Ref. [109], (e)-(i) from Ref. [314])

### 3.3 Enhanced Light Emission by Optical Resonance

When a fluorescent emitter is embedded into an optical nanocavity, the spontaneous emission of gain materials can be modified via the Purcell effect depending on effective mode volume $V_{eff}$ and the quality factor (Q-factor) of nanocavity, as described in section 2. Thus, one can realize the strong PL emission by coupling different optical resonators with 1L-TMDCs.

### 3.3.1 PL Enhanced by Fabry Perot Resonance

Perhaps the simplest optical resonator is the Fabry-Perot resonator, built by stacking multilayer thin film. SiO$_2$/Si substrate is a commonly used substrate for 2D materials because it gives a distinct color contrast that helps to identify the location of 2D materials[315]. Also, it was demonstrated by Lien et al. that light outcoupling of 1L-WSe$_2$ could be engineered by modulating the thickness of SiO$_2$[316] when monolayer semiconductor is put on SiO$_2$/Si substrate (Fig.5a). The PL intensity of WSe$_2$ on 90 nm SiO$_2$ reaches the maximum value (Fig.5b), which is 11 times of PL for WSe$_2$ on 185 nm SiO$_2$. Raman intensity of $E_{2g}^1$ peak for WSe$_2$ on 90 nm is enhanced to most, 30 times of Raman intensity on 260 nm SiO$_2$/Si substrate. Such a large tunability on the PL and Raman is well explained by the substrate interference (Fig.5c), enabling the collective engineering of light absorption and light emission (Fig.5d). Multiple reflections happening at different interfaces results in constructive interference or destructure interference. For example, when SiO$_2$ thickness decreases from 185 nm to 90 nm, substrate interferences transits from constructive interferences (Fig.5e), leading to one order enhancement of PL intensity. Also, further simulation results indicate the amplitude of electric field at the interface WSe$_2$/SiO$_2$ has a maximum value for 90 nm SiO$_2$, but reduces to a minimum for 185 nm SiO$_2$ substrates when the incident wavelength is 532 nm or 752 nm, again confirming the PL enhancing or quenching at this thickness. PL emission can also be modulated by a capping layer on the top of 1L-TMDC while the SiO$_2$/Si substrate is kept. Carefully choosing the capping layer with a suitable thickness in principle maintains the PL intensity on the one hand and simultaneously protect the monolayer semiconductor from degrading on the other hand[317]. Such an interference effect is also studied by Zhang et al. [318]. The strong Raman signal induced by constructive interferences from 90 nm SiO$_2$/Si substrate also allows evaluating the layer number of MoS$_2$ thin film[319]. A monotonic increase relationship has been found between the intensity ratio of A$_{1g}$/Si (or E$_{2g}^1$/Si) and the number of layers, which suggests that the thickness of MoS$_2$ can be identified by just checking the Raman spectrum.

In addition to the multilayer structure SiO$_2$/Si, TiO$_2$/metal substrate has been used to tune the PL emission of MoS$_2$[320]. When the thickness of TiO$_2$ is more than 10 nm, the modulation of PL intensity is mainly governed by the Fabry-Perot interference. For TiO$_2$ with a thickness of less than 10 nm, metal's surface roughness plays a vital role in enhancing PL. The PL intensity of MoS$_2$ on 41 nm TiO$_2$/Au is amplified 15 times compared to MoS$_2$ directly sitting on Au substrate. However, PL lineshape does not change too much. The enhanced PL arises from the increase of absorption after the spacing layer is introduced. The PL enhancement factor can be further improved to 20 by replacing Au with Ag because Ag has a larger reflection in the visible wavelength range than Au. When the top layer TiO$_2$ is reduced to 3.5 nm and the metal underneath has a rough surface, the PL emission intensity is four times of PL for MoS$_2$ on the substrate without spacer layer. Besides, the emission spectrum is dominated by neutral exciton. It deserves noting that a rough surface is crucial to realize PL enhancement here. When the oxidation layer is less than 10 nm, the PL intensity for 1L-MoS$_2$ on a rough Al substrate is more than 22 times than on a smooth Al substrate. All these results seem to suggest a different mechanism from the thicker spacer layer. Indeed, localized surface plasmon resonance is excited in MoS$_2$/3.5 nm TiO$_2$/Cu, resulting in the strong confinement of the electric field at the edges of a rough surface. Finally, it is necessary to point out that the high refractive index of TiO$_2$ also contributes to the larger PL enhancement than SiO$_2$ because the Fabry-Perot resonance has a larger Q factor.

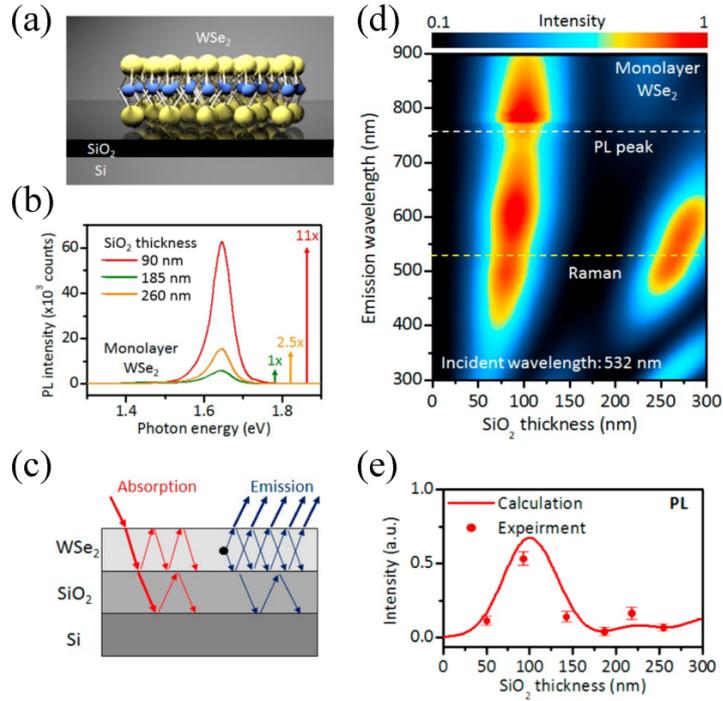

Figure 5. PL emission engineered by Fabry-Perot resonance. (a) Schematic of 1L-WSe$_2$ on SiO$_2$/Si substrate. (b) PL spectrum of 1L-WSe$_2$ on the substrate with different thickness of SiO$_2$. (c). Schematical illustration of the destructive or constructive inference induced by multiple reflections at interfaces for absorption and emission processes. (d) Light outcoupling strength as functions of the emission wavelength and the SiO$_2$ thickness. (e) Measured and calculated PL intensity at different SiO$_2$ thickness. ((a)-(e) from Ref. [316])

### 3.3.2 PL Enhancement by Plasmonic Resonance

Benefiting from the extreme electric field confinement in the vicinity of structures, plasmonic nanostructures made of noble metals have been widely used in boosting light-matter interaction of 2D 1L-TMDCs, such as light absorption and emission. Typically, light emission from 1L-TMDC involves two physical processes. Firstly, a laser with photon energy larger than the semiconductor bandgap is absorbed to excite the electrons transiting from the VB to CB, forming electron-hole pairs. Secondly, electrons in the conduction band and holes in the valence band are radiatively recombined to re-emit photons. Therefore, PL emission simultaneously depends on the light absorption at the excitation wavelength and field enhancement at the emission wavelength. Plasmonic nanocavity is regarded as a powerful platform to modulate the emission of 1L-TMDCs, which have a relatively low intrinsic absorption and QY.

Researchers from Rice University demonstrated that the PL intensity of 1L-MoS$_2$ on gold (Au) nanoantenna array enhanced by 65% with plasmonic mediated pumping[321]. Kern et al. experimentally demonstrated that the PL emission intensity could be improved by more than one order of magnitude when single plasmonic nanoantennas are placed on the top of 1L-TMDCs[322], which is shown in Fig.6a-b. The anisotropy of the Au nanorod gives rise to polarization-dependent PL intensity, which can be found in Fig.6c-e. The strong PL enhancement is fundamentally rooted in the coupling of plasmonic resonance and 1L-TMDC, where the electric field is confined within the interface of nanoantennas and emitters (Fig.6f). The broadband nature of plasmonic resonance simultaneously enhances light absorption at the excitation wavelength and the radiative rate at the emission wavelength via Purcell effects. The aniosotropic property of PL spectrum can be lifted by patterning a square lattice of silver (Ag) nanodisk on the top of 1L-TMDCs[323]. Under such circumstances, the PL emission intensity shows 12 times enhancement compared to that of monolayers on substrates when appropriate structure parameters are chosen, as shown in Fig.6g.

Instead of using an Ag nanodisk array, Lee et al. proposed to apply the silver bow-tie nanoantennas arrays to tailor the optical properties of 1L-MoS$_2$[324]. Owning to the lattice-coupled localized surface plasmons, both PL and Raman intensity experiences more than one order of magnitude enhancements. They also found that the emission spectrum can be modified by controlling the coupling strength of exciton and plasmonic resonance through changing the size of bow-tie antennas and the period of lattices. Like the bow-tie antennas, plasmonic dimers with sub-10 nm gaps can also confine the electric field at the nanoscale. They are incorporating 1L-MoS$_2$ into plasmonic dimer results in 160 fold enhancement on PL intensity[325]. Besides, the anisotropic polarization-dependent PL and Raman spectrum are displayed from MoS$_2$/dimer systems. This is originated from the dimer configuration parallel aligning along the x-axis. Apart from 1L-MoS$_2$, plasmonic nanostructures are also used to manipulate the PL of 1L-WS$_2$, 1L-MoSe$_2$, and 1L-WSe$_2$. Chen et al. reported that PL of MoSe$_2$ on Au nanoantenna array can be tuned from enhancing three folds to quenching one fold with and without a spacer layer[326]. Cheng experimentally realized three times PL enhancement in the complex system Ag nanowires-WS$_2$-Ag film[327]. Johnson et al. demonstrated that the enhancement factor for defect-bound exciton emission could be up to ~200 when CVD-grown WSe$_2$ is transferred onto Ag nanotriangle arrays[145].

Note that the aforementioned PL enhancement factor is limited to a few tens of folds because those structures only support a single plasmonic resonance aligned to either the excitation or the emission wavelength. It can be optimized to thousands or even more if an optical cavity supports two spatially overlapped modes: one is at the exciton wavelength, and the other matches the emission wavelength. In other words, each mode should have a small mode volume and significant field enhancement and simultaneously have a large linewidth that well overlaps the broad emission spectrum from 2D emitters. Akselrod et al. reported a 2000-fold enhancement in the PL intensity of 1L-MoS$_2$ by integrating Ag nanocube on the top of Au substrates[151] (Fig.6h-j), between which is an ultrathin dielectric spacer layer (less than 10 nm). Such a significant enhancement factor is well interpreted as the dual plasmonic resonance with ultrasmall mode volumes, where fundamental mode corresponds to the excitation wavelength of the laser and the first harmonic mode is located at the emission wavelength of the exciton. The small volume is manifested by the fact that the electric field is strongly confined within the sub-10 nm region (Fig.6j). Consequently, the light absorption of MoS$_2$ at the interface is significantly improved while the radiative rates are increased via Purcell effects due to the extensive local density state of photons. Giant PL enhancement of 1L-WSe$_2$ is also experimentally observed in the WSe$_2$-Au plasmonic hybrid structures[150]. The Au plasmonic structure is made of the Au thin film with sub-20 nm trenches (Fig.6k), defined by the template-stripping method and then transferred on the top of SiO$_2$/Si substrates. Unlike previous work that WS$_2$ is sandwiched between the substrate and plasmonic structures, the WSe$_2$ is transferred on the top of the Au substrate with trenches that supports lateral gap plasmons. Proper choose of structure parameters allows for simultaneously tuning the plasmonic resonances to the pumping wavelength 633 nm of the laser and the emission wavelength. As a result, strong light absorption at 633 nm is induced. Besides, the ultra-strong electric field confinement within the tiny trench leads to dramatic increases in the radiative rate and quantum efficiency through Purcell effects. After considering the small area occupied by the trenches, the maximum PL enhancement factor reaches 20,000 (Fig.6l).

In addition to strong PL enhancement, plasmonic structures were also applied to achieve directional emission of 1L-TMDC. Since the PL spectrum has random polarization at room temperature due to the valley decoherence, Han et al proposed to integrate 1L-WS$_2$ onto an anisotropic plasmonic metasurface-Ag sawtooth slits array to enhance the valley coherence[328]. Through tuning the dielectric environment and geometry parameters, resonant transmission modes stemming from surface plasmon resonance are strongly coupled to the exciton of WS$_2$ monolayer at both excitation and emission wavelength, leading to the directional emission. The experiment results demonstrated that such anisotropic transmission resonant modes enable a giant linear dichroism (LD) of PL emission (up to 80%), which is defined as LD=(I$_{TE}$-I$_{TM}$)/(I$_{TE}$+I$_{TM}$).

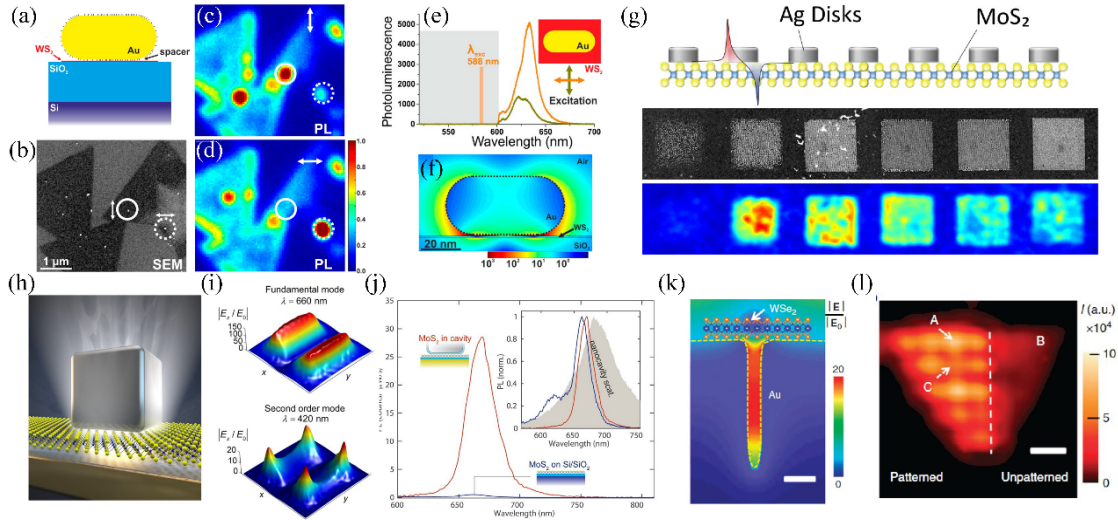

Figure 6. Enhanced PL emission of 1L-TMDC by plasmonic resonance. (a) Schematic of the hybrid system. (b) SEM image of 1L-WS$_2$ with vertical and horizontal oriented Au nanorods on top. (c-d) Normalized PL mapping of the triangle region under the laser illumination with vertical and horizontal polarization. (e) PL spectrum under 588 nm laser excitation with horizontal and vertical excitations. (f) Calculated near field map at 612 nm. (g) Top panel: schematics of Ag nanodisks array on 1L-MoS$_2$. Middle panel: SEM image of fabricated nanodisk arrays with different diameters. Bottom panel: PL mapping of 1L-MoS$_2$ on above-fabricated samples. (h) Schematic of the sample, which consists of Ag nanocube on 1L-MoS$_2$/HfO$_2$/Au film. (i) Spatial mapping of the field enhancement of two plasmonic resonances located at 660 nm and 420 nm. (j) PL spectra from 1L-MoS$_2$ with plasmonic nanocavity on top and MoS$_2$ on SiO$_2$/Si substrate. (k) Calculated electric field distribution of gap plasmons with 1L-WS$_2$ on top of the trench. (l) PL intensity mapping of 1L WSe$_2$ on Au trench array. A dashed line indicates the boundary between the patterned and unpatterned area. ((a)-(f) from Ref. [322], (g) from Ref. [323], (h)-(j) from Ref. [151], (k)-(l) from Ref. [150])

### 3.3.3 PL Enhancement by Mie Resonances and Guided Mode Resonances

Analogous to plasmonic resonances in metallic structures, high index semiconductors (i.e., Si, Ge, GaAs) support electric and magnetic Mie resonances[271,329], enabling large electric field confinement at the nanoscale. These resonant features have the advantage that semiconductors' intrinsic loss is much smaller compared to noble metals. This is beneficial to realize high-performance photonic devices, such as metalenses[277]. Due to these two unique properties, dielectric nanostructures with a high refractive index provides an alternative platform to boost the light-matter interactions through engineering the Mie resonant modes[271]. For example, the PL intensity can be significantly improved for dielectric metasurface made from direct bandgap semiconductor (i.e., III-V GaAs). Such an approach of enhancing light-matter interaction can be transplanted to 1L-TMDCs, where the PL intensity and radiation properties from 1L-TMDC can be engineered by integrating them with single dielectric nanocavity or dielectric nanoantenna arrays via multipolar interferences.

For example, silicon metagratings have been introduced to enhance the emission from monolayer by Chen et al. [330]. They support multiple waveguide modes, as shown in Fig.7a. Carefully tuning the structure parameters can align the wavelength of waveguide modes to both excitation and emission wavelength (Fig.7b), leading to the three folds of PL enhancement (Fig.7c). Besides, the dispersion properties of these modes allow for tailoring the polarization and directivity of emission. Bucher et al. exploit Mie resonance metasurface made of silicon nanocylinder to control the exciton emission[331]. The experiments results show that the PL spectrum does not only enhance in magnitude but also become broad in line width, which is caused by the synergistic effect of the photonic modification density by Mie resonance and unintentional doping from the substrate resulting in the formation of trion. Giant PL enhancement ($>10^4$) of 1L-WSe$_2$ has been demonstrated recently by Sortino et al. [153]. As shown in Fig.7d, when 1L-WSe$_2$ is placed on the top of the gallium phosphide (GaP) nanoantenna, its PL intensity shows 50-fold enhancement compared to that of a monolayer on GaP thin film. Compared to the resonant dielectric metasurface, high refractive index dielectric nanoantenna, such as GaP nanoantenna, supports confined optical modes with reduced mode volume (Fig.7e-f). The marriage of 1L-WSe$_2$ as emitter and GaP nano-antenna as optical resonator gives rise to PL's enhancement factor up to almost $10^4$ for 1L WSe$_2$ and $4\times10^4$ for 2L WSe$_2$, as displayed in Fig.7g. The huge enhancement on PL is induced by the combination of Purcell enhancement, efficient radiation above the substrate and improved light absorption at excitation wavelength due to optical modes' excitation.

Another example of enhanced emission is based on coupling the 1L-MoS$_2$ with the Fano resonances supported by a photonic crystal slab[146]. The enhancement factor of PL is 1300, which arises from the simultaneous excitation of two resonance modes at both pump and emission wavelength of MoS$_2$. Unidirectional PL emission within 5° is realized via Fano resonance engineering. Note that the control of emission properties of 1L-TMDC by Mie resonance is not limited to the case of the resonant all-dielectric metasurface. It can also be tailored by the single dielectric NPs[332]. Ma et al. demonstrated that PL is quenched, broaden and shows redshift when 1L WS$_2$ is integrated with single Si NPs. On the contrary, the PL from multilayer WS$_2$ is greatly enhanced in the presence of Si nanoparticles. The opposite trend is ascribed to the heating and strain effect for the case of single-layer and the enhanced coupling strength between the exciton resonance of multilayer WS$_2$ and magnetic dipole resonance from Si NP. Cihan et al experimentally demonstrated the highly directional emission from 1L-MoS$_2$ integrated with Si nanoantenna[333]. Since 1L-MoS$_2$ can be viewed as an electric dipole emitter, interference between multipoles from Si nanowire and emitter dipole is explored to achieve the unidirectional emission by simply tuning the radius of the nanowire. The forward-to-backward ratio is up to 20, suggesting 25 fold directionality enhancement comparing to the ratio of bare MoS$_2$ on the substrate.

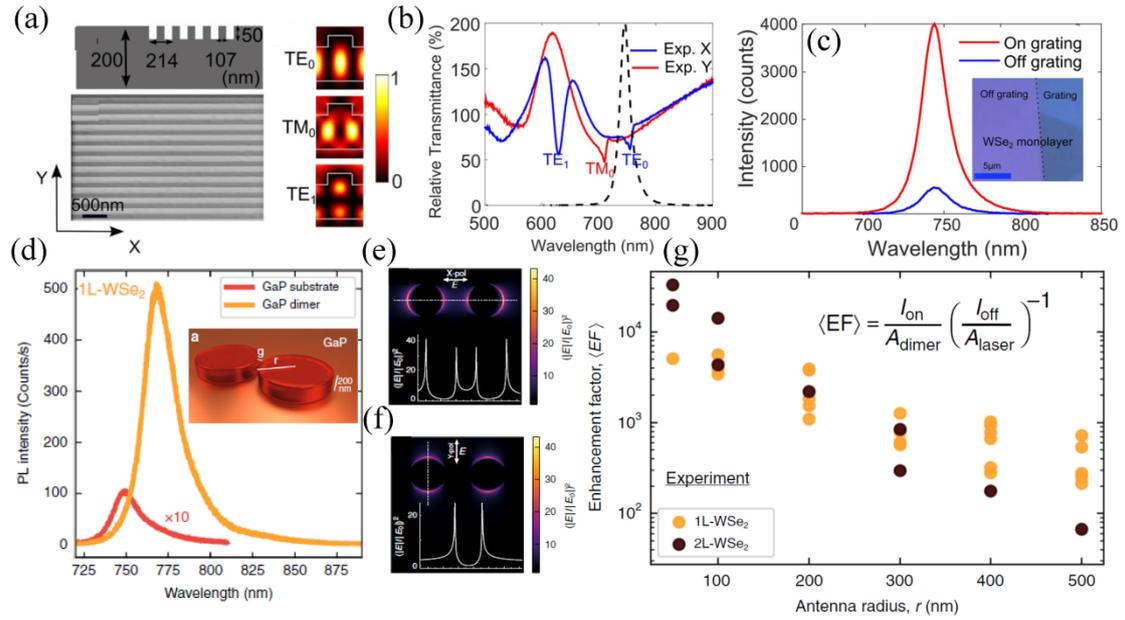

Figure 7. (a) Top left panel shows the structure parameters of the silicon grating structure. The bottom left is the top view SEM image of the grating structure. Right panels display the electric field distribution for three guided more resonance under normal incidence. (b) Measured transmittance of the grating structure under the illumination of incidence with x and y polarization. (c) PL spectrum of 1L-WSe$_2$ with and without grating. (d) PL spectrum of 1L-WSe$_2$ on top of GaP nano-antennas and planner GaP substrate. inset shows the schematic of the GaP dimer antenna. (e-f) Calculated electric field intensity enhancement at 685 nm is scattered by a GaP dimer under the plane-wave excitation with horizontal (e) and vertical directions (f), respectively. (g) Measured PL enhancement factor of 1L- and 2L-WSe$_2$ as a function of the antenna radius. ((a)-(c) from Ref. [330], (d)-(g) from Ref. [153].)

**3.4 Lasing in TMDC Monolayer Supported by a Resonant Cavity**

When the gain material is embedded into a high-Q microcavity, the spontaneous emission rate is significantly enhanced via resonant coupling to cavity modes. When the pumping power is above the threshold, lasing behavior occurs. 1L-TMDC can serve as the thinnest optical gain materials due to its nature of direct bandgap semiconductors. With their large binding energy, 1L-TMDC is regarded as an alternative candidate for realizing laser across a broadband wavelength range as its bandgap spans from visible to near-infrared wavelength range. In this section, we review the recent progress of laser-based on 1L-TMDC or multilayers.

**3.4.1 Lasing by Photonic Crystal Cavity**

In 2000, Noda demonstrated that the Q-factor of a cavity resonance can be improved to 45,000 by employing a modified L3 type cavity[334]. Since this pioneering work, the Q-factor record has been refreshed and pushed to a million scale by fine-tuning the structure parameters[335,336]. As the key component of realizing laser is high-Q cavity mode, the photonic crystal cavity becomes a strong candidate to implement lasing in 1L-TMDC. Wu et al. experimentally demonstrated lasing in 1L-WSe$_2$ by integrating gain material onto L3 type GaP photonic crystal cavity[202](Fig.8a-b). 1L-WSe$_2$ was chosen as a gain material because it has a relatively high QY. As laser hallmark feature, a distinct kink effect was found in the light-light (L-L) curve with the ultralow threshold of 27 nanowatts at 130K due to high-Q cavity mode, while the fraction of spontaneous emission into the cavity mode is β=0.19. In

2017, Li et al. experimentally demonstrated room-temperature CW lasing from 1L-MoTe$_2$ with silicon photonic crystal cavity[337] (Fig.8c-d). Compared to other 1L-TMDC, 1L-MoTe$_2$ has an exciton emission around 1.1eV, which is slightly below the silicon bandgap. As the most mature and abundant semiconductor, the fabrication of a high-Q cavity has been well established based on silicon photonic crystal cavity[338]. One dimensional photonic crystal cavity was adapted in their design to ensure theoretical Q-factor height to $6.5\times10^4$ at 1132 nm. The lasing feature is manifested by the nonlinear kink L-L curves with a threshold of 6.6W/cm$^2$ at room temperature and a shrunk effect of linewidth from 0.4 nm to 0.202 nm. The real Q-factor is obtained as 5603 for laser at 1132.25 nm. The spontaneous emission factor is fitted as β=0.10. The lasing emission wavelength was further pushed to the window from 1260 nm to 1360 nm while multilayer MoTe$_2$ is used as gain materials[339]. Although multilayer MoTe$_2$ has an indirect bandgap, its PL intensity is still moderate. Laser-like emission in multilayer MoTe$_2$ is still feasible due to the following two reasons: (1) the absorption coefficients of MoTe$_2$ drops by two orders from 1130 nm to 1300 nm. (2) the loss of silicon reduces from 6.5dB/cm at 1130 nm to 0.01dB/cm. Fang et al. demonstrated optical pumped MoTe$_2$ laser-like emission at 1305 nm with a threshold of 1500 W/cm2[339]. Except for the L3-type photonic crystal cavity, researchers from the same group proposed integrating sandwiched structure hBN-MoTe$_2$-hBN onto silicon single-mode resonator to enable the laser-like emission at 1309 nm[340].

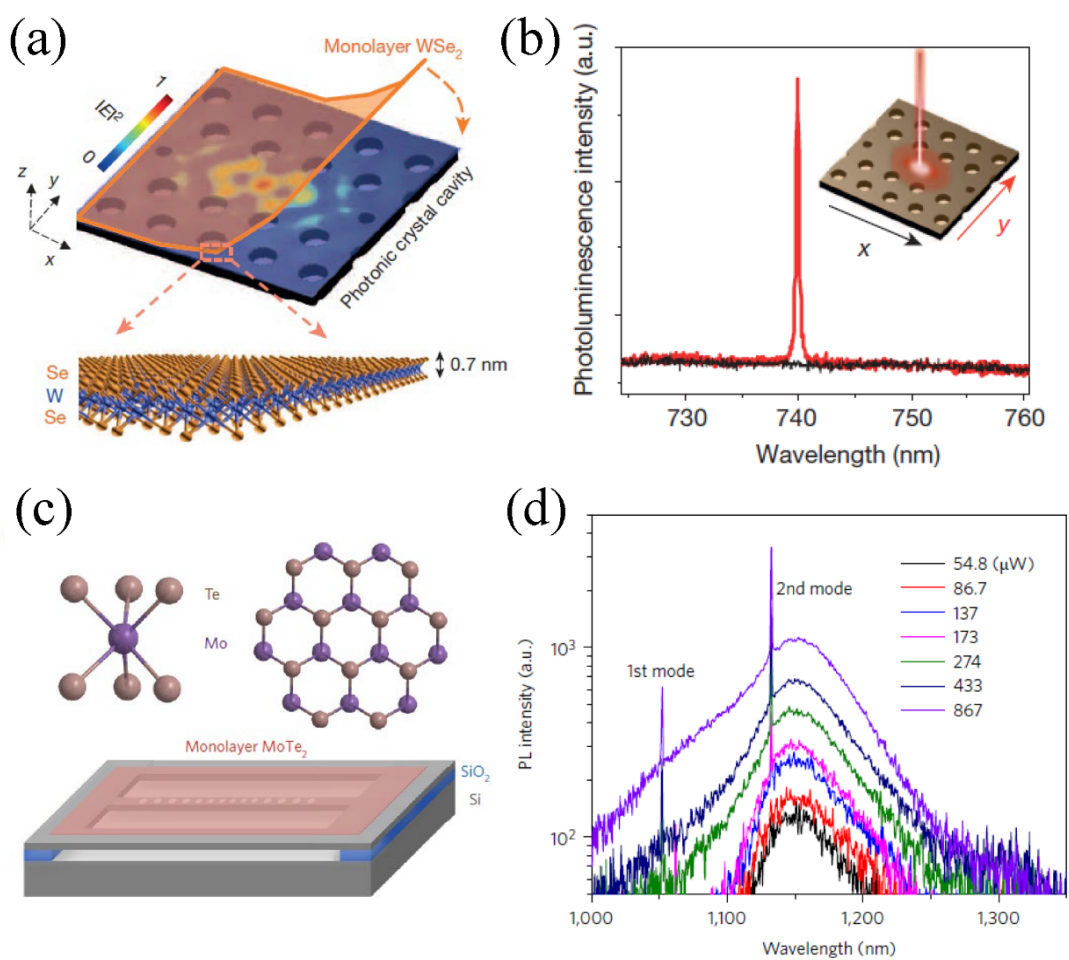

Figure 8. (a) Schematics of WSe$_2$-photonic crystal cavity-based nanolaser, where colorful part illustrates the electric field distribution of cavity. (b) PL spectrum of 1L WSe$_2$ on PCC cavity at 80K under the laser pumping with x and y polarization. (c) Top panel shows the schematic drawing of crystal structure from MoTe$_2$ while the bottom one is schematic of the device, which consists of the silicon photonic crystal nanocavity structure with a 1L-MoTe$_2$ on top. (d) PL spectra from 1L-MoTe2 on silicon nanocavity as a function of pump power at room temperature. ((a)-(b) from Ref. [202], (c)-(d) from Ref. [337])

### 3.4.2 Lasing by Whisper Gallery Cavity

Another promising candidate to realize high-Q factors is the whispering-gallery-mode (WGM)[341]. Thus, it is natural to explore lasing by incorporating 1L-TMDC into whispering-gallery microcavities. Ye et al. reported an excitonic laser of 1L-WS$_2$ in a microdisk resonator at 10K[342]. Unlike most of the previous designs that the monolayer directly sits on the photonic crystal cavity, 1L-WS$_2$ is sandwiched between the Si$_3$N$_4$ and hydrogen silsesquioxane, as shown in the inset of Fig.9a. On the one hand, it enables a considerable overlap between the cavity mode and gain materials. On the other hand, it protects 1L-WS$_2$ from oxidation by air and ensures long-time stability. The microdisk resonator fabricated by electron beam lithography has a diameter of 3.3um and thickness of 180 nm for Si$_3$N$_4$ and 240 nm HSQ. Due to the high-Q factor (2,604) of cavity mode, excitonic laser occurs at 612 nm above 22.4MW/cm$^2$ (Fig.9b). The linewidth shrinking was also observed from 0.28 nm to 0.24 nm. The spontaneous emission factor $\beta$[203] is fitted to be $\beta$=0.5, indicating a relatively large coupling between spontaneous emission and cavity modes. Another optically pumped lasing in 1L-MoS$_2$ was reported at room temperature by Salehzadeh et al. [343], where the 2D MoS$_2$ is put at the interface of freestanding microdisk and microspheres (see the inset of Fig.9c). Multiple laser peaks show up across the wavelength range of 600 nm and 800 nm due to the WGMs supported by microdisk (Fig.9c). The threshold is low to 5μW under the CW 514 nm laser excitation, as shown in Fig.9d. Later, Zhao et al demonstrated large scale lasing based on CVD grown MoS$_2$ thin film by integrating SiO$_2$ microsphere onto monolayer semiconductor, which can operate in a broad temperature range from 77K to 400K[344]. Besides, Galy et al. integrated 1L-WSe$_2$ with high-Q (Q>10$^6$) optical microdisk cavities to measure the cavity quantum yield ($\approx$10$^{-3}$) by efficient near field coupling based on a tapered optical fiber[345].

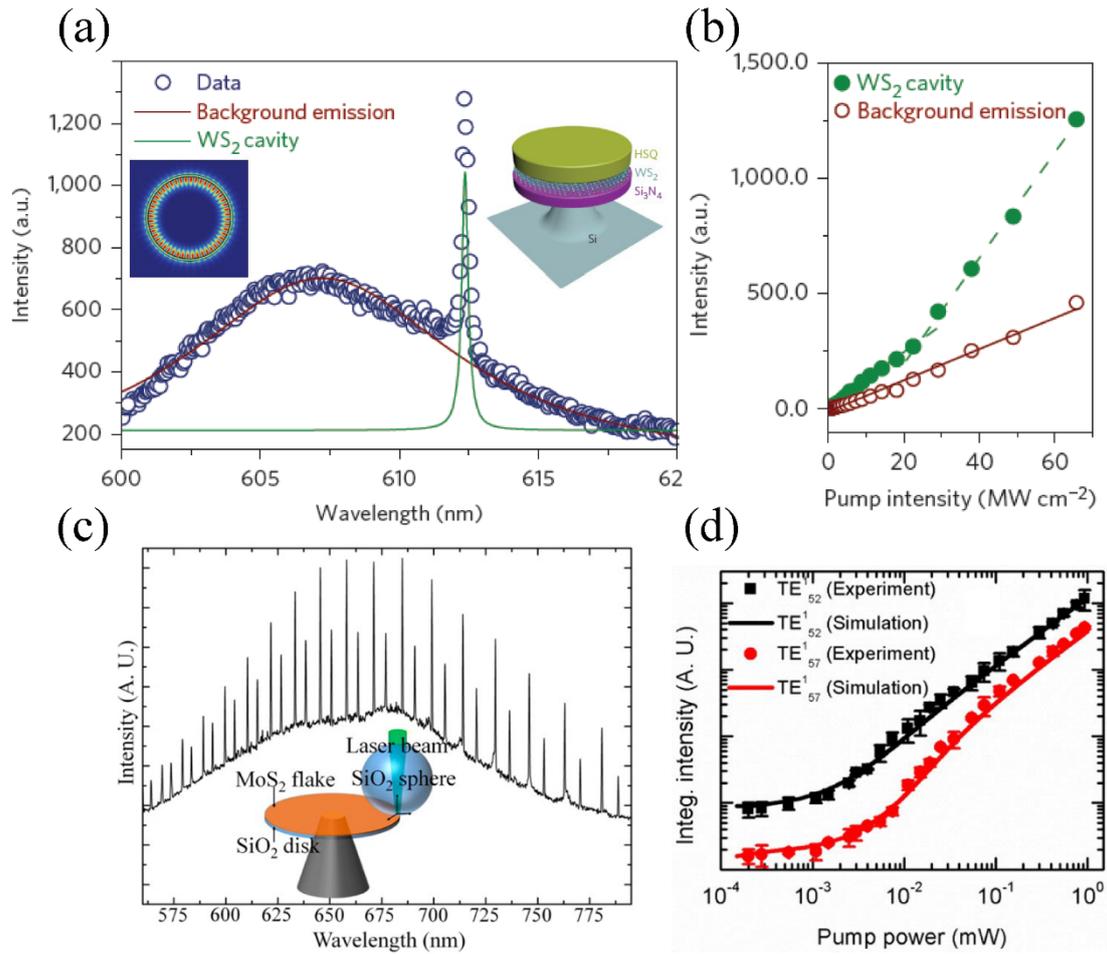

Figure 9. (a) PL spectrum of 1L-WS$_2$ under the laser pumping with the intensity of 65.7 MW cm$^{-2}$. Inset shows the schematic drawing of exciton laser based on a hybrid WS$_2$-microdisk system. (b) PL background and cavity emissions from 1L-WS$_2$ versus pump intensity. (c) Measured PL spectrum of the laser device with an excitation power of 30 μW at room temperature. Inset shows the schematics of the laser device, which consists of sandwiched 1L-MoS$_2$ between SiO$_2$ microsphere and SiO$_2$ microdisk. (d) Measured and simulated L-L curve of the integrated PL intensity versus excitation power for two modes TE$_{52}^1$ and TE$_{57}^1$. ((a)-(b) from Ref. [342], (c)-(d) from Ref. [343])

### 3.4.3 Lasing by Other Cavities

Shang et al. reported the experimental demonstration of ultra-low threshold lasing in 1L-WS$_2$ at room temperature via integration of distributed Bragg reflector-based microcavity with 1L-WS$_2$[346] (Fig.10a-d), where the schematics of the structure is shown in Fig.10a. The single high-Q mode is formed by sandwiching a half-wavelength thick cavity into an upper and bottom distributed Bragg reflectors (DBRs). 1L-WS$_2$ is put at the antinode of the cavity to maximize the light-matter interaction. The light-light curve in Fig.10c indicates the threshold is as low as 0.44W/cm$^2$ and the spontaneous emission coupling factor is β=0.77±0.1. Moreover, because of the asymmetric reflection of top and bottom DBR, such a laser is highly directional and is known as a vertical-cavity surface-emitting laser.

Another way of realizing directional emission is to utilize the Fano resonance in the photonic crystal. Ge et al realized photonic crystal surface-emitting laser by using 1L-WS$_2$ as the gain material[347]. The high-Q Fano resonance is enabled by the *bound state in the continuum* (BIC)[243,348,349] at the Γ-point that also guarantees the vertical field confinement. The lateral field confinement is achieved by using a mode gap and mode mismatch at different bands. The threshold is 89W/cm$^2$ due to the high-Q factor

2500. Angle dependence emission measurement indicates that such a laser emission only occurs within ±3° of surface-normal directions. More recently, Liao et al reported the ultra-low threshold laser in 1L-MoS$_2$ with micro/nano fibers[350]. The schematics of the device are presented in Fig.10e. The micro/nano fibers fabricated by the taper-fiber drawing process introduces high-intensity oxygen dangling bonds by simple photoactivation methods. These reactive oxygen atoms either fill the sulfur vacancy or bridge the neighboring sulfur atoms in 1L-MoS$_2$, constituting the nonradiative recombination channels. Consequently, the QY of 1L-MoS2 grown on the surface of fibers increases to a value ranged from 30% to 1% even when the pumping intensity increases from $10^{-1}$ W/cm$^2$ to $10^4$ W/cm$^2$. Moreover, the micro/nano fiber itself can serve as an optical microcavity. The combination of high QY and high-Q cavity mode makes the realization of ultralow threshold laser become possible. Fig.10f shows the PL spectra under the CW laser pumping with different power. It is found that the threshold is only 5W/cm$^2$ and the coupling factor is β=0.5 due to the high Q-factor 1177 and large QY (QY≈19% at threshold).

It is necessary to note that the laser reported above has a large discrepancy for the threshold, although the Q-factor for cavity mode is in order of thousands. The discrepancy may be attributed to the different quantum yields of 1L-TMDC or different coupling factor. Further investigation is required to clarify the fundamental physics behind this discrepancy. Moreover, a recent review paper by Reeves et al pointed out that lasing behavior can be recognized only when threshold behavior, linewidth narrowing, polarization, and coherence are together satisfied[351], which was highlighted in Nature Photonics by Samuel et al. to recognize lasing in organic semiconductor gain materials[352].

The majority of works on lasing in 1L-TMDC have addressed the four criteria, except for coherence, which is also challenging to measure. In a more recent study, the coherence property of interlayer exciton laser has been demonstrated in van der Waals heterostructure 1LWSe$_2$/MoSe$_2$[169], which is discussed in Section 9.

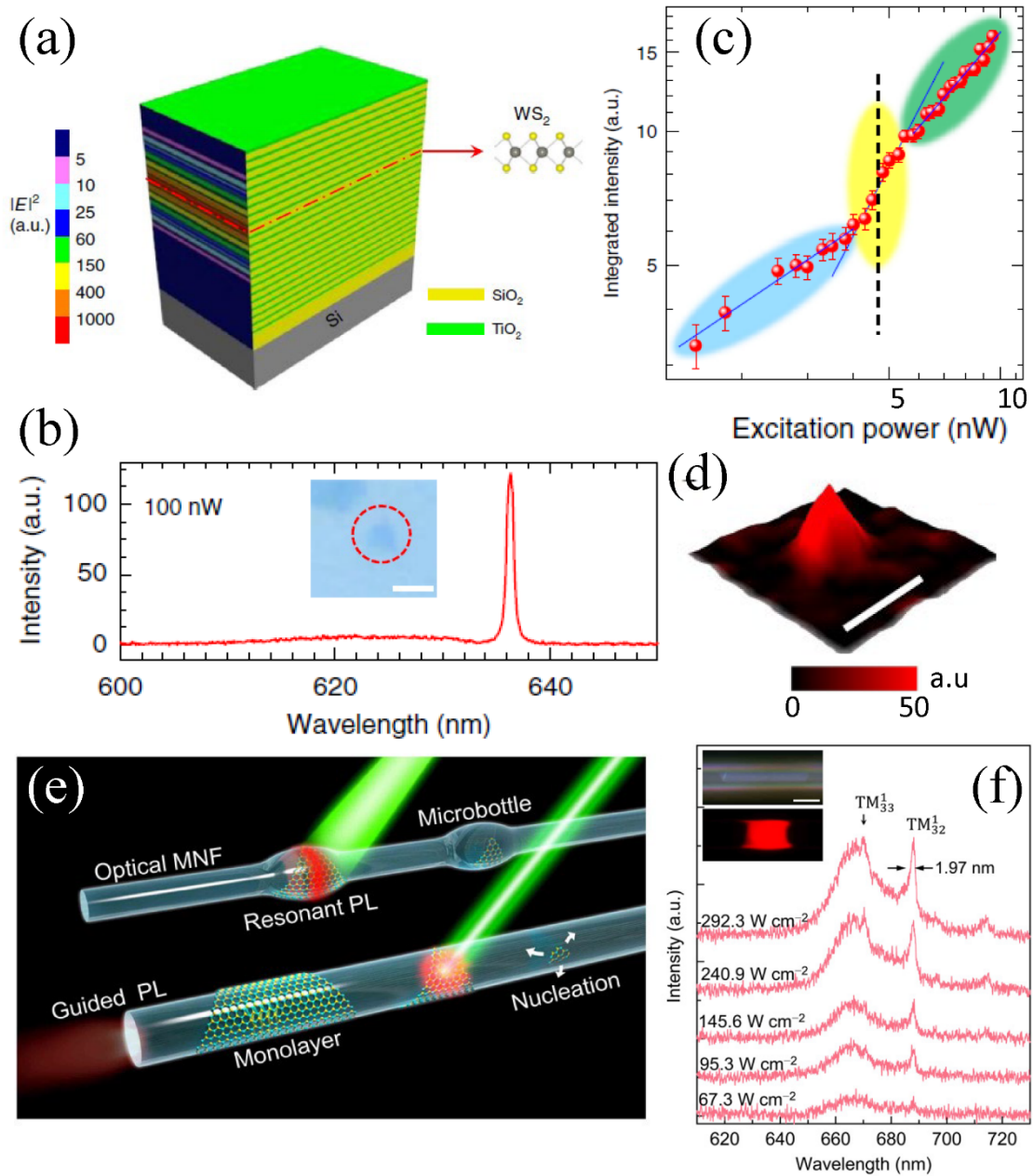

Figure 10. (a) Schematic drawing of laser devices, which consists of a 2D semiconductor embedded into DBR based microcavity. (b) The PL spectrum at the pump power of 100 nW. Inset shows the optical image of 1L WS$_2$ on the half cavity. (c) L-L curve of output PL intensity as a function of excitation power. (d) PL mapping of 1L-WS$_2$ on cavity. (e) Schematical illustration of 1L-MoS$_2$ grown on micro/nanofibers with micro bottle structures. (f) PL spectra of photoactivated MoS$_2$ monolayer/microfiber at different pump by the 532-nm CW laser. ((a)-(d) from Ref. [346],(e)-(f) from Ref. [350])

## 3.5 Light-emitting diode

A light-emitting diode (LED) is an essential optical component for optics interconnect and plays an vital role in applications like lighting, display, and sensor. Benefiting from the direct bandgap within the visible wavelength range, 1L-TMDC is viewed as a strong candidate for developing next-generation atomically thin LED. The first experiment on electroluminescence (EL) was demonstrated in the 1L-MoS$_2$ field-effect transistor by Sundaram et al. [353]. The EL is observed only in the region of contact because it occurs via the hot carrier process. However, such a device has relatively low efficiency. Jo et al. developed 1L-, and 2L-WS$_2$ LED based on ambipolar ionic liquid FET, which enables simultaneous electron and hole injection at opposite contacts to promote electron-hole recombination to emit the photons in the channel of FET[354]. It is well known that the p-n junction is a fundamental functional component of realizing efficient EL because electron and hole injection leads to electron-hole recombinations to emit photons. Three works independently reported the LEDs based on 1L-WSe$_2$ lateral p-n junction[54,65,355]. Shown in Fig.11a is the schematic drawing of devices. The electrically doping in WSe$_2$ can be separately controlled by two palladium gate electrode[65]. 10 nm hBN sheet is sandwiched between 1L-WSe$_2$ and gate electrode, acting as the smooth substrate to suppress the nonradiative decay and the high-quality gate dielectrics. When gating voltage $V_{g1}=-V_{g2}=8V$, the device functions as a p-n junction. Consequently, bright EL is observed in the WSe$_2$ area between two electrodes, as shown in Fig.11b. Nevertheless, EL only happens in the local area across the junction interface because complete depletion in p-n junction limits the efficient carrier injection to the entire monolayer region.

Later, Wither et al introduced single (or multi) quantum well (QW) based on van der Waals heterostructure to expand the effective EL area[356]. For example, a single quantum well is built by vertically stacking hBN/Graphene/hBN/WS$_2$/hBN/Graphene/hBN. Such a vertical stacking heterostructure allows for reduced contact resistance, higher current densities and luminescence from the whole WS$_2$ area. The quantum efficiency, defined as $\eta=Ne/I$(I is current, e is the electron charge, and N is the emitting photon number), is 1% for single QW and 5% for multi QW, which are respectively one order and two orders higher than that of the lateral p-n junction and Schottky barrier devices. Moreover, experiment results indicate that the external QE (EQE) increases with the decreasing temperature. In 2015, researchers from the same group demonstrated that the EQE shows abnormal behavior for WSe$_2$ based QW structure. The EQE increases with the rising temperature and can reach 5% at room temperature[104]. This is different from MoS$_2$ and MoSe$_2$, where the EL is reduced by a factor of 10-100 when the temperature is varied from 6K to 300K. The difference is attributed to the sign inversion of spin-obit splitting in conduction bands, resulting in the formation of dark exciton with the lowest energy. The WSe$_2$ based LED with high EQE holds great promise in real applications. The remaining challenge is to produce the multi QW devices with fined control of the hBN tunneling barrier on a large scale.

Lien et al have realized the large scale and bright electroluminescence in monolayer semiconductors by applying AC voltage between the gate and 1L[357]. The light emission is enabled by bipolar carrier injection at the transition mode. In addition to EL from 1L-TMDC, Li et al developed LED in multilayer MoS$_2$ based on two different heterojunctions GaN-Al$_2$O$_3$-MoS$_2$ and GaN-Al$_2$O$_3$-MoS$_2$-Graphene[358]. Compared to those mentioned above vertically stacked vdWs heterostructure, thicker MoS$_2$ could prevent the rapid carrier leakage and promote the efficient recombinations of injected electron and hole pairs. Although multilayer MoS$_2$ has an indirect bandgap, the observed EL efficiency is comparable to or even higher than that of the monolayer. This surprising finding is explained by the carrier redistribution from the lowest energy state to a higher energy state under strong vertical electric field's pumping.

Since the emission is strongly dependent on electromagnetic environments, EL can also be engineered by optical microcavity. Liu et al. demonstrated the cavity-enhanced EL in van der Waals heterostructure Graphene/hBN/WSe$_2$/hBN/Graphene by integrating them onto the photonic crystal cavity[359](Fig.11c-f). Unlike the polarization-independent EL from the continuous film, the emission is enhanced by a factor of four at cavity resonance and is highly polarized along the L3-cavity direction.

By embedding vdW heterostructure Graphene/hBN/WS$_2$/hBN/Graphene into monolithic microcavity made from the Ag mirror on the top and DBR mirror on the bottom (See Fig.11g), Gu et al. realized room temperature polariton LED in 1L-WS$_2$[360]. The combination of vertically stacked heterostructure and microcavity allows us to tune the EL peaks by tuning the excitation angle when electron and hole carriers are injected into 1L-WS$_2$. Strong coupled between exciton and cavity mode is confirmed by both PL and EL measurements (Fig.11h), leading to the formation of exciton-polariton. The Rabi splitting retrieved from EL measurement is 34 meV. Additionally, EL is highly directional due to the cavity dispersion. The emission direction can be controlled by tuning the cavity resonance. The extracted EQE is around 0.1% and may be further improved by employing a cavity with a higher Q-factor. Another work on monolithic microcavity enhanced EL was reported by Pozo-Zamudio etal[361]. The high-Q factor of two cavity modes enhances EL peaky intensity by a factor between 35 and 100. Similar to the previous work, the spectrum of cavity-induced EL can be tuned by incident angle.

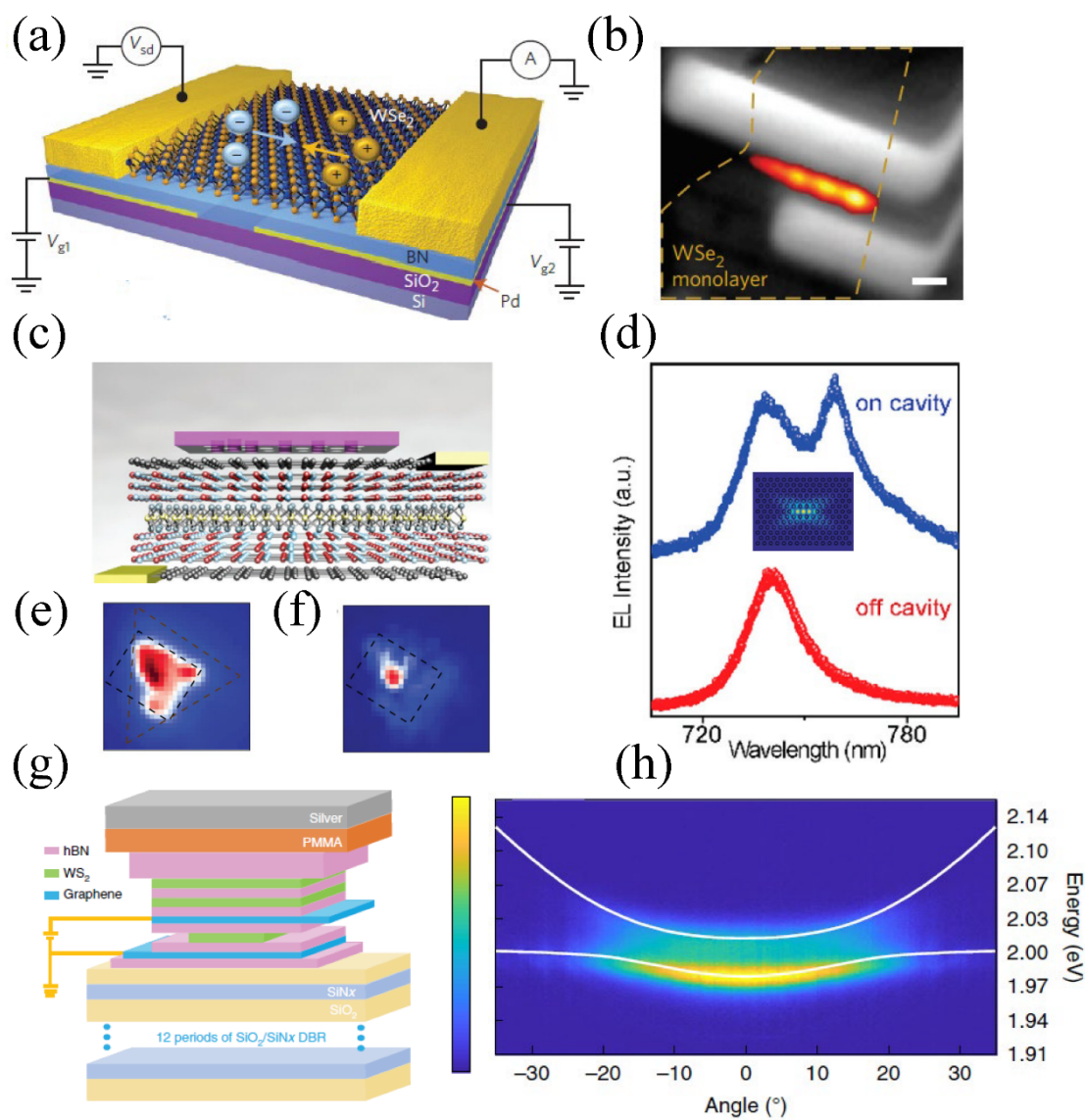

Figure 11. (a) Cartoon of 1L-WSe$_2$ p-n junction devices. (b) Electroluminescence mapping superimposed on the device's optical image, where a dashed line highlights the monolayer region. (c) Schematics of device architecture. (d) Electroluminescence of 1L-WSe$_2$ on (red) and off (blue) cavity area. (e-f) Spatially 2D mapping of cavity-enhanced electroluminescence without (e) and with polarization (f) at $V_b$=2.6V. (g) Schematic drawing of polariton LED devices. (h) Angle-resolved EL mapping under carrier injection with a density of 0.28uA/cm$^2$. ((a)-(b) from Ref. [65], (c)-(d) from Ref. [359], (e)-(f) from Ref. [360])

### 3.6 Nonlinear harmonic generation

When TMDC materials are thinned down to the monolayer limit, we do not only observe a transition to direct bandgap but also the emergence of a large second-order optical nonlinearity due to inversion symmetry broken, which may render TMDCs ideal candidates for developing the wavelength-tunable coherent light source[362–364]. For a few-layer TMDC material, which is schematically shown in Fig.12a, the symmetry properties are different for even and odd layer thicknesses, where the former belongs to the $D_{3d}^3$ space group and the latter belongs to the noncentrosymmetric $D_{3h}^1$ space group. The broken inversion symmetry in 1L-MoS$_2$ is confirmed by three orders of magnitude enhancement of optical second harmonic generation (SHG) comparing to the even layer and bulk MoS$_2$[365–367], as displayed in Fig.12b. Due to $D_{3h}$ symmetry, the SHG electric field is proportional to $\cos(3\theta+\theta_0)$ or $\sin(3\theta+\theta_0)$ for parallel or perpendicular polarization of analyzer, where $\theta$ is the angle between laser polarization and mirror plane and $\theta_0$ is the initial crystal orientation. Then, the SHG intensity follows the six-fold symmetry as a function of angle $\theta$ (Fig.12c). The initial crystal orientation can be easily obtained as $\theta_0 = 0$.

SHG may serve as an effective and noninvasive method of detecting the crystal orientation, thickness uniformity and layer stacking. Kumar et al. successfully applied this technology to explore the crystal orientation of CVD-grown MoS$_2$[367]. Note that the above SHG measurement was conducted by a pulse laser with a fixed center wavelength. Later, Wang et al demonstrated that SHG intensity strongly depends on the laser excitation energy[368]. When the two phonon energy of laser matches the exciton resonance, SHG efficiency is enhanced by three order of magnitude owing to unusual electric and magnetic dipole transitions. Seyler et al also found that second-order optical nonlinearities of 1L-WSe$_2$ can be modulated by external gating in a field-effect transistor[369]. The electrostatic doping leads to a tunability by order of magnitude for SHG intensity at A-exciton resonance at 30K, which arises from the strong exciton charging effect in 1L-WSe$_2$. It is also necessary to mention that second-order nonlinear optics is not just limited to the flake of 1L-TMDC. Yin et al. reported the observation of 1D nonlinear optics edge states in 1L-MoS$_2$[370], which is correlated to the broken translation symmetry at the edge of 2D crystals. This finding allows researchers to image the atomic edge and boundaries. Except for SHG in 1L-TMDC, there are the third harmonic and fourth harmonic generations in 1L-MoS$_2$, as reported by Saynatjoki et al. [371].

Despite large second-order nonlinearity in 1L-TMDC, the total nonlinear conversion efficiency is limited by the subnanometer thickness, which remarkably shortens the light-matter interaction lengths. It is well established that light-matter interaction can be significantly enhanced when 2D crystals are embedded into an optical cavity environment. Therefore, various resonant photonic structures have been utilized to resolve the weak light-matter interaction strength. For example, Day et al. reported an order enhancement of SHG intensity in 1L-MoS$_2$ when it is embedded into the DBR based microcavity which supports an optical resonance matching the pumping wavelength[372]. The enhancement factor is further improved to 3,300 when 1L-MoS$_2$ is integrated with an optical cavity[373], which supports double resonance at both fundamental and second harmonic frequencies. Shown in Fig.12d are schematics of the optomechanical system which enables a voltage tunable Fabry-Perot cavity. The nonlinear light-matter interaction can be dynamically engineered by the Fabry-Perot cavity length by applying a voltage between the top mirror and indium tin oxide sitting on the DBR mirror. Carefully tuning the voltage allows two resonances to happen at both pumping wavelength and second harmonic wavelength, thus maximizing the SHG intensity (Fig.12e). Moreover, such an optomechanical system can be switched on/off by merely tuning the voltage (Fig.12f).

The second type of resonant structure of interest for this purpose is noble metal-based plasmonic metasurface. By exploiting the efficient in-plane coupling between surface plasmon resonator and 1L-WS$_2$, Shi et al realized an enhanced SHG by a factor of 400 in a hybrid WS$_2$-Ag grating system[374] (Fig.12g-i). Such a giant enhancement factor is related to two factors: (1) strong in-plane electric field within the V-shaped nanogroove resonator at both fundamental and second harmonic frequencies; (2) plasmonic resonance aligns with the C-exciton resonance, as shown in Fig.12h. Wang et al report almost 7000 fold SHG enhancement in WSe$_2$ by transferring monolayer onto the Au thin film with sub-20 nm-wide trench, as compared to SHG of WSe$_2$ on sapphire substrates[375]. The giant enhancement factor is attributed to the lateral gap plasmon resonance at pumping wavelength, ensuring the electric field's confinement within the trenches.

Other than plasmonic metasurfaces, a single Ag cube on a dielectric layer Al$_2$O$_3$ with Ag film underneath[376] has been proposed to enhance the SHG by Han et al by 300 folds when 1L-WS$_2$ is sandwiched between Ag cube and a dielectric spacer layer. The augmented SHG is enabled by tuning the surface plasmon resonance to SHG wavelength, which guarantees the extreme field confinement with the tiny gap between cube and mirror.

The third type of frequently used resonant photonic structures is the dielectric microcavity and dielectric metasurface. Li et al demonstrated a 20 fold enhancement of optical SHG in 1L-WS$_2$ with SiO$_2$ dielectric microsphere dispersing on top[377]. Simulation results indicate that strong electric field locates at the interface and thus results in improved SHG. Chen et al's integration of 1L-MoSe$_2$ with silicon metagrating leads to the five-fold enhancement of SHG intensity due to the increased interaction length through coupling evanescent field with monolayer[378]. The enhancement factor can be improved to more than 1000 by incorporating 1L-WS$_2$ into silicon metasurface which supports high-Q mode (also called as quasi BIC) at fundamental frequency[379], as demonstrated by Bernhardt et al. Guo et al also explore the guided surface wave to realize the efficient frequency mixing on 1L-WS$_2$[380]. The conversion efficiency increased by three orders in counter-propagating excitation configuration while the SHG signal is highly collimated with controlled emission direction and polarization.

Despite weak second-order nonlinearity due to inversion symmetry in bilayer TMDCs, strong SHG in bilayer WSe$_2$ is realized by a back gate field and can only be enabled by charge accumulation instead of charge depletion at different gating bias[381]. Wen et al. reported SHG in bilayer WSe$_2$ by plasmonic hot carrier injection that breaks the inversion symmetry and induces second-order nonlinearity[382]. Pump-probe experiments reveal that the hot carrier injection but not optical field enhancement is the main factor causing the generation of SHG by inducing the separation electric field. Transient absorption measurement signifies that charge transfer happens within less than 2ps.

Nonlinear harmonic generation also carries the valley information. Unlike the linear case, SHG with left (or right) circular polarization can only be excited by the excitation with the same-handedness due to the two-photon optical selection rules[369,383]. Hu et al demonstrated that such nonlinear valley-locked chiral emission can be well controlled by a synthetic Au-WS$_2$ metasurface[384]. Also, nonlinear wavefront can be effectively engineered with monolayer TMDC. For example, Dasgupta et al. reported the generation of optical vortex beam, Airy beam, holograms and Hermite–Gaussian beams at the second harmonic frequency with patterned 1L-WS$_2$[385,386].

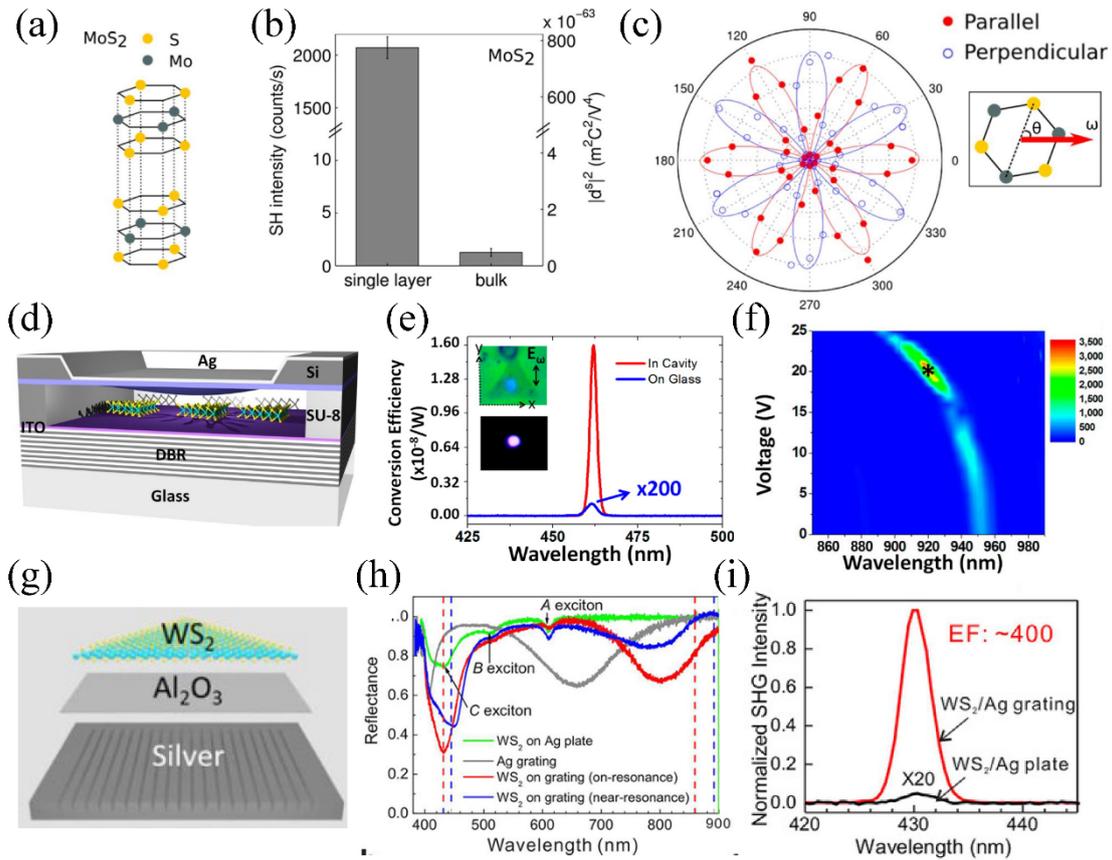

Figure 12. (a) Schematics of 2H-stacked $MoS_2$ with repeating two-layer units. (b) SHG intensity of 1L- and bulk $MoS_2$. (c) SHG intensity as a function of the azimuthal angle θ of monolayer crystal. The right panel shows the definition of a crystal's azimuthal angle. (d) Schematics of device architecture. (e) SHG spectrum from 1L-$MoS_2$ on and off cavity region. (f) Enhanced SHG intensity factor from cavity region compared to $MoS_2$ on a glass substrate as a function of voltage and wavelength. (g) Schematics of the $WS_2$–plasmonic hybrid metasurface structure for enhanced SHG. (h) Measured reflectance spectrum from the Ag grating structure with and without $WS_2$ on top. (i) Normalized SHG from 1L-$WS_2$ on Ag film and Ag grating, respectively. ((a)-(c) from Ref. [365], (d)-(f) from Ref. [373], (g)-(i) from Ref. [374])

## 4. Strong Light-Matter Coupling between a TMDC Monolayer and a Nanocavity

In the previous section, the light-matter coupling between 1L-TMDC and different cavity types (i.e. photonic crystal cavity) has been limited to the weak coupling regime, for which the coupling rate between cavity photons and induced dipoles is slower than their average dissipation rate. In the weak coupling regime, the spontaneous emission rate can be modified via the Purcell effect, where the local photon density of states is amplified due to the high-Q cavity modes with small mode volume. When the coherent energy exchange rate is faster than the dissipation rate of both quantum emitter and cavity mode, it enters the strong coupling regime[185,387]. The direct consequence of strong coupling is the formation of exciton-polaritons (part-matter/part-particle) and Rabi splitting. The unique features of exciton-polaritons allow for studying fundamental problems in physics (i.e., Bose-Einstein condensation and superfluid) and promise many exciting applications, such as low threshold polariton lasing. As an essential branch of solid-state cavity electrodynamics, exciton-polariton was first experimentally demonstrated in GaAs-quantum well microcavity systems[388] and later was generalized to other materials, such as GaN, ZnO and organic materials.

Nevertheless, exciton-polaritons in these materials may only be observed at cryogenic temperature due to the relatively small binding energy of exciton. Recently, 1L-TMDC has been regarded as an excellent platform for studying strong light-matter coupling due to its exceptional electronic and optical properties[10,389]. Firstly, it has considerable exciton binding energy because of the quantum confinement and reduced dielectric screening at reduced dimensions. Secondly, it has a uniform surface free of dangling bonds that removes the restriction of lattice matching. This part briefly reviews the effort made in implementing the strong coupling in 1L-TMDC or few layers.

**4.1 Strong coupling by DBR based microcavity**

The key ingredient to realizing strong coupling is to integrate an optical microcavity within 1L-TMDC. The simplest and most typical optical microcavity is the 1D photonic crystal cavity, where a defect is introduced in the DBR. The first experimental demonstration of strong coupling in 1L-$MoS_2$ was done by Liu et al [390], where the 1L-$MoS_2$ is sandwiched inside the DBR based microcavity with Q-factor around 200. The Rabi splitting was 46 ±3 meV and coupling strength was 25±2 meV, obtaining from angle-resolved reflection spectroscopy and PL spectrum. The criterion of strong coupling is satisfied by comparing the Rabi splitting and average line width of exciton and cavity resonance. Later, strong coupling was demonstrated in the 1L-$WS_2$-microcavity system across a broad temperature range[161]. Fig.13a shows the schematical drawing of such a hybrid system. Coherent coupling exciton-polariton was verified by a large Rabi splitting 40 meV and splitting-to-linewidth ratio (>3.3) at 110K, retrieving from angle-resolved reflection spectrum and PL spectrum as shown in Fig.13b-c. Moreover, the polariton composition (percentage of exciton and photon) can be dynamically modulated by temperature, i.e. switching the low branch's polariton from more photon-like to more exciton-like when temperature increases from 130K to 230K.

Dufferwiel et al. reported the observation of exciton-polariton in single quantum well $MoSe_2$/hBN and double quantum well $MoSe_2$/hBN/$MoSe_2$/hBN embedded in a tunable optical microcavities[391], as shown in Fig.13d. Instead of controlling the cavity resonance by incident angle, the cavity resonance in such an open cavity is externally tuned by piezo voltages to match the exciton energy. Fig.13e shows the PL spectrum of $MoSe_2$/hBN. By fitting the PL spectrum, the upper and lower polariton branches are obtained. A clear anticrossing can be found between two branches with Rabi splitting 20 meV. When double quantum well was incorporated into such tunable open cavity, the Rabi splitting was further enhanced to 29 meV. It is interesting to point out that weak coupling is induced between negative charged exciton and cavity mode with coupling strength 8.2 meV.

In fact, the DBR based microcavity can simultaneously interact strongly with neutral exciton and negative charge exciton, as demonstrated by Dhara et al. [392]. Anomalous band inversion for the lower polariton branch (trion polariton) was observed. This was explained by electron-mediated interaction between the exciton-polariton branch and trion-polariton branch that is much stronger than trion-cavity interaction, leading to the effective level attraction. Silder et al. demonstrated active control over the weak coupling and strong coupling between cavity mode and exciton and negatively charged exciton by external gating[393]. Repulse and attractive exciton-polaron resonances were simultaneously observed.

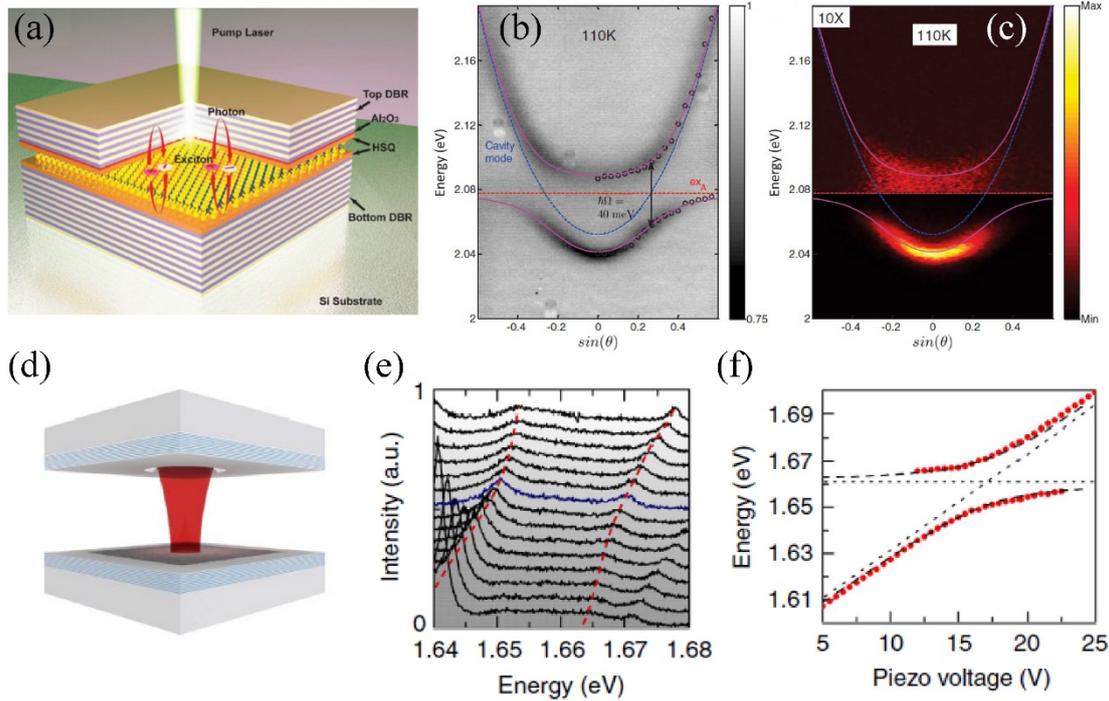

Figure 13. (a) Schematic of WS$_2$-microcavity structure. (b) Angle-resolved reflectivity mapping at 110K. (c) Angle resolved PL mapping at 110K. (d) Schematic illustration of 1L-MoSe$_2$ embedded into a tunable hemispherical cavity. (e) PL spectra of 1L-MoSe$_2$ for exciton–photon detunings varied from Δ=16 meV (bottom) to Δ=-12 meV (top). (f) The upper and lower polariton branches with Rabi splitting being 20 meV. ((a)-(c) from Ref. [161],(d)-(f) from Ref. [391].)

**4.2 Strong coupling with plasmonic nanocavity**

Apart from microcavities based on DBR, metallic nanocavities can serve as an alternative candidate to realize strong plasmon-exciton coupling since they sustain the surface plasmons that can confine the electric field below the diffraction limit. As a typical plasmonic nanocavity, two independent groups have employed a single metallic nanorod to successfully demonstrate the strong coupling between localized plasmonic resonance and exciton in 1L-TMDC.. For instance, Zheng et al realize strong plasmon-exciton coupling in WSe$_2$-Ag nanorod system[394], as shown in Fig.14a. In this case, the plasmonic resonance with field distribution shown in Fig.14c is tailored to cross the exciton peak of 1L-WSe$_2$ by deposition of ultrathin alumina with different thicknesses. Mode splitting was observed in dark-field scattering spectra (Fig.14b), and anticrossing can be found by extracting the information of two resonance (Fig.14d). Following the simple coupled oscillator model, the Rabi Splitting was retrieved as 49.5 meV, larger than the average dissipation rate of WSe$_2$ exciton (43 meV) and plasmon resonance (98 meV), satisfying the criterion of strong coupling. In another independent work by Wen et al, a single Au nanorod was used to achieve the coherent plasmon-exciton coupling in 1L-WS$_2$[224]. Unlike the Ag nanorod, the plasmonic resonance is tuned by changing the aspect ratio of the Au nanorod. The Rabi splitting in such a system is high to 106 meV, also falling within the strong coupling regime. Further theoretical analysis indicated that only a small number of the exciton (N=5 ~18) was involved in the strong coupling. Moreover, they also demonstrate active control of strong coupling via temperature with Rabi splitting 110 meV.

Interestingly, strong coupling between a plasmonic nanocavity and 1L-TMDC not only can induce the generation of plasmon-exciton but also lead to the formation of plasmon-trion. Cuadra et al reported the experimental observation of strong coupling between localized surface plasmon and exciton and trion in hybrid 1L-WS$_2$-Ag nanoantenna systems[66]. From dark-field scattering spectra shown in Fig.14e, it can be found that strong interaction only occurs between plasmonic resonance and exciton at room temperature, yielding the Rabi splitting 120 meV and the formation of plasmon-exciton (Fig.14f). When the temperature is cooled down to 77K or even lower, surface plasmon interacts with both exciton and trion, leading to intermixed plasmon-exciton-trion state (Fig.14g). The Rabi splittings between the up and middle branch, middle and bottom branches are 81 meV and 77 meV (Fig.14h-i), respectively. Both of them meet the criterion of strong coupling.

It is worth noting that a single metallic nanorod is not the only candidate for realizing strong coupling. Researchers have developed strong coupling with an array of plasmonic nanostructure and 1L-TMDC. For instance, Wang et al studied strong plasmon-exciton coupling in 1L-WS$_2$ with two different types of plasmonic nanocavity-metallic Fabry-Perot cavity and plasmonic nanohole arrays[395] (Fig.14j-k). The principle of achieving strong coupling, however, is fundamentally different. While the former metallic resonator relies on two mirrors made from Ag thin film and 1L-WS$_2$ is put in the maximum of the electric field, the latter can support surface plasmonic resonance, which can be tuned to A exciton by simply adjusting the period. Anticrossing can be easily found between two branches, which is obtained from angle-resolved transmission (or reflection) spectrum (Fig.14k). The coupled oscillatory model gives the Rabi splitting 101 meV and 60 meV, respectively. When 1L-MoS$_2$ is transferred to the Ag NPs array, strong exciton-plasmon coupling occurs between the exciton of MoS$_2$ and localized surface plasmon resonance, plasmonic lattice resonance[396]. The strong coupling was confirmed by angle-resolved reflection spectroscopy with Rabi splitting 58 meV at 77K. It was also demonstrated that the exciton-plasmon coupling strength and dispersion could be effectively modulated by geometry parameters of plasmonic lattices, as shown in Fig.14l-o. Apart from plasmonic nanocavity above, Deng et al proposed a new strategy of realizing strong coupling between exciton in 1L-WS$_2$ and surface plasmon polariton propagating on the Au surface via Kretschmann-Raether configuration and local surface plasmon resonance[397]. The Rabi splitting can even exceed 120 meV for a complex system, where 1L-WS$_2$ is sandwiched between Au film and Ga nanosphere.

Although most published works focus on strong coupling in 1L-TMDC, there is still some work studying the strong coupling in multilayer TMDC materials. Despite weak emission in TMDC few layers (>1), the exciton feature is still prominent in either the absorption or reflection spectrum. For example, Kleeman et al report the strong coupling in few-layers WS$_2$ when it was integrated with the plasmonic nanocavity[398]. Unlike the metallic resonators discussed previously, such a plasmonic nanocavity made of Au NPs on Au thin film could only induce the strong near field confinement of E$_z$ component perpendicular to the plane of the thin film. Since the exciton's dipole moment for 1L-TMDC is located in the layer plane, cavity mode in such a structure cannot couple effectively to exciton resonance in monolayer. However, when the layer number increases, out-of-plane exciton dipole increases up to 25% of in-plane exciton dipole, allowing for strong coupling between WS$_2$ multilayer (>7) and cavity modes with Rabi splitting >140 meV.

Another example of strong coupling in a few layers WS$_2$ was uncovered by Liu et al. [399]. Ag NP array was fabricated onto the few-layer WS$_2$. Except for normal exciton-polariton branches arising from strong coupling, a new collective polariton mode emerges and is accompanied by a complete polariton gap. These collective modes were attributed to excitons' cooperative coupling to the diffractive modes in plasmonic lattices. Furthermore, active control was realized by integrating a hybrid 1L-TMDC-plasmonic lattice system with the field-effect transistor[399,400]. It has been found that the polariton gap can be turned on and off via external gating voltages. Besides, nanostructured TMDC multilayer can act as a Mie resonator due to its high refractive index and low loss. In more recent work, Zhang et al observed the strong coupling in a hybrid system that consists of patterned $WS_2$ multilayer with thickness 20 nm on Au substrate[401]. In such a heterostructure, the exciton of multilayer $WS_2$ interacts strongly with the Mie resonance of dielectric grating and plasmonic mode in dielectric-Au systems. The Rabi splitting between the upper and lower polariton branch is up to 410mV.

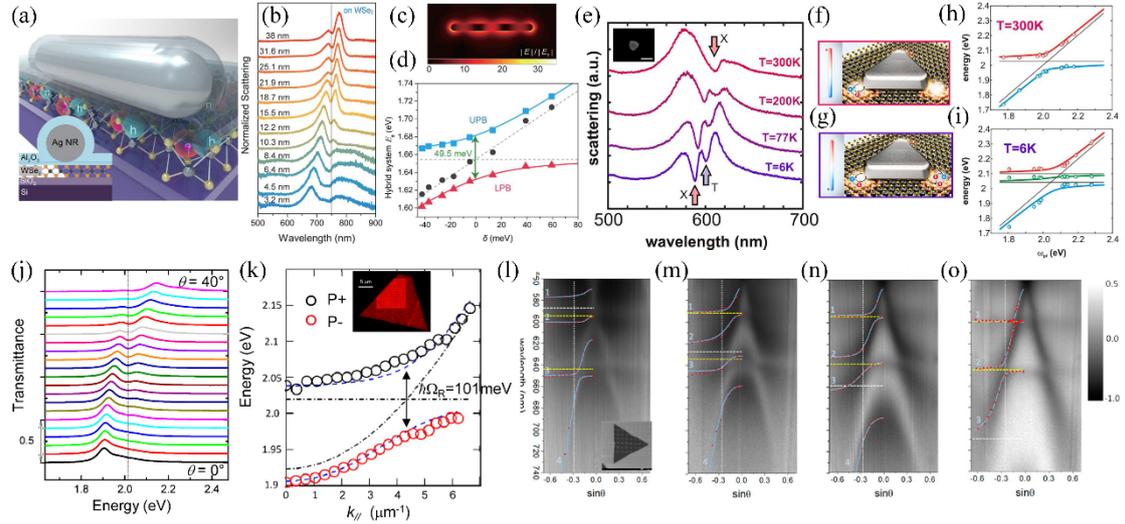

Figure 14. (a) Schematic diagram of structure architecture, which consists of a single Ag nanorod on a 1L-$WSe_2$. (b) Measured scattering spectra of 1L-$WSe_2$ with Ag nanorod on top. (c) Calculated electric field distribution for plasmonic mode (n=3) in Ag nanorod. (d) Extracted dispersion relation from scattering spectrum. (e) Measured scattering spectra at T = 300, 200, 77, and 6 K by darkfield microscope. (f-g) Cartoon depiction of plasmon-exciton at 300K and plasmon-exciton-trion at 6K. (h) Up and low polariton energies versus plasmonic resonance extracted from dark-field scattering spectrum at 300K. (i) Upper, middle, and lower polariton energies versus plasmonic resonance extracted from dark-field scattering spectrum at 6K. (j) Angle-resolved transmission spectra for 1L-$WS_2$ integrated with FP cavity. (k) Polariton dispersion extracted from transmission spectra. (l-o) Angle-resolved differential reflectance spectra of Ag nanodisk arrays with diameters of 70 nm, 100 nm, 120 nm, 150 nm on 1L-$MoS_2$. ((a)-(d) from Ref. [394], (e)-(i) from Ref. [66], (j)-(k) from Ref. [395], (l)-(o) from Ref. [396])

## 4.3 Strong coupling by dielectric nanocavity

Although plasmonic nanocavities enable subwavelength nanoscale field confinement, they are also associated with considerable intrinsic loss that results in heat generation, which may deteriorate exciton-polariton-based devices' performance. An alternative candidate for realizing strong coupling is to take advantage of Mie resonances or guided-mode resonances supported by a dielectric nanocavity, which also allows for subwavelength field confinement. Compared to metallic nanoresonators that only support low-order (i.e., electric dipolar or quadrupolar resonance) plasmonic resonances, high-index semiconductors support multipolar Mie resonances, which provide unprecedented flexibility in manipulating light at the nanoscale.

The simplest example of strong coupling between exciton and a cavity mode supported by a dielectric nanocavity is based on a Si nanosphere integrated with 1L-TMDC [269], as shown in the inset of Fig.15a. Resonant coupling was demonstrated between the exciton of 1L-$WS_2$ and magnetic dipole mode of Si NP [269]. Fig.15a shows the scattering cross-section of single Si NP and heterostructure $WS_2$-Si NP and absorption spectrum of the $WS_2$ shell structure. The magnetic dipole mode interacts strongly with the exciton, leading to the mode splitting in the scattering spectrum. By varying the diameters of Si NP, the magnetic resonance can be tuned to match the exciton resonance (see Fig.15b). After putting 1L-WS$_2$ shell and Si NP, anticrossing with mode splitting 43 meV can be clearly seen from the scattering mapping shown in Fig.15c. In a similar work by Lepeshov et al., they demonstrate the strong coupling of 1LWS$_2$-Si NP with Rabi splitting 110 meV[83]. The Rabi splitting can be further enhanced to 208 meV if such heterostructure is embedded into water. Experimental results show that the mode splittings are 49.6 meV and 86.6 meV when the background medium is air and water.

From a practical application perspective, it might be more appealing to transfer 1L-TMDC onto an array of dielectric nanoresonators[402–406], such as dielectric metagrating or metasurface. For example, Zhang et al. experimental realized the strong coupling between exciton in WS$_2$ and WSe$_2$ and guided-mode resonance in dielectric grating[402] (Fig.15d-f). Such a simple grating structure enables relatively high-Q resonance (≈270). The strong interaction between exciton and guided-mode resonance leads to Rabi splitting 17.6 meV and 22 meV for WS$_2$ and WSe$_2$, respectively, retrieved from both angle-resolved reflection spectrum and PL spectrum (Fig.15e-f). More recently, Chen et al. demonstrated exciton-polariton by strong coupling of exciton in 1L-TMDC to the guided mode resonance in Si$_3$N$_4$ metasurface with Rabi splitting 18 meV[403]. They also found that the Rabi splitting, dispersion, and far-field emission property of exciton-polariton can be tailored by engineering the pattern of optical meta-atoms. In fact, recent theoretical work demonstrates that Rabi splitting may be improved by putting the 1L-TMDC in the center of dielectric grating[406].

Since the key of realizing strong coupling is to utilize an optical resonance with high Q-factor and strong field confinement, optical bound states in the continuum provide an alternative way of realizing enhanced light-matter interaction[389]. Note that the bound state in the continuum corresponds to the leaky modes with infinitely large Q factor. The strong coupling can be achieved by a quasi bound state in the continuum[407,408]. Kravtsov et al. applied the concept of bound states in the continuum to engineer strong coupling in 1L-MoSe$_2$[407]. Bound state in the continuum was realized through a photonic crystal slab (Fig.15g). Due to its high-Q property (Q≈900), strong coupling was achieved with a large Rabi splitting 27 meV (Fig.15h-i), leading to a smaller polariton linewidth <3 meV and a large splitting-to-linewidth up to 9. The narrow polariton linewidth allows for studying the exciton-exciton interaction with the strength g≈1.0 μeVμm$^2$.

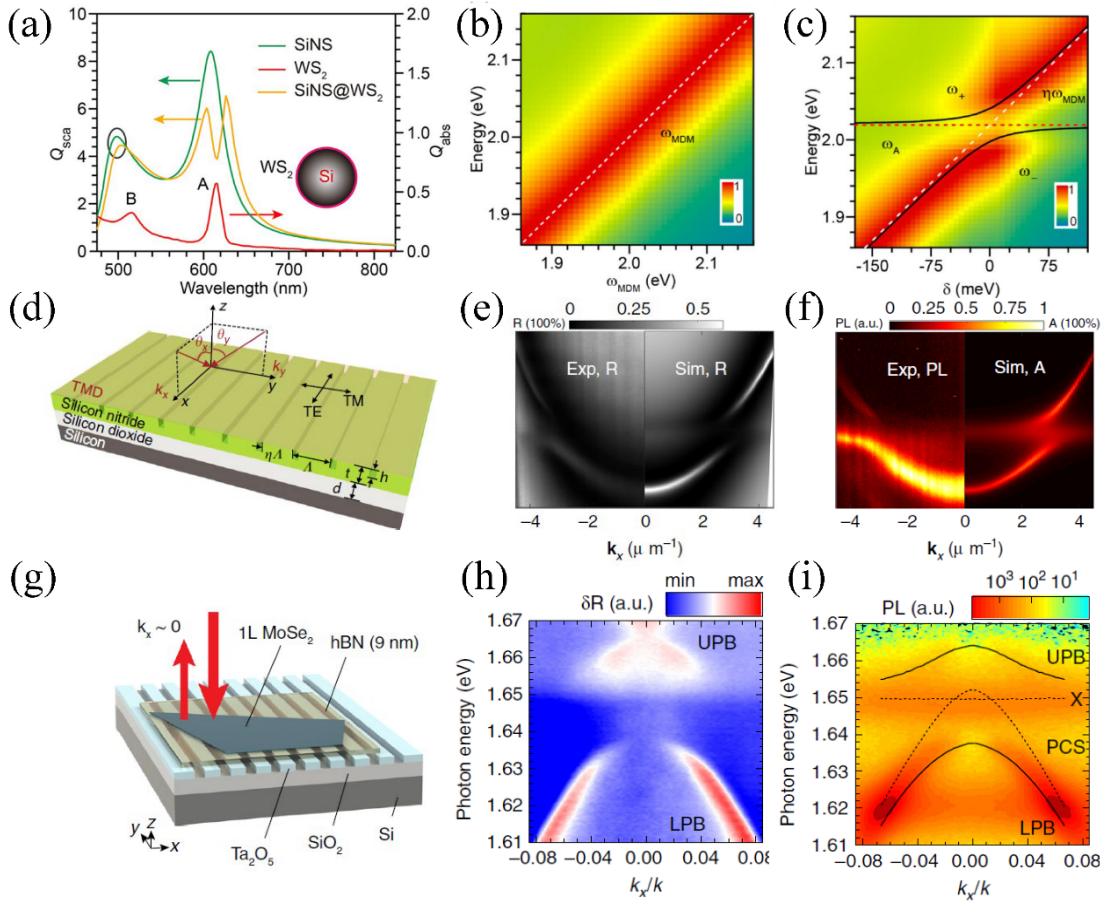

Figure 15. (a) Calculated scattering and absorption spectrum for pure Si sphere, $WS_2$ film, and $WS_2$-Si sphere. (b) Normalized scattering mapping of a silicon nanosphere as a function of the photon energy of magnetic dipole mode and incident wavelength. (c) Normalized scattering mapping of heterostructure as a function of the photon energy and incident wavelength. (d) Schematic of the device structure consists of 1L-TMDC on top of $Si_3N_4$ grating on $SiO_2$/Si substrate. (e-f) Simulated and measured reflection (e) and PL(f) mapping as a function of in-plane wavenumber and wavelength. (g) Schematic diagram of device architecture consisting of van der Waals heterostructure hBN/1L-$MoSe_2$ on $Ta_2O_5$ grating/SiO2/Si substrate. (h-i) Measured differential reflection (h) and PL (i) mapping as a function of incidence photon energy and in-plane wavenumber $k_x$. ((a)-(c) from Ref. [269], (d)-(f) from Ref. [402], (g)-(i) from Ref. [407].)

## 5. Valleytronics in TMDCs Monolayers

The broken inversion symmetry of 2D TMDCs hexagonal crystalline structure and the strong spin-orbit coupling leads to in equivalent valleys K and K' with antisymmetric spin states at the corners of the Brillouin zone[205]. To distinguish valley states, one can introduce the Berry curvature, $\Omega_n(k)$, which describes the geometric properties of the electronic bands and their topology-related effects. The Berry curvature [$\Omega_n(k)$] at K and K' can have opposite signs due to the broken inversion symmetry. In the k-space, the Berry curvature acts as an *effective magnetic field,* and in the semiclassical equations of motion gives rise to the anomalous velocity [$\dot{\mathbf{k}} \times \Omega_n(k)$]:

$$\dot{\mathbf{r}} = \frac{1}{\hbar} \frac{\partial E_n(k)}{\partial \mathbf{k}} - \dot{\mathbf{k}} \times \Omega_n(k), \quad (21)$$

$$\hbar \dot{\mathbf{k}} = -e\mathbf{E} - e\dot{\mathbf{r}} \times \mathbf{B}, \quad (22)$$

and the Berry curvature $\mathbf{\Omega}_n(k)$ can be defined in terms of the Bloch functions, $\mathbf{\Omega}_n(k) = \nabla_k \times \mathbf{A}_n(k)$, where $\mathbf{A}_n(k) = i\int u_n^*(r,k)\nabla_k u_n(r,k)d^3r$ is the Berry connection, and $u_n(r,k)$ stands for the periodic part of the Bloch electron wavefunction in the n-th energy band[409]. $E_n(k)$ is the n-th band energy dispersion, $\mathbf{E}$ and $\mathbf{B}$ are external electric and magnetic fields.

The motion Eqs. (21) and (22) must remain invariant under the system symmetry. Therefore, the system possesses the time-reversal then $\mathbf{\Omega}_n(k) = -\mathbf{\Omega}_n(-k)$, since the Berry curvature has the quasi-vectorial nature. If the system would also have the inversion symmetry, $\mathbf{\Omega}_n(k) = \mathbf{\Omega}_n(-k)$, then we would conclude $\mathbf{\Omega}_n(k) = 0$. Thus, the valley-contrasting phenomena can appear only in systems with broken inversion symmetry as we have in single-layer TMDCs. For example, graphene possesses the inversion symmetry and hence has the valleys of the same Berry curvature that washes away its valley polarization and allows having nonzero valley currents only at edges[410], defect lines[411], and strain[412]. However, in ferromagnetic graphene (or, more generally, in 2D Dirac materials), these effects are possible[413]. As another example, in 2L-TMDCs, the inversion symmetry is restored, and the valley-polarization effects (and second-order nonlinearity) vanish.

Substituting (22) to (21), it can be seen that if an in-plane electric field ($\mathbf{E}$) is applied to the single-layer TMDC, then a nonzero Berry curvature leads to an *anomalous electron velocity perpendicular to the field*. Thus, the velocity of electrons that belong to opposite valleys have an opposite sign and move to different directions, effectively locking the spin and valley degrees of freedom (pseudospin)[125,126,414]. This fascinating property gives rise to various effects, including valley-selective photoexcitation[124–128], valley Hall effect[129,130], valley-tunable magnetic moment[131], and valley-selective optical Stark effect[132,133].

All these features also relate to excitons, which inherit the valley polarization from the electrons and holes they are composed of. If allowed by the selection rules discussed above, these valley polarized excitons' excitation can be realized with $\sigma^-$ and $\sigma^+$ light[125,126]. For example, after $\sigma^+$ light excitation, the 2D TMDC acquires the K valley excitons also polarized clockwise, and this valley polarization with be maintained for a while. This behavior contrasts with that of III-V or II-VI quantum wells, where excitation with the circularly polarized light usually results only in the spin polarization of the charge carriers. The valley polarization time significantly depends on temperature, material, and other factors. The annihilation of the valley polarized exciton (or valley polarized electron and hole) is accompanied by a circularly polarized photon's emission. A system with valley polarized particles demonstrates optical activity because they interact differently with photons of opposite polarization.

The degree of the valley polarization can be quantitatively described by the valley polarization contrast $\eta$ defined as[127,128,415]

$$\eta = \frac{|PL(\sigma^+) - PL(\sigma^-)|}{PL(\sigma^+) + PL(\sigma^-)} \quad (23)$$

where $PL(\sigma^+)$ and $PL(\sigma^-)$ are PL intensity of right and left circularly polarized light, respectively.

The dynamics of exciton valley polarization in 2D WSe$_2$ has been experimentally investigated in Ref.[416] by the mean of pump-probe Kerr rotation technique. The reduction of the valley depolarization time from ~6 ps to ~1.5 ps upon the temperature increase from 4 K up to 125 K is demonstrated. These results have been backed up by *ab initio* simulations, which took into account various valley depolarization mechanism[417]. Remarkably, the valley polarization lifetimes are orders of magnitude longer for dark excitons[415,418], interlayer excitons in TMDC heterostructures[419] and resident carriers in doped 1L-TMDCs[420,421].

Thus, pristine 2D TMDCs preserve valley polarization for a short time, making them barely useful for real-world applications. Nevertheless, recent works report on valley polarization in 2D TMDCs coupled to wisely designed nanostructures and metasurfaces[127,204,205,208,210,211,384]. These devices for valley-polarization enhancement can either separate valley degrees of freedom in space, e.g., using surface plasmon waves, or separate the valley-polarized PL in momentum-space (*photonic Rashba effect*)[210].

Valley polarized charged carriers (electrons, holes, trions) can be detected as a Hall voltage that appears at different points of 2D TMDC when the in-plane electric field is applied[85,130,422], giving rise to optoelectronic devices like valleytronic transistors[423]. The valley-polarization effects can also be detected via the valley Zeeman effect when an external magnetic field lifts the valley pseudospin's degeneracy[424]. Accessing different valley states through CP lights has been experimentally demonstrated with several TMDCs including molybdenum disulfide ($MoS_2$)[125,126,414], molybdenum diselenide ($MoSe_2$)[425], tungsten sulfide ($WS_2$)[426], and tungsten diselenide ($WSe_2$)[69,427]. Today, valley polarization is considered as another alternative carrier of information[17,18,85,428] in addition to currents in electronics and spin in spintronics and holds a great deal of promise for future classical and quantum computers[12,127,128,205,429–431].

The valley polarization is preserved in pristine 1L-TMDCs at room temperature for a concise period, preventing valley pseudospins in practical on-chip valleytronic devices. Recent studies show that resonant optical nanostructures such as plasmonic waveguides, metasurfaces, and *Panchatantra-Berry* (PB) phase metastructures pave the way to get around this obstacle. In a recent work[204], it has been demonstrated that coupling of a few-layer TMDC with a plasmonic (Ag) nanowire allows achieving high-efficiency photonic *spin-orbit interaction* (SOI) via coupling with the SPP modes. The possibility of valley-polarization contrast of $\eta = 0.7$ in such a simple system at room temperature in the 5-L $WS_2$ has been predicted in simulations, while the experimentally measured was ~0.35. This reduction in valley polarization is explained by the defects in Ag wire and TMDC and the imperfect experimental conditions.

The use of plasmonic metasurfaces for tailoring photonic SOI effect for TMDC valley polarization has been suggested and realized at room temperature in Ref.[208]. This design exploits the PB geometric phase giving rise to different phases that the SPP wave squares depending on $\sigma^+$ or $\sigma^-$ polarized exciton. The work also reports the strong coupling regime of the valley-polarized excitons with the SPPs of the metasurface. Recent results show that in the strong-coupling regime, the TMDC exciton-polaritons become less sensitive to the disorders and defects, resulting in long-lived valley polarization[431,432]. Ref.[208] reports the persistence of the valley polarization even after 200 ps. In Ref.[433], a PB plasmonic metasurface has been theoretically suggested for spontaneous exciton valley coherence enhancement in 2D TMDCs and valley separation in k-space (emission direction).

In another study Ref.[127], the valley polarized excitons sorting and spatial separation at room temperature by coupling a 1L-$MoS_2$ with a tailored asymmetric groove metasurface has been demonstrated (Fig. 16a). The absence of mirror symmetry of this structure in the plane perpendicular to the metasurface grooves, combined with spin-valley locking, enables the spatial separation of excitons with opposite valleys, as shown in Fig. 16b-c. The device can further sort photons with opposite helicity from valley polarized excitons towards different directions, operating as a powerful interface between photonic and valleytronic devices. More broadly, this approach can be extended to a wide range of material platforms, enabling the practical use of the valley degree of freedom for information processing and storage at room temperature.

Nonlinear optical response steered by valley excitons in TMDCs coupled with metasurface interface has also been investigated[384] (Fig. 16d-e). It was shown that the SHG can be achieved via spin-direction locking in a similar TMDCs–metasurface interface. In this TMDCs–metasurface system, the fundamental-frequency light was separated by the space-variant aperture units into different directions in real space, which interacted with different TMDCs valleys to achieve a coherent SHG process. It was found that the second-harmonic nonlinearity of WS2 has been increased by one order of magnitude.

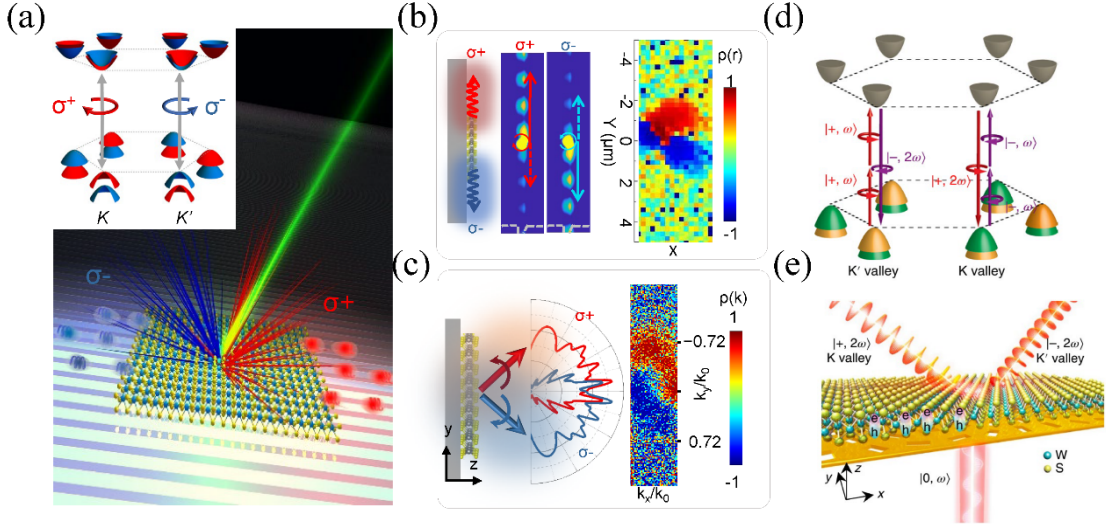

Figure 16. Metasurfaces for valley-polarization enhancement. (a) Schematics of valley excitons in 1L-TMDC by a metasurface. The inset shows the band structure and optical selection rules of valley exciton in monolayer MoS$_2$; (b) The left panels are schematic illustration of valley exciton separation and simulated electric field distribution of a 1L-MoS2/asymmetric Ag grating hybrid system which is illuminated by σ$^+$ and σ$^-$ dipole. The right panel is measured valley polarization contrast in real space for 1L-MoS$_2$ on the asymmetric groove. (c) The bottom panel is the experimental observation of separated valley exciton emission in momentum space. (d) Schematic drawing of valley-exciton locked optical selection rule for SHG in monolayer WS$_2$. (e) Schematic of hybrid Au–WS2 metasurface that can steer the nonlinear chiral valley along different directions. ((a)-(c) from Ref. [127], (d)-(e) from Ref. [384])

## 6. *Tunable photonic devices based on TMDCs*

Active control on the optical properties of 1L- or a few-layer TMDC is highly desired in real applications, such as electrical-optical modulators and optical-mechanical resonator. This section briefly reviews several approaches to realizing tunable photonic devices based on TMDC materials.

**6.1 Chemical composition tuning**
Alloy semiconductors with different bandgaps can, in principle, engineer the bandgap of bulk semiconductors. In order to explore full potential and expand the functionality of 1L-TMDC, it is desirable to have the flexibility of modifying the physical and chemical properties by controlling the compositions. The first experiment demonstration on 2D alloy Mo$_{1-x}$W$_x$S$_2$ is done through alloying exfoliated atomically thin MoS$_2$ and WS$_2$ by Chen et al. [434]. Tunable bandgap is verified from both the density function theory calculation and the tunable PL emission from 1.82eV (x=0.2) to1.99eV (x=1).

The optical bandgap of alloys satisfies E=(1-x) $E_{g-MoS2}$+x·$E_{g-WS2}$-bx(1-x), where b is bandgap bowing parameters. It is better to develop the direct growth of large scale ternary TMDC alloys from a practical application perspective.

Tremendous efforts have been devoted to growing the high-quality composition modulated layered atomically thin semiconductors[435–441]. For instance, Gong et al reported the single-step synthesis of 1L- and 2L-$MoS_{2(1-x)}Se_{2x}$ on $SiO_2$[435]. The tunable composition was made by controlling the ratio of sulfur and selenium powder, which are mixed as chalcogen source during synthesis. The PL spectrum and atomic resolution Z-contrast imaging demonstrated that optical bandgap could be continuously tuned by more than 200 meV. A similar study on the growth of monolayer alloy $MoS_{2(1-x)}Se_{2x}$ was conducted by Mann et al.[437]. At the same time, Li et al proposed the single-step chemical vapor deposition synthesis of ternary $MoS_{2(1-x)}Se_{2x}$ alloy with an arbitrary composition by tuning the sulfur's reaction temperature and selenium power with $MoO_3$ instead of changing the ratio of S/Se[438]. Both the PL and Raman spectra signify that the optical properties of alloy can be modulated by the composition while the optical bandgap of alloy can be smoothly tuned from 1.856eV (pure 1L $MoS_2$) to 1.56eV (pure 1L $MoSe_2$), as demonstrated in Fig.17a-b.

Following a similar approach, Duan et al. demonstrated the controlled synthesis of alloy $WS_{2x}Se_{2(1-x)}$ nanosheets with tunable chemical compositions and optical properties[439]. The optical bandgap can be smoothly varied from 626.6 nm (pure 1L WS2) to 751.9 nm (pure 1L $WSe_2$) (Fig.17c-g). Besides, electric transport measurement signifies that the alloy $WS_{2x}Se_{2(1-x)}$ can be switched from n-type semiconductor for the sulfur-riched alloy to p-type semiconductor for selenium-riched alloys. Both tunable electric and optical properties of alloy $WS_{2x}Se_{2(1-x)}$ may provide more freedom in constructing optoelectronic devices, such as solar cells and LED. Also, a composition graded alloy $MoS_{2(1-x)}Se_{2x}$ along a single triangle nanosheet was grown by improved chemical vapor deposition approach[440], where sulfur and selenium source are moved during the growth process. The composition in a single triangle strongly depends on the position, where x increases from 0 at the center to 0.68 at the edge. The tunable composition also continuously redshifts the emission wavelength from 680 nm (pure 1L $MoS_2$) to 755 nm. Later, Wu demonstrated the synthesis of spatially composition modulated alloy 1L-$WS_{2x}Se_{2(1-x)}$[441], where the composition can be tuned in one way or both way from the center to edge.

Kobayashi et al. successfully grew the lateral and vertical heterostructure based on bandgap tunable 1L-$Mo_{1-x}W_xS_2$ by sulfurization on the two spatially separated patterned $WO_3$ and $MoO_3$ thin films[442]. Zhang et al. report the one-step CVD growth of ternary monolayer alloy $Mo_{1-x}W_xS_2$ with a mixture of $MoO_3$ and $WO_3$ powders and fine sulfur powders[443]. In 2016, Li et al. reported the synthesis of composition modulated lateral heterostructure $MoS_2$-$MoS_{2(1-x)}Se_{2x}$ via layer-selected atomic substitution on the pregrown $MoS_2$[444], where bilayer triangle is located at the center of monolayer $MoS_2$. The Se substitution temperature of the bilayer (810°C) is higher than that of the monolayer (740°C). The large difference of substitution temperature suggests an effective way of realizing lateral heterostructure $MoS_2$-$MoS_{2(1-x)}Se_{2x}$ by selecting substituting Se in the peripheral area of the monolayer region and remaining bilayer region unaffected. Moreover, the bandgap of $MoS_{2(1-x)}Se_{2x}$ can be easily tuned by varying the reaction time, verified by both PL and Raman spectroscopy.

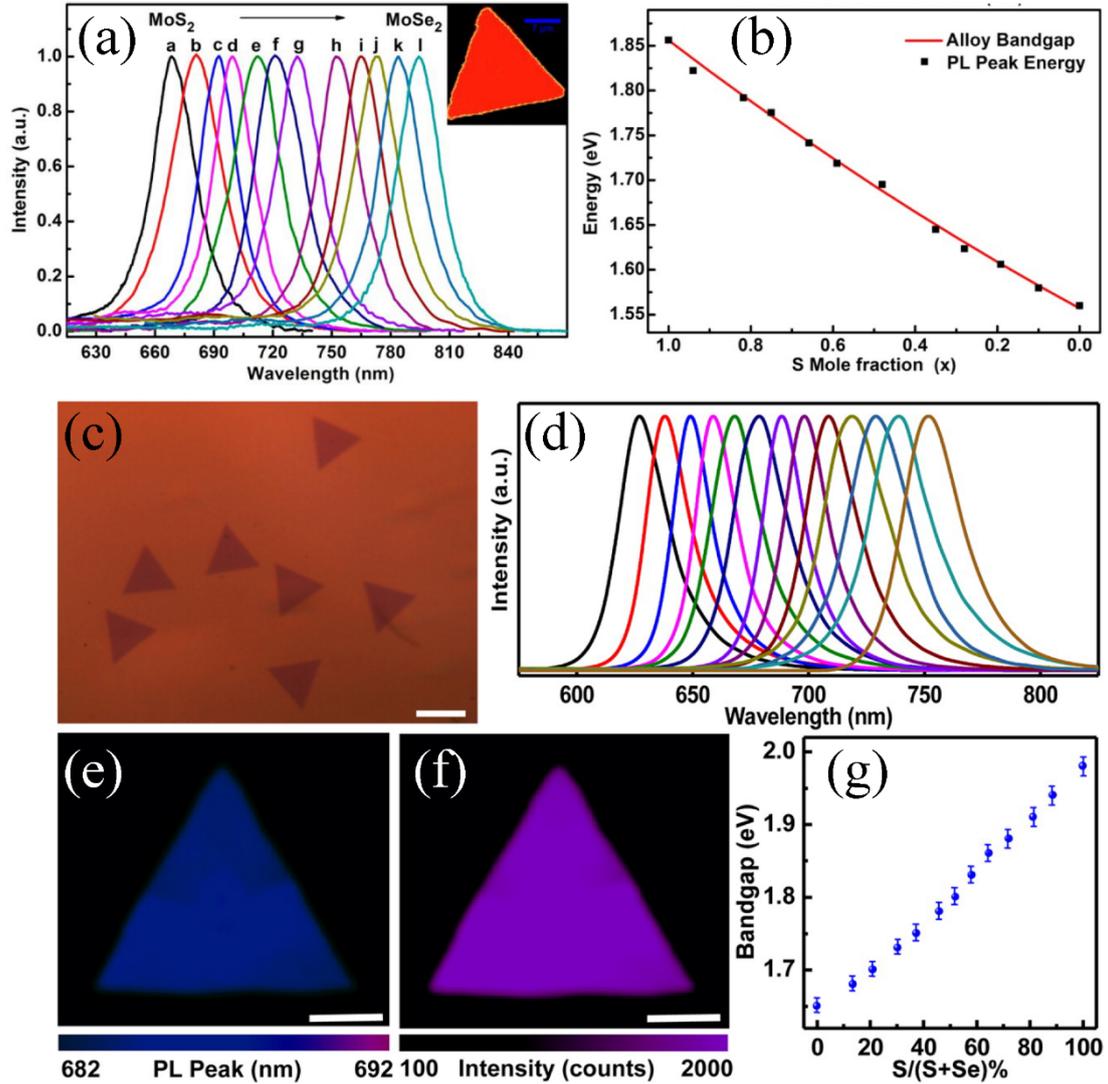

Figure 17. (a) PL spectra of alloy 1L-MoS$_{2(1-x)}$Se$_{2x}$. (b) Extracted peak value from PL spectra as a function of S mole fraction. (c) Optical image of synthesized WS$_{2(1-x)}$Se$_{2x}$ nanosheets. (d) PL spectra of tunable MoS$_{2(1-x)}$Se$_{2x}$ nanosheets. (e-f) PL peak position and intensity mapping for 1L-WS$_{1.044}$Se$_{0.956}$. (g) Extracted optical bandgap from PL spectra as a function of sulfur ratio. ((a)-(b) from Ref. [438], (c)-(g) from Ref. [439])

## 6.2 Mechanical tuning by strain

Due to the atomical thickness of 2D materials, the strain can effectively modulate their electronic band structure and thus modify both electronic and optical properties by altering the lattice constants. Conley et al. investigated the effect of uniaxial tensile strain on the phonon spectrum and bandgap of 1L- and 2L-MoS$_2$[445]. Fig.18a shows the schematics of the beam bending apparatus that is used to produce the tensile strain on MoS$_2$. After applying strain, the phonon mode E$_{2g}$ for 1L-MoS$_2$, as shown in Fig.18b, splits into two mode E$_+$ and E$_-$ with the redshifts by 4.5 cm$^{-1}$ and 1cm$^{-1}$/% strain while A$_{1g}$ almost remain the same. A similar phenomenon also occurs on the bilayer MoS$_2$. Simultaneously, optical bandgap shows almost linear decreases with the increasing strain in Fig.18c, with the slope being ~45 meV/% strain for monolayer. For bilayer MoS$_2$, the shift rates for direct and indirect band gap are 53 meV and ~129 meV/% strain, respectively. Moreover, the PL intensity of 1L-MoS$_2$ shows pronounced decrease for considerable strain, indicating that material crossover from direct bandgap to indirect bandgap semiconductor.

The other two groups conducted similar studies on both monolayer and bilayer $MoS_2$[446,447]. Linear dependence is also observed for the phonon mode $E_{2g}$ and optical bandgap. The retrieved shift rates per strain are comparable to the value reported above. Besides, Zhu et found that valley PL polarization decreases monotonically with the increased strain for 1L and 2L, and even almost vanishes at 0.8% strain[447]. Similar to the case of 1L-$MoS_2$, the tensile strain also induces the transition of $WS_2$ from direct bandgap semiconductor to indirect bandgap semiconductor[448].

In contrast to the monotonic decrease of PL intensity for both monolayer and bilayer $MoS_2$, Desai et al. observed a significant enhancement in the PL intensity for multilayer $WSe_2$ when uniaxial tensile strain up to 2% is applied[449]. The enhancement factor of PL is up to 25 for bilayer $WSe_2$ at 1.51% strain, and PL intensity is even comparable to monolayer counterparts without strain (Fig.18d). Simultaneously, the full width half maximum experiences decrease from 110 meV to 62 meV, close to monolayer's value of 61 meV. The giant PL enhancement is attributed to the indirect to direct bandgap transition for strained bilayer $WSe_2$, as verified by the density functional calculation (Fig.18e). The opposite trend of multilayer $MoS_2$ and $WSe_2$ is due to the small energy difference of direct and indirect bandgap in multilayer $WSe_2$. The applied tensile strain allows for bandgap crossover.

Niehues et al performed the absorption and PL spectroscopy to study how the linewidth and lineshape of excitonic resonance in 1L-TMDC are affected by the applied strain[450]. They found that for 1L-$MoSe_2$ and 1L-$WSe_2$ the lineshape of exciton resonance changes from Fano profile to Lorentz shape while the linewidth becomes narrower with the increasing strain. However, the linewidth for 1L-$MoS_2$ increases while it remains almost unchanged. This intriguing phenomenon is mainly caused by the strain-altered electron-phonon coupling. Instead of applying tensile strain, Hui et al fabricated electromechanical devices to study the effect of compressive strain on trilayer $MoS_2$[451]. PL measurement indicates that the direct bandgap shows blueshift by 300 meV/% strain. Also, the PL intensity increases with the strain by 200%, while the full width half maximum reduces by 40% at a strain level of 0.2%.

Strain-induced band gap engineering does not only happen in 1L-TMDC or a few layers on flexible polymer substrate but also works for suspended monolayer[452], as demonstrated by Lloyd et al. Since the CVD-grown 1L-$MoS_2$ is impermeable to gas, the biaxial strain can be generated by applying the pressure difference across the suspended membrane. The optical bandgap of suspended 1L-$MoS_2$ can be tuned by 500 meV when a considerable biaxial strain is applied. The optical bandgap's shift rate is 99 meV /% strain, twice as that of the supported monolayer by uniaxial strain.

Strain also plays a critical role in controlling the optical properties of as-grown 1L-TMDC on the substrate, such as $SiO_2$ and sapphire substrate. Usually, tensile strain is induced in as grown TMDC monolayer in the fast cooling process from the growth temperature because of the huge difference in the substrate and monolayer's thermal expansion coefficient. Liu et al. found that a non-uniform tensile strain is created in CVD grown 1L-$MoS_2$, and non-uniformity is confirmed by the variation of PL intensity and peak position[453]. Such strain can be released after transferring $MoS_2$ onto other substrates. Besides, the intrinsic strain is evaluated by comparing the PL of the controlled sample with applied strain to that of the as-grown sample. Three-dimensional finite element analysis and density functional theory calculation reveal that only ~10% tensile strain can be transferred from flexible substrate polydimethylosiloxane while substrate with high Young's modulus enables the reduction of scattering loss. The less transfer efficiency of strain is induced by the weak van der Waals force interaction between 2D materials and polymer substrate, leading to decoupling and interlayer slippage. To overcome this limitation, Li et al proposed a simple strain engineering approach to efficiently modulate the bandgap of 1L-TMDC[454], including $MoS_2$, $WS_2$, and $WSe_2$. The large strain-transfer efficiency is realized by encapsulating 2D materials in the flexible polyvinyl alcohol (PVA) through the spin-coating method because of collective contribution from large Young's modulus of PVA (~10GPa) and strong interaction force between 2D materials and PVA. The experiment results indicate that by applying the uniaxial strain on the 1L-$MoS_2$, the optical bandgap change and bandgap shift rate is high to 300 meV 136 meV/% strain, twofold of previous best results. The strong interaction between 1L-$MoS_2$ and PVA with negligible slippage is confirmed by multicircle load-unload test. Such a PVA spin-coating method is also generalized to other TMDC materials (i.e. $WS_2$ and $WSe_2$) with improved bandgap tunability compared to conventional methods.

The knowledge of strain-induced bandgap engineering allows developing growth techniques with controlled bandgap tuning. Zeng et al. applied a spherical diameter engineering strategy to tune the bandgap of 1L-$MoS_2$[455]. The large tunability of bandgap (~360 meV) is enabled by the strain of $MoS_2$ crystal introduced by self-reshaping of the liquid glass substrate on a non-wettable substrate. They also established a one-to-one correspondence between the bandgap of 1L-$MoS_2$ and the spherical diameter, allowing for mass production of 2D materials with a controlled bandgap. Except for pure TMDC materials, strain engineering was also demonstrated able to modulate excitonic behavior by changing the interlayer coupling strength in CVD grown $MoS_2$/$WS_2$ van der Waals heterobilayers[456,457]. Both the in-plane vibration phonon modes and emission strongly depend on the applied strain, signifying that interlayer interaction has been changed due to modification of band structure in heterobilayers.

Finally, as an example of applications, strain engineering has been successfully applied to develop a high-responsivity $MoTe_2$ photodetector at 1550 nm for silicon photonic integrated circuits[458]. As demonstrated previously, the applied strain can decrease the direct bandgap of TMDC materials. Fig. 18f shows schematics of silicon microring resonator integrated $MoTe_2$ photodetector. The 2D film is intentionally wrapped around a non-planarized waveguide to introduce the local tensile strain (Fig.18g), which creates a graded bandgap of a few layers $MoTe_2$ below the telecommunication wavelength. The role of silicon microring resonator is to provide resonant modes at 1550 nm that is well overlapped with $MoTe_2$, thus maximizing the light-matter interaction. The photoresponsivity is measured as 0.5$Am^{-1}$ at -2V, while the dark current is only 13nA (Fig.18h). The high responsivity of $MoTe_2$ based photodetector at 1550 nm is ascribed to the enhanced absorption arising from the strain-induced bandgap shrink and the enhancement of photon lifetime in the resonator (Fig.18i).

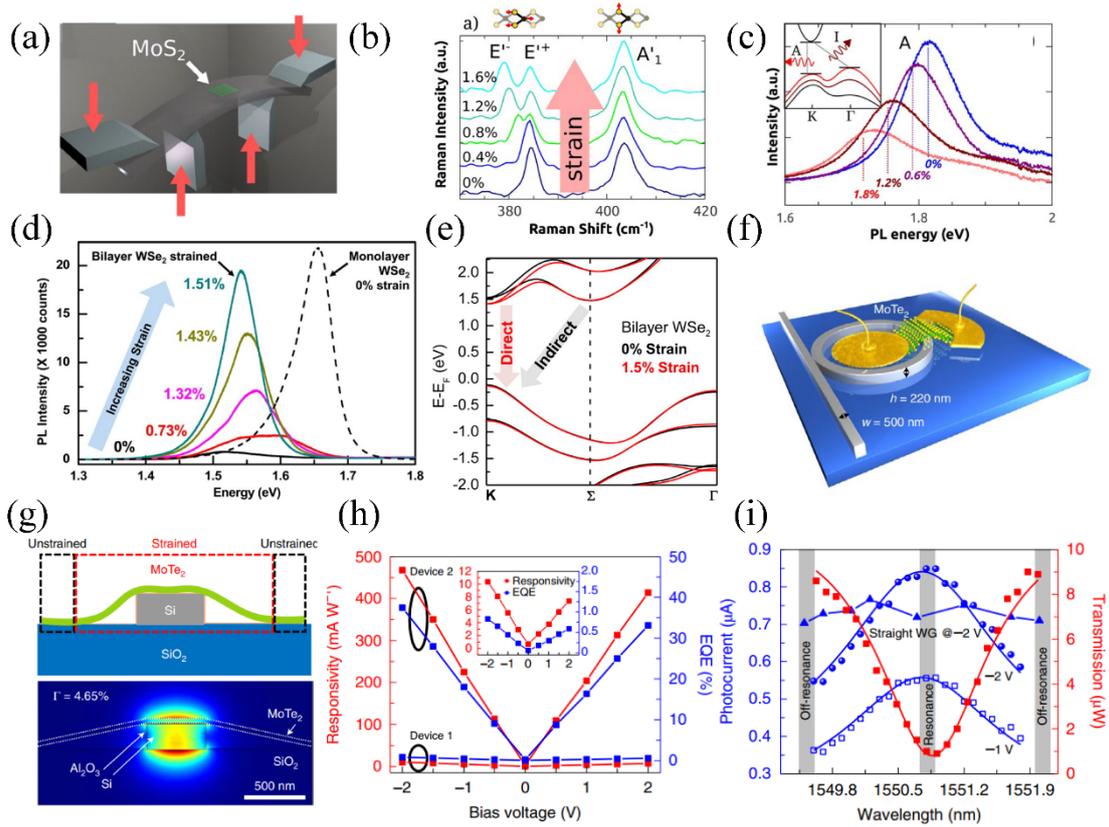

Figure 18. (a) Schematic of device structure used to apply strain on $MoS_2$. (b-c) Raman (b) and PL (c) spectra of 1L-$MoS_2$ under different levels of strain. (d) PL spectra of $WSe_2$ bilayer at the different strains. PL spectrum of 1L-$WSe_2$ is also given as a reference. (e) Calculated band structure evolution at the different strains. (f) Schematic of 1L-$MoTe_2$ photodetector integrated onto microring resonator. (g) Top panel shows the schematic of 2D $MoTe_2$ on top of the non-planarized waveguide. The local strain is introduced in the vicinity of the waveguide. The bottom panel shows the calculated electric field distribution of transverse magnetic mode. (h) The measured responsivity of photodetectors at different bias voltage. Inset is zoom in responsivity of device 1. (i) Measured photocurrent of microring integrated photodetector at a different wavelength. ((a)-(c) from Ref. [445], (d)-(e) from Ref. [449], (f)-(i) from Ref. [458])

## 6.3 Electric tuning

Effective tuning refractive index electrically constitutes the foundation for realizing photonic devices with dynamic tuning such as electro-optic modulator and phase array etc. However, the performance of conventional semiconductors like InP, GaAs, Ge, and Si is restricted by their limitation of intrinsic electro-optical tunability (~1%) upon charge injection.

Owing to the atomically thin nature of 2D TMDCs, their optical properties are susceptible to the electric field. In particular, the charge density can be tuned in a vast range through gating, which provides an ideal playground for efficient electro-optical modulations. Two research groups independently demonstrated that the optical properties[63,64], such as absorption and PL, can be electrically tuned by back gating voltages. The carrier injection from gating can effectively dope electrons (or holes) to 1L-$MoS_2$ (or 1L-$MoSe_2$) and thus results in the formation of negatively (positively) charged exciton (also named as trion). The large binding energy of trion (20 meV for 1L $MoS_2$ and 30 meV for 1L $MoSe_2$) in 1L-TMDC is reflected from the neighbor resonance peak close to exciton resonance in both absorption and PL spectrum. With the increasing doping level from the neutral one, the trion not only broadens and redshifts the spectrum but also reduces the overall absorption and PL intensity. In 2015, Chernikov found that the exciton binding energy of 1L-WS$_2$ and quasiparticle bandgap can be continuously tuned over a range of 100 meV from the reflection spectrum by applying external gating voltage on the 1L WS$_2$ based field effect transistor[75]. The large tunability is ascribed to the electrically controlled free charge carrier injection into 1L-WS$_2$ with density up to $8\times10^{12}$cm$^{-2}$. The essence of electric tuning the reflection or absorption spectrum in 1L-TMDC is the effective refractive index tuning. In 2017, Yu et al. fabricated the CMOS-compatible 1L-TMDCs gating devices and measured their refractive index under different gating voltages[459]. They found that the refractive index of 1L-WS$_2$ can be dynamically tuned by external gating. The imaginary (real) part shows >60 (>20%) tunability near the exciton resonance, as shown in Fig.19a-b. Such a large tunability is attributed to the spectrum broadening and interconversion between neutral exciton and trion populations at different level of charge injections. Theoretical simulation suggests that the giant tuning of the refractive index allows for modulating the absorption of 1L-WS$_2$ from 40% to 80% by integrating such a gating configuration onto the resonant photonic structure (i.e. GaN metagrating).

The electrical tunability of refractive index (or dielectric permittivity) also enables the dynamic control of the strong light-matter interactions in 1L-TMDC[399,400,460–462]. For instance, Chakraborty et al. demonstrate that the strong exciton-polariton coupling can be modulated between 1L-WS$_2$ and DBR based microcavity via electrically induced gating at room temperature[460]. When the 1L-WS$_2$ is tuned from neutral to more n-type under gating, light-matter interaction transits from a strong coupling regime with Rabi splitting 60 meV to weak coupling. Such a transition is well explained by the reduced oscillatory strength of exciton caused by free charge carrier injection by electrostatic gating. Electrically tunable exciton-plasmon coupling is also achieved between the exciton of 1L-WSe$_2$ and the plasmonic crystal cavity by Dibos et al. [461]. In their work, hBN/WSe$_2$/hBN heterostructure was placed into a novel plasmonic crystal cavity with Q-factor 550, which consists of a dielectric photonic crystal cavity sitting on the single-crystalline Ag substrate. The external gating allows for the modulation of strong coupling between exciton and cavity on/off. This phenomenon can also be explained by the reduction of Columb interaction in the presence of free charge carrier injection via gating. More recently, Munkhbat et al realized an electrically tunable charged exciton-plasmon polariton by integrating of 1L-WS$_2$ with plasmonic nanoantenna operating not only at cryogenic temperature but also at room temperature by tuning the oscillatory strengths of neutral and charged excitons[462]. The electrical control via gating on such a hybrid system allows for tuning the strong coupling from neutral-dominated exciton-plasmon to trion dominated-plasmon coupling.

Except for dynamically tuning the strong coupling, Yan et al. demonstrated the electric control of PL emission and scattering through inducing electric doping and tensile strain via gating voltage when the 1L-WS$_2$ and bilayer are incorporated into Si Mie resonator[463]. Kravets et al demonstrated that the light absorption of 1L-MoS$_2$ could be modulated by 10% in a hybrid structure that consists of 1L-MoS$_2$ transferred onto the plasmonic-metal-dielectric waveguides[464]. Zhou et al demonstrated that the optical response of 1L-MoSe$_2$, including PL and reflection[465], can be dynamically manipulated by electrical gating while the atomically thin semiconductor is suspended on a metallic mirror. The electrostatic gating can effectively change the distance between 1L-TMDC and mirror, and thus enable the dynamical control on the light-matter interaction via engineering the local photonic density of states. It has been found that from PL and reflection spectrum both the intensity and linewidth of exciton resonance can be modulated. Moreover, due to the system's electromechanical nature, the maximum deflection of the heterostructure was observed when the electrical driving frequency resonates with the suspended heterostructure's mechanical resonance.

Note that electric tuning does not only happen around the exciton resonance but also can generalise to telecommunication wavelength. In contrast to large absorption at a wavelength near exciton resonance, the absorption is negligible at the near-infrared region for 2D TMDC, which is of great interest in optical communications. Based on this low loss feature, Datta et al. investigated the electric tunability of optical response of 2D TMDC in NIR region[466] and demonstrated a highly efficient electro-optic phase modulator based on 1L-$WS_2$. The strong electro-refractive tunability leads to considerable change in the real part of refractive index $\Delta n = 0.53$, but with only a minimal change in the imaginary part $\Delta k = 0.004$. The electro-optic response of 1L-$WS_2$ in this work is first characterized by integrating the 1L-$WS_2$ with a micro-ring resonator cavity as shown in Fig. 19c. The ionic gating is applied for electrostatic doping into the 1L-$WS_2$. The results shown in Fig.19d demonstrate that the shift of resonance peak from effective phase modulation depends on charge injection. The high Q factor of resonance peaks is remained without broadening, demonstrating the low loss features. The final TMDC–$HfO_2$–ITO capacitor-based composite $Si_3N_4$-$WS_2$ waveguide phase modulator demonstrates $\Delta n/\Delta k \sim 125$ for 1L $WS_2$, with performance exceeding the best Si and hybrid III-V -on-Si devices. The mechanism of how NIR optical response for TMDCs can be tuned via charge injection is still elusive, and future studies are required for a more detailed understanding of this below gap light-matter interactions, to facilitate optimization of 2D TMDC photonic devices for different applications.

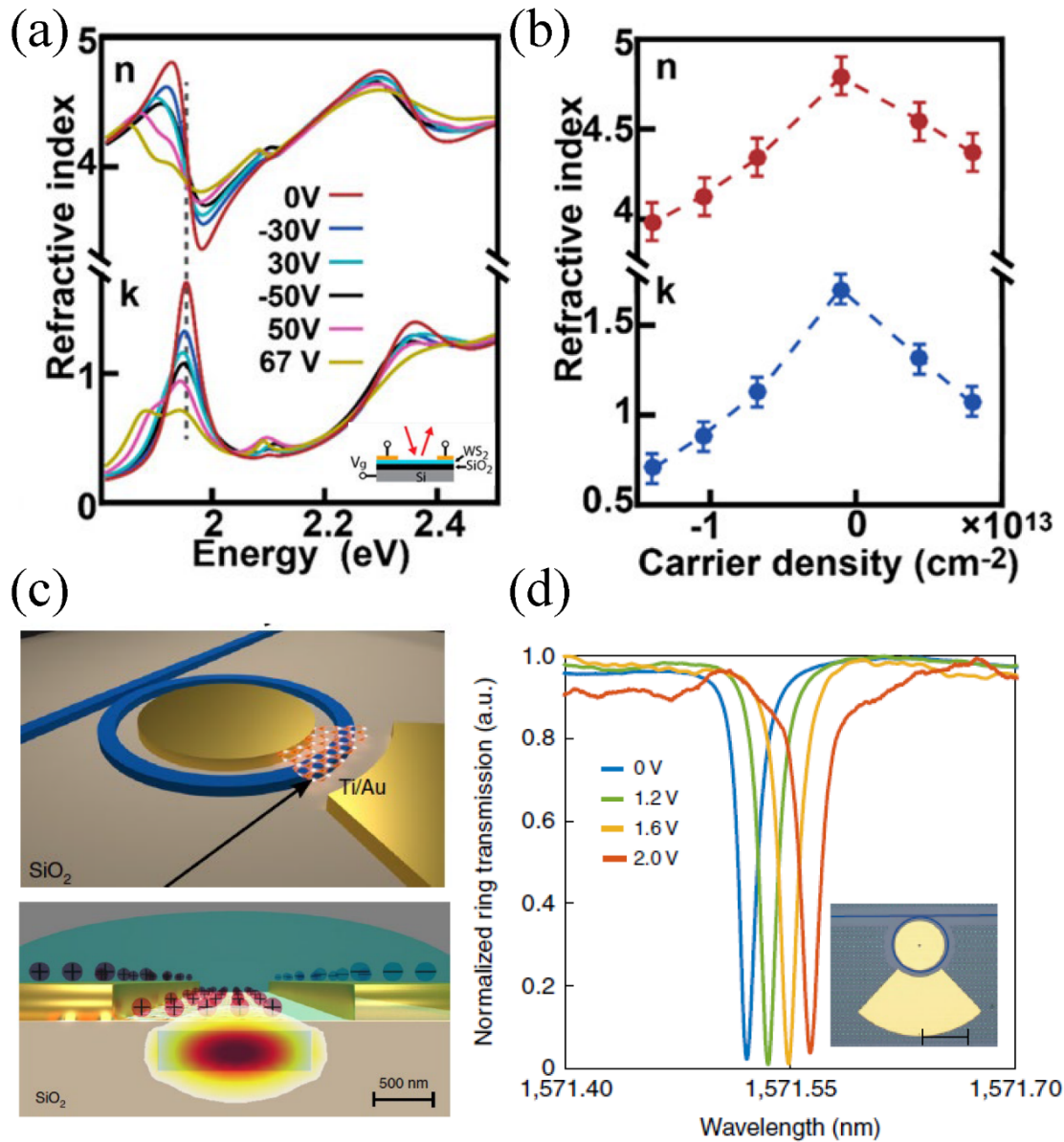

Figure 19. (a) Real and imaginary part of refractive index retrieved from reflection measurement at different gating voltages. (b) The peak value of real and imaginary part of the refractive index versus carrier densities. (c) Top panel is a schematic drawing of a hybrid microresonator-WS$_2$ monolayer and patterned Ti/Au electrode, and the bottom panel is the mode profile in the Si$_3$N$_4$ waveguide. (e) Normalized transmission spectra of microring resonator at different voltages. ((a)-(b) from Ref. [459], (c)-(d) from Ref. [466])

## 6.4 Optical tuning

The rich many-body effects in 1L-TMDC make it possible to actively control their optical properties (i.e., emission/absorption) with light. Aside from the electrical doping by external gating, 1L-TMDC can be doped via an optical way. Mitioglu et al. demonstrated the active tuning on the ratio of trion and exciton emission in 1L-WS$_2$ by simplifying tuning the power of laser excitation at 4K[467]. They attribute this to the photoionized carrier trapped at the donor level with laser illumination, generating an electron-hole pair. Moreover, by fixing the above bandgap excitation, a second subband gap excitation is introduced to generate the excessive density of electrons so that the trion emission can be independently modulated.

Except for the tuning of trion emission, biexciton is formed under the high power excitation[122,468]. Fig.20a shows the PL spectrum for exfoliated 1L-WS2 at room temperature under different illumination powers. With the increasing power, trion emission's contribution becomes dominant, clear sign of optical doping. By extracting the peak and intensity of PL, they found the integrated intensity from exciton and trion emission shows a sublinear and linear dependence on the excitation power (Fig.20b), respectively. When simultaneously cooling 1L-WS$_2$ to 4.2K and increasing excitation power, in addition to exciton and trion, localized states and biexciton emerge (Fig.20c). Different from the other three types of emission, biexciton emission $I_{xx}$ shows superlinear behavior with respect to the excitation power $P(I_{xx} \sim P^{1.9})$ (Fig.20d), which is the key feature of biexciton. Its binding energy is found as 45 meV at 4.2K but vanishes at room temperature.

Similar work on identifying many-body effect[469], such as exciton, trion and biexciton of 1L-WS$_2$, was also done by Plechinger et al., where PL spectrum is measured at a different temperature, gating voltage and excitation power of the laser. The formation of biexciton is also found in 1L-WSe$_2$ at 10K and 50K by You et al. [69]. The biexciton state is identified by its superlinear strength of emission versus exciton emission ($I_{xx} \sim I_x^\alpha$, $\alpha=1.3$). Typically, biexciton has $\alpha=1.2-1.9$ in the quantum well system due to lack of equilibrium between states. Time-resolved PL measurements signify that biexciton's lifetime is 31 ps, same order as the lifetime of the exciton (14ps) and trion (25ps). The binding energy of biexciton is obtained as 52 meV, an order larger than that of a conventional quantum well system. Pei et al. demonstrated biexciton's existence with binding energy 60 meV in free-standing MoSe$_2$ bilayer at both cryogenic and room temperature[470], which is attributed to the reduced dielectric screening and large trion density at high incident laser power. Moreover, they found that the biexciton emission intensity increases with the increased trion emission but decreases with the exciton emission, indicating that the nature of biexciton belongs to the excited state biexciton (a charge+charged trion) instead of grounded state biexciton (exciton+exciton). Lee et al. found that biexciton can even survive at room temperature in 1L-MoS$_2$ on SiO$_2$/Si substrates[471]. Kim et al. reported the biexciton emission from edge and grain boundaries in as grown 1L-WS$_2$[123]. Charged tunable biexciton was reported in 1L-WSe$_2$ by Barbone et al. [472]. Additionally, benefiting from the large exciton binding energy, the formation of electron-hole liquid was independently demonstrated in 1L-MoS$_2$[473,474] and 1L-MoTe$_2$[475] by Yu et al. and Arp et al., respectively.

Optical doping can also be introduced by the integration of noble metallic single NP or array with 1L-TMDC. Li et al. demonstrated the active control of absorption and PL spectrum when a single Au NP is transferred onto the 1L-MoS$_2$[476] (Fig.20e-f). The fundamental mechanism behind this is that plasmonic hot electronics enable n-doping in MoS$_2$, which allows for modulating the dielectric permittivity of materials and thus redshift of both absorption and PL spectrum while laser power increases monotonically. Femtosecond pump-probe spectroscopy was performed to investigate the ultrafast dynamics in Au nanoantenna/MoS$_2$ hererostructure[477]. Experimental results show that plasmonic hot-electron transfer takes place within 200fs. Moreover, pump power-dependent PL measurements reveal that PL can be enhanced by improving the exciton-plasmonic coupling.

Zu et al. realized the active control of surface plasmon resonance in the hybrid MoS$_2$-Ag NP array systems[478]. Due to the strong near-field coupling between plasmonic resonance and exciton, the absorption spectrum from hybrid structure displays redshift with the increasing excitation power. The hot electron injection from the Ag s makes the trion become dominant and thus changes the dielectric permittivity of MoS$_2$ effectively. Deng et al. demonstrated the efficient near field energy transfer in hybrid system MoS$_2$-Au metasurface[479]. The optical properties, such as Raman, absorption, and PL, can be tailored by the laser power due to the strong near-field coupling. Another promising platform for amplifying the optical tuning effect is integrating photonic crystal fano resonance with 1L-MoS$_2$, as reported by Zhang et al. [480]. Except for the linear response, Taghinijad et al. realized the extensive modulation (up to 55%) of second harmonic generation by controlling the photocarrier generation[481].

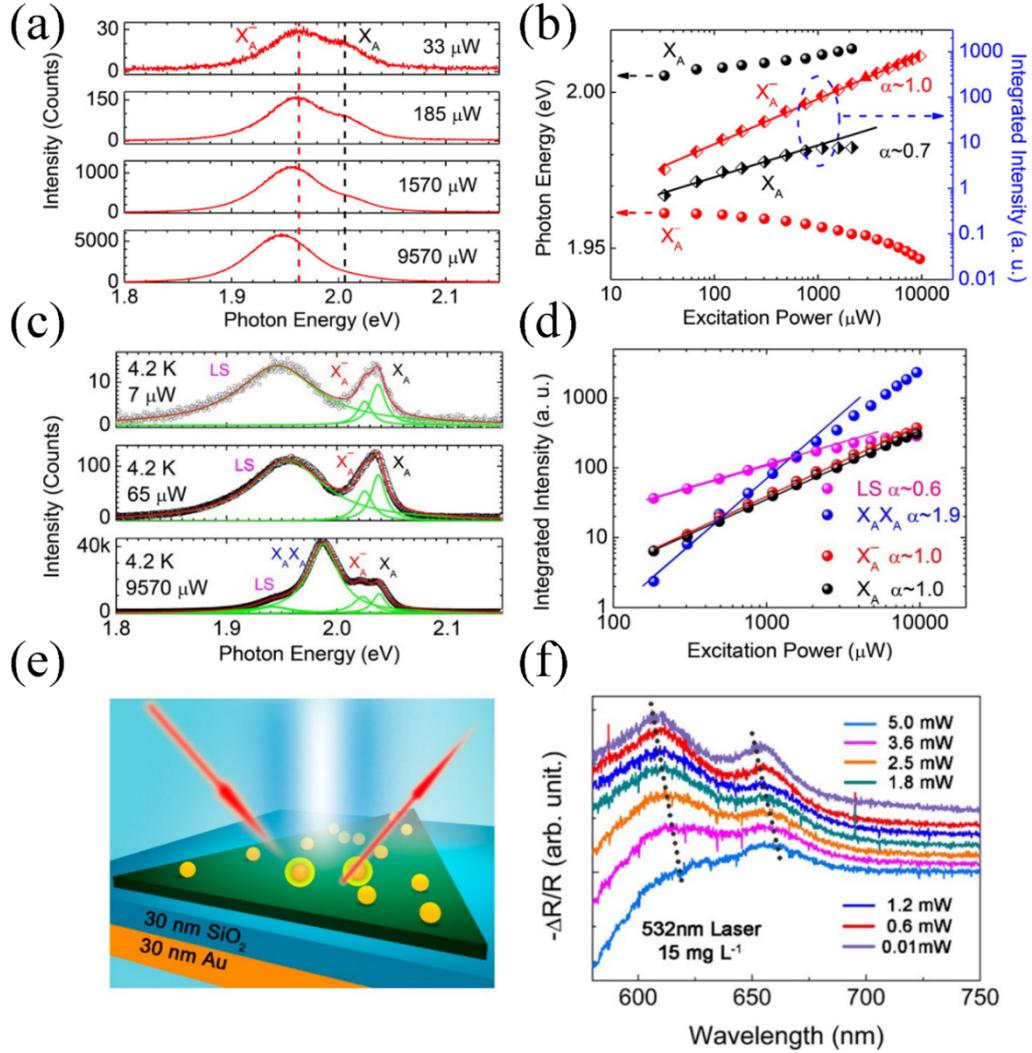

Figure 20. (a) PL spectra of exfoliated 1L-WS$_2$ at different excitation power. (b) Extracted exciton and trion energies as a function of excitation power. (c) PL spectra of as-grown 1L-WS$_2$ at 4.2K while the excitation power is 7μW, 65μW, 9570μW. (d) Extracted photon energies of the exciton, trion, biexciton and localized state as a function of excitation power. (e) Schematic diagram of 1L-MoS$_2$ with Au NP on top. (f) Measured differential reflection spectra at different excitation power. ((a)-(d) from Ref. [122], (e)-(f) from Ref. [476])

## 7. Narrowband and broadband absorbers based on monolayer and few-layer TMDCs

1L-TMDC can interact strongly with light due to their unique electronic and optical properties, such as direct-bandgap and large binding energy. However, its atomic thickness significantly limits its absorption to less than 10% in a broadband wavelength range. In some real applications like solar cells and photodetectors, it is desirable to have large absorption either in a narrowband or broadband wavelength range. Thus, much effort has been made to achieve perfect light absorption with 1L-TMDC or few layers in the past few years.

The first example of enhanced absorption in 1L-TMDC was proposed by Liu et al. [482], where they proposed integrating the 1L-MoS$_2$ with a dielectric spacer and a resonant back reflector, made of either one-dimensional photonic crystals or a metallic thin film. The absorption at the exction peak can be improved to 35% due to constructive interference. The dielectric spacer's critical thickness is found to gurantee the maximum field enhancement at the interface of MoS$_2$ thin film and spacer layer. Lu et al. theoretically demonstrated near-perfect absorption of 1L-MoS$_2$ with multilayer photonic resonant structure [483], as shown in Fig.21a, where the MoS$_2$ is sandwiched between a dielectric Bragg grating, dielectric spacer, and thin metal films. Simulaton results indicate that the absorption can be up to 96% by exciting the Tamm plasmon modes that enable strong electric field confinement. Moreover, the operating wavelength can be tuned by the thickness of the spacer layer, as demonstrated in Fig.21b.

The experimental verification of near-perfect absorption in 1L-TMDC was done by Epstein et al.[484]. In their work, they introduce WS$_2$-based vdW heterostructures-based cavity consisting of hBN/1L-WS$_2$/HBN on the Au substrate (Fig.21c), which provides 91% absorption for exciton and 41% for singlet and triplet trion states (Fig.21d). Such large absorption was attributed to the narrow linewidth of exciton and balance between radiative, nonradiative, and dephasing rate, which is made by controlling the temperature. In other words, as such a multilayer structure can be viewed as a Fabry-Perot cavity, the perfect absorption was achieved at a given temperature due to the critical coupling between optical resonator and incidence. Besides, strong absorption also makes biexciton PL observable under ultralow continous laser powers down to a few nWs. Another independent experiment on perfect absorption by atomically thin crystals was reported by Hoang et al [485]. Like the previous one, MoSe$_2$ is also encapsulated inside hBN layers, sitting on the flat DBR mirror. Perfect absorption of MoSe$_2$ was realized by tuning temperature, pulse laser exciton, and distance between mirror and monolayer, allowing for engineering exciton-phonon, exciton-exciton exciton-phonon interactions to realizing critical coupling.

When the layer number of TMDC increases from one to a few layers, enhanced broadband absorption becomes accessible for heterostructure TMDC few-layer/metals. This concept was originally demonstrated experimentally in ultrathin Ge film (<25 nm) on Au and sliver substrate by Kats et al. [486]. Unlike the previous absorbers requiring the thickness of the dielectric layer equal to $\lambda/4$, highly absorbing materials with thickness reduced to few nanometers still enable interferences when it directly contacts with metals (i.e., Au and Ag), leading to strong absorption in the visible range. As TMDC few layers are also highly lossy material, it is possible to follow the same strategy to realize the broadband absorption. Unlike other conventional photovoltaic materials, TMDC materials have an atomically smooth surface free of dangling bonds, rendering them good solar cell candidates due to the suppressed nonradiative recombination loss. Jariwala et al. experimentally demonstrated near-unity absorption with sub-15 nm thickness of TMDC materials (WSe$_2$, WS$_2$, and MoS$_2$) on Ag substrate[487]. Fig.21e-g show the schematic drawing, optical image and AFM image for such a hybrid structure. The broadband absorption was experimentally realized in Fig.21h when multilayer WSe2 is in intimate contact with an Ag back reflector. Moreover, broadband absorption is highly insensitive to the incident angle, which is of particular interest in photovoltaic applications. Since broadband absorption strongly depends on the thickness of absorbing materials but shows weak dependence on the composition, Wong fabricated a broadband absorber based on vdW heterostructure hBN/Graphene/WSe$_2$/MoS$_2$/metal[488]. The light absorption is above 70% between 500 nm and 700 nm for 0.6 nm Graphene/4 nm WSe$_2$/9.5 nm MoS2/Ag. The superior absorption allows for developing high-performance vdW based photovoltaic devices. The external quantum efficiency(EQE) >50% and internal quantum efficiency in the active layer>70% was demonstrated in pn WSe$_2$/MoS$_2$ heterojunction. Such a large EQE is the direct consequence of strong light absorption and efficient electronic carrier collection. The enhanced absorption of 1L-TMDC by Fabry-Perot cavity can also improve photodetector's performance with external photo gain and specific detectivity of the photodiode up to $5.8×10^4$ and $2.6×10^{10}$, as demonstrated by Wang et al. [489].

Except for the Fabry-Perot cavity, resonant photonic structures, such as dielectric grating, metasurface, and photonic crystal slab, have also been utilized to enhance the absorption of 1L-TMDC[490–496]. These structures support leaky mode resonance or guided-mode resonance. The unique strengh of such structures is that one can easily get any desired Q-factor by tuning the structure parameters, which is vital in realizing a perfect absorber with 2D monolayers. For example, Huang et al. expermentally demonstrated strong light absorption (>70%) with atomically thin materials (≤4L)in both narrowband and broadband wavelength range[495](Fig.21i-j). The narrowband absorption was made by integrating three layers onto the GaN grating structure backed by an Ag mirror (Fig.21i). The structure configuration for the broadband absorber is slightly different from the case of the narrowband absorber, 4L-MoS$_2$ was conformally grown on the GaN NP array. The designing principles build upon coupled leaky mode theory, which turns the perfect absorber design to find out the leaky modes with the required Q-factor at the given wavelength. It was also demonstrated that the optimal absorption wavelength could be tuned by merely tuning the structure parameters, such as period and width of GaN nanowire.

The enhanced light absorption has also been realized in monolayer graphene at 1507 nm and MoS$_2$ at 538 nm with fano resonant photonic crystal by Valentine's group[496]. Interestingly, Zhang et al. found that perfect absorption can also be realized from patterned MoS$_2$ grating on TiO$_2$ waveguide layer on silica backed by Au reflector while the line width of absorption can be as narrow as 0.1 nm[497]. Such ultra narrowband perfect absorption is mainly due to the excitation of guided resonances that are critically coupled to the incidence. Broadband absorption of 1L-MoS$_2$ was also theoretically studied by Piper et al. [498]. In their design, 1L-MoS$_2$ directly sits on the TiO$_2$ photonic crystal slab, backed by a perfect electric conductor mirror. Following the light trapping theory, the structure parameters of the photonic crystal slab are carefully tuned to realize broadband absorption in the visible range, where the average absorption is 51% from 400 nm to 690 nm.

Other examples of enhanced absorption in 1L-TMDC take advantage of plasmonic nanocavities[499,500]. For instance, Bahauddin et al. employed a complex structure 1L MoS$_2$/Ag NPs/NiO$_x$/Al to improve the absorption of 1L-MoS$_2$ to 37% across a broadband wavelength (400 nm-700 nm)[499]. Zhou et al. combined a 1L-MoS$_2$ with an array of Ag NPs backed by DBR mirror to realize polarization-independent nearly perfect absorption from 400 nm to 560 nm[500]. Here, it is worth noting that the Ag NP also contributes to the absorption due to its lossy nature. After deducting the absorption contributed by Ag NPs, the absorption of 1L-MoS$_2$ is still high to 55.2% at 467.8 nm and 84.8% at 557.8 nm. The improved absorption is attributed to the localized surface plasmon resonance.

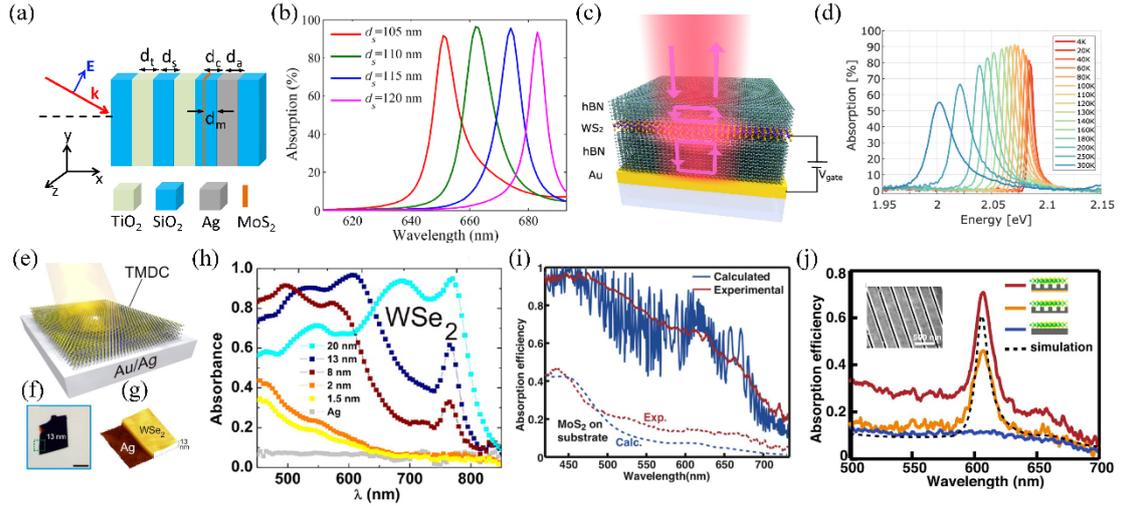

Figure 21. (a) Schematics of MoS$_2$ absorber, where 1L-MoS$_2$ is sandwiched between Bragg grating and Ag mirror. (b) Tunable light absorption of 1L-MoS$_2$ by varying the thickness ds. (c) Schematic of vertical van der Waals heterostructure hBN/WS$_2$/hBN on the Au substrate. (d) Measured absorption spectra of devices at different temperatures. (e) Schematic diagram of TMDC broadband absorber, which consists of TMDC few layers on Au/Ag substrate. (f) Optical image of 13 nm WSe$_2$ on the Ag substrate. (g) AFM image of 13 nm WSe2 film on Ag. (h) Measured absorption spectrum for exfoliated WSe$_2$ multilayer on Ag substrate. (i) Narrowband absorber of 3L-MoS$_2$ on GaN grating with and without Ag mirror. (j) Broadband absorber of 4L-MoS$_2$. ((a)-(b) from Ref. [483], (c)-(d) from Ref. [484], (e)-(h) from Ref. [487], (i)-(j) from Ref. [495])

## 8. Light Scattering Manipulation by TMDCs

In the past few decades, it has been demonstrated that light scattering can be arbitrarily controlled by plasmonic or dielectric metamaterials[271,501,502]. For example, Fano resonance can be generated in either plasmonic nanostructures and dielectric nanostructures supporting Mie resonance[501,502]. The marriage of TMDCs and metamaterials provides unprecedented flexibility to engineer the light scattering as TMDCs can interact strongly with the resonant structures. Besides, as a high index semiconductor, 1L-TMDC or few layers can be patterned as a dielectric metasurface, which serves as either waveguiding at angstrom scale or metalens that focus light with several nanometers. In this section, we discuss how TMDCs enables the manipulation of light scattering.

### 8.1 Fano resonance enabled by TMDC monolayer and a few layers

Exciton resonances supported by 1L-TMDC could be coupled to other types of optical resonances, such as plasmonic resonances [93,324,478,503] and Mie resonances, to produce Fano resonances. In 2015, Lee et al. systematically studied the optical properties of complex systems, consisting of chemically grown 1L-MoS$_2$ and Ag bow-tie nanoantennas array[324]. Fig.22a shows the schematics of the Ag bowtie array system sitting on the 300 nm SiO$_2$/Si substrate, while Fig.22b presents the SEM images of Ag bowtie lattice on the as-grown MoS$_2$ triangle supported by substrates. It has been well established that the Ag bowtie nanoantenna array supports surface lattice plasmonic resonances, where the electric field is strongly confined in the vicinity of bow-tie structures. Benefitting from the extreme field localization, more than one order magnitude enhancement has been demonstrated for both PL and Raman of 1L-MoS$_2$ integrated with the metallic nanoantenna arrays compared to bare MoS$_2$ on SiO$_2$/Si substrate. By cooling down the temperature to 77K, the reflection spectrum displays the typical asymmetry Fano line shape, which can be explained by the interference between sharp exciton resonance (high-Q resonance) of MoS$_2$ and broadband continuum of plasmon resonance (low-Q resonance), as shown in Fig.22c-d. Moreover, tunable optical properties (i.e., emission or reflection spectrum) can be easily realized by simply varying the bowtie array's lattice constant. In fact, Fano resonance can also be formed by coupling the exciton resonance with localized surface plasmon resonance supported by single metallic NPs. Wang et al. reported the experimental observation of Fano resonance in the hybrid systems consisting of stacked Au nanotriangles and 1L-WS$_2$[93]. Fano resonance is attributed to the interaction between dark K-K excitons in 1L-WS$_2$ and the localized surface plasmon resonances. When such hybrid systems are embedded in water, Fano resonance's linewidth becomes narrower than the whole structure in air. Zhao et al. conducted a comprehensive study on the light-matter interaction of 2D MX$_2$ (M=Mo/W, X=S/Se) monolayer and Ag NPs[503]. The plasmonic resonance supported by Ag NPs does not only contribute to the significant enhancement on PL of 1L-TMDC but also leads to the exciton-induced transparency dip at room temperature, which indicates coherent dipole-dipole coupling between 1L-MX$_2$ and LSP in the strong-coupling regime. The scattering spectrum of 1L-TMDC can be dramatically changed by the integration of 1L-TMDC with single silicon NPs.

Except for the exciton resonance, a guided optical mode (also called 2D exciton-polariton) was developed below exciton resonance with a 1L-TMDCs, as theoretically demonstrated by Bhurgin et al. [504]. Such an exciton-polariton mode in the visible and near-infrared wavelength ranges is confined within a few micrometers while its propagation length is larger than hundreds of micrometers. 1L-TMDC or few layers also support quasi-guided resonance known as Wood's anomalies when they are patterned as photonic crystal slab[505], as shown in Fig.22e. The monolayer or few layers TMDC film supports waveguide modes only when the difference between the top and bottom cladding layer is very small. The patterning of TMDC thin film as photonic crystal slab folds the mode dispersion below the light cone into the first Brillouin zone, making the guided mode transisting to quasi-guided resonance. This type of optical resonances was verified by transmission spectrum of 1-4L suspended photonic crystal slab (Fig.22f-h), which shows good agreement with the simulation results. Such a quasi-guided resonance can be also represented by the complex effective mode index. The real part complex effective mode index is slightly larger than the refractive index of air, while the confinement is the same scale of the visible wavelength. Guided mode resonance also can serve as a second-order coupler that couple PL emission from suspended continuous monolayer film to the region of photonic crystal slab. These results seems suggest that the atomically thin film itself can act as optical resonator and may find many applications in boosting light-matter interactions, such as sensor, enhanced light absorption and enhanced second harmonic generation.

Also, single layer MoSe$_2$ encapsulated into hBN was demonstrated to realize electrical tunable resonant mirrors at cryogenic temperature by two independent groups[42,43](Fig.22i-j). The high reflection approaching 90% is mainly induced by destructive interference between the incident light and optical field of exciton at resonance. The upper limit of reflection depends on the exciton's radiative decay rate ratio and non-radiative decay rate. At cryogenic temperature, the contribution of nonradiative decay rate is quenched to a minimum but non-zero. Such an electric tunable mirror can be used to realize active photonic devices, such as modulators and metasurfaces. Besides, 1L-TMDC has also been demonstrated as a nonlinear mirror[506].

Bulk TMDC materials have a high refractive-index in the visible and near the infrared range that is comparable to or even greater than semiconductors like Si, GaAs, and GaN. It is well established that the high index dielectric structure supports the Mie resonance (also named as leaky mode resonance)[329,507–509]. This makes it possible to construct resonant photonic devices based on subwavelength Mie resonators. Besides, similar to the 1L-TMDC film, they also support the exciton resonance. Through varying the size and aspect ratio of nanodisk with SEM images shown in Fig.22k, Verre et al. experimentally realized the strong light-matter interaction between anapole and exciton[510] (Fig.22l), which is evidenced by the formation of anapole-exciton polariton with Rabi-Splitting 190 meV. The anapole corresponds to a dark state which has suppressed scattering in the far-field but strong field enhancement in the near-field (Fig.22m). It is the direct consequence of destructive interference between toroidal dipole and electric dipole[240]. Besides, strong enhancement of second harmonic generation can be realized by TMDC nanodisk resonators[511]. More recently, TMDC metamaterials have been successfully fabricated with atomic precision and may hold promise for applications in optoelectronic devices requiring atomically sharp edges[512].

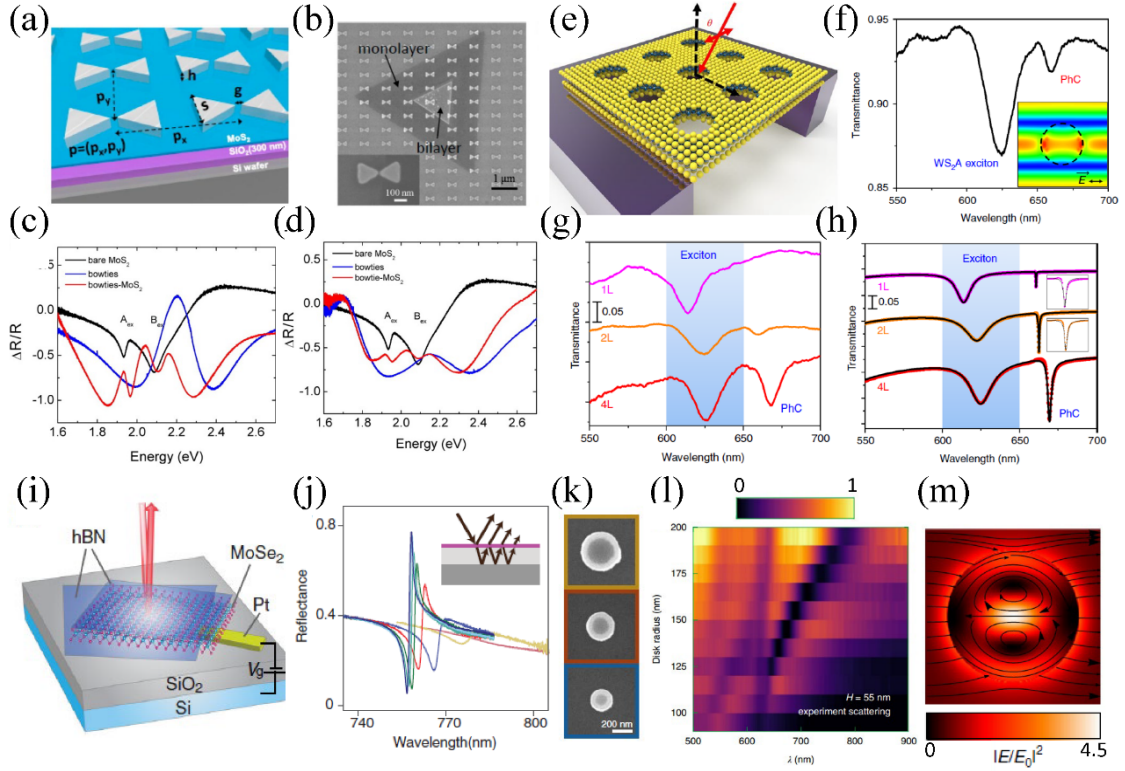

Figure 22. (a) Schematic of Ag bowtie array on 300 nm-SiO$_2$/Si substrate. (b) SEM image of fabricated Ag bowtie array on 1L-MoS2. (c-d) Measured differential reflection spectra of bare MoS$_2$, bare bowtie, and MoS$_2$-bowtie structure under the illumination of TE (c) and TM(d) polarization incidence. (e) Schematic of suspended photonic crystal slab based on few-layers WS$_2$. (f) Measured transmission spectrum of bilayer WS$_2$ photonic crystal slab. (g-h) Meausured (g) and simulated (h) Transmission spectra of 1L, 2L and 4L WS$_2$ photonic crystal slab. (i) Schematic of the atomically thin reflective mirror, which consists of MoSe$_2$ is encapsulated into hBN on SiO$_2$/Si substrate and patterned Pt platinum electrode. (j) Reflection spectra of 1L-MoSe$_2$ at different temperatures. Inset shows the multireflection process in vertical van der Waals structure on SiO$_2$/Si substrate. (k) SEM image of a fabricated WS$_2$ nanodisk with varied diameter. (l) Experimentally measured scattering spectra mapping for WS$_2$ nanodisk with varied diameter. (m) Calculated electric field distribution of anapole mode. ((a)-(d) from Ref. [324], (e)-(h) from Ref. [505], (i)-(j) from Ref.[43], (k)-(m) from Ref. [510])

## 8.2 Wavefront shaping

TMDC materials also promise great potential in manipulating the wavefront of the impinging electromagnetic wave due to the enhanced light-matter interactions, which allows for the miniaturization of optoelectronic devices. Featured by their high refractive index in the visible and near-infrared wavelength range, 1L-TMDC and few layers enables the giant optical path lengths[513], as demonstrated by Yang et, al. As shown in Fig. 23a, the optical path length is up to 40 nm at 532 nm for 1L-MoS$_2$ on 275 nm SiO$_2$/Si substrates, which is almost more than 50 times larger than the physical length of 1L-MoS$_2$ and one order magnitude larger than that of graphene. Such giant optical path lengths are caused by the interferences at the interface of air-MoS$_2$ and MoS$_2$-SiO$_2$ and the Fabry-Perot resonance in 275 nm SiO$_2$/Si thin film structure. Due to extensive and layer-dependent optical path lengths, one can realize the wide range of phase control with atomically thin thickness. By etching the 9Ls MoS$_2$ into the bowl shape, an atomically thin focusing lens (6.3 nm) was experimentally realized with the focal length being -248um (Fig.23b) while the operation wavelength is 532 nm. Here, it is quite necessary to emphasize that large optical path length is not only limited at 532 nm. It can be realized in a broadband wavelength range by simply varying the thickness of the SiO$_2$ layer.

More recently, van de Groep et al experimentally demonstrated a tunable atomically thin lens based on 1L-WS$_2$[45] (Fig.23c-d). Unlike metalenses constructed from plasmonic resonators or Mie resonators, the WS2 zone plate lens takes advantage of strong exciton resonances enabled by hundreds of meV binding energy of exciton room temperature, an order magnitude larger than in conventional materials. Although the focusing efficiency for such a flat lens is relatively low (~0.08%), dynamic modulation on the focusing efficiency with 33% maximum modulation depth was demonstrated by turning on and off exciton resonance via external gating voltages. The low focusing efficiency is mainly attributed to the low quality of 1L-WS$_2$, which is supported by the broad linewidth of exciton resonance (75 meV). Previously discussion on atomical thick mirror suggests the focusing efficiency may be further improved by encapsulating high-quality 1L-TMDC into HBN[42,43]. Another way of increasing the focusing efficiency is to generate local scattering media inside a monolayer by femtosecond laser direct writing[514]. By converting 1L-WSe$_2$ into WO$_x$ NP through photochemical effect, Lin et al. experimentally realized 3D focusing with high efficiency and diffraction-limited image. The most recent work by Qin et al. demonstrated a 1L-MoS2 supercritical lens with sub-diffraction limited focal spots[47]. The working principle is based on the loss-assisted singular phase near the critical coupling,which enables π phase jump.

Because of the relatively high indices of 2D materials (i.e., hBN and MoS$_2$), bulky 2D materials can also be patterned to form Mie resonators, which may serve as the building blocks for constructing metadevices. Liu et al. experimentally demonstrated a metalens using ultrathin vdW materials whose thickness can be reduced to ~λ/10[515]. The designing method is based on incomplete phase modulation that allows for maintaining the high focusing efficiency with limited phase coverage. Such metalens enbale not only the near diffraction-limited focusing but also strain-induced tunable focusing by transferring vdW materials onto flexible substrates due to weak interaction among vdW layered materials. Wang et al realized an atomically thin photon sieve with multilayer MoS$_2$. Thanks to the strong exciton resonance, the efficiency of 2D nanosieve holograph is enhanced by ten folds compared with the non-2D counterpart. Also, they demonstrated 2D metalens, which enables focusing beyond the diffraction limits[46]. Except for TMDC materials, other 2D materials are developed to build the atomically thin photonic devices[516]. For example, Wang et al realized binary atomic thick meta-optics by exploiting the topological darkness of structured PtSe$_2$ films with a thickness of only 4.3 nm. Black phosphorus with high anisotropy was experimentally demonstrated as an optical waveplate that can rotate polarization of 0.05° per atomic layer at 520 nm[517], which is comparable to the commercial bulky birefringence materials like CaCO$_3$.

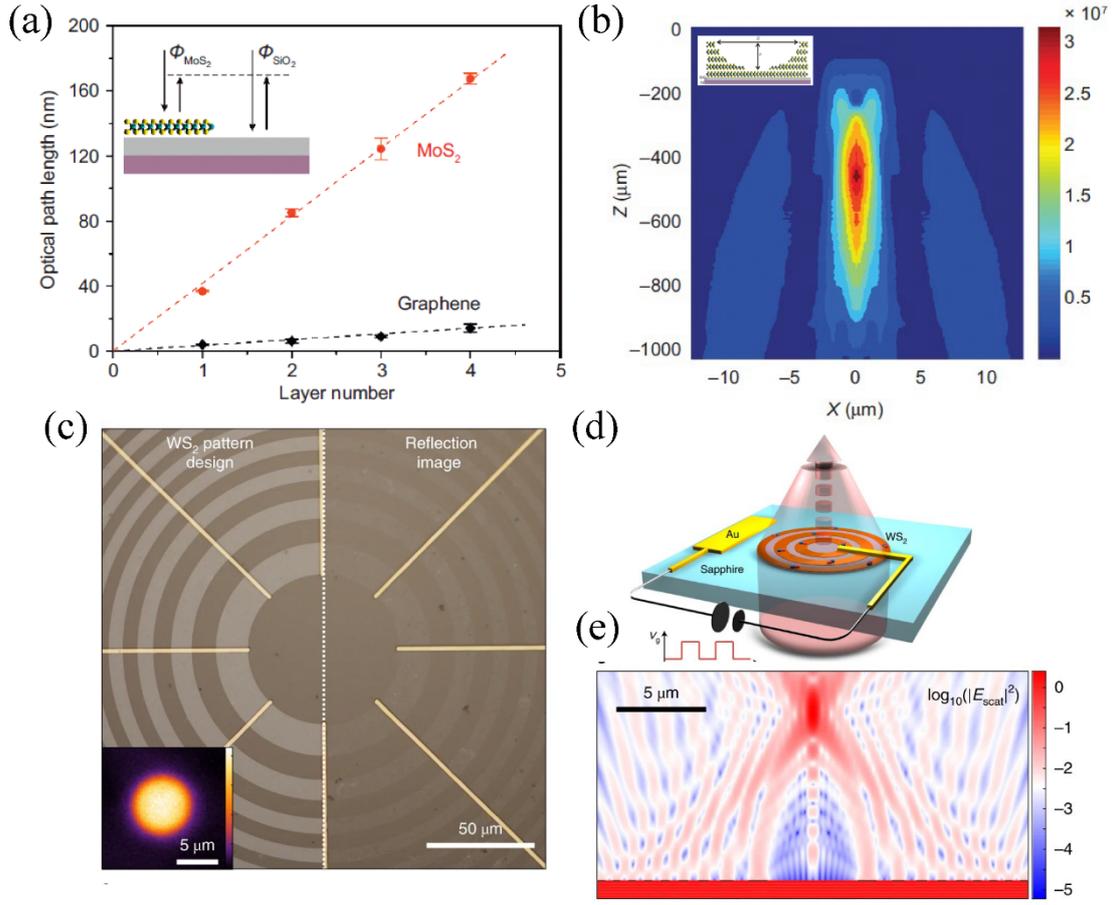

Figure 23. (a) Measured optical path length of 1L-4L MoS$_2$ and graphene by the phase-shifting interferometer. (b) Spatial mapping of the intensity of MoS$_2$ micro-lens measured by scanning optical microscopy. Inset is a schematic drawing of the multilayer MoS$_2$ micro-lens. (c) Optical image of fabricated zone plate lens, where left is fabricated lens and right is WS$_2$ patterned overlaid. Inset is x-y scanned focus taken at 2mm above the pattern area at 620 nm. (d) Schematic drawing of the WS$_2$ zone plate lens. (e) Scattered field intensity behind a zone plate lens with a focal length f=10μm on sapphire at 620 nm. ((a)-(b) from Ref. [513], (c)-(d) from Ref. [45])

## 9. *Light-Matter Interactions in van der Waals Heterostructure*

As the surface of 1L-TMDC is free of dangling bonds and the interaction between layers is dominated by the vdW force, a heterostructure can be built by directly assembling dissimilar 1L-TMDC vertically without resorting to lattice matching. TMDC-based heterostructures provide more freedom to engineer the optical and electronic properties of quantum materials than in 1L-TMDC. In this section, we briefly review the recent progress in light-matter interaction of TMDC based heterobilayers. One may find more comprehensive reviews from Rivera et al[163] and Jin et al. [164].

**9.1 Optical Properties of TMDC-Based Heterostructures**

First principle calculations predict that type II band alignment is induced when MoX$_2$ (X=S and Se) is stacked on WX$_2$ (X=S and Se)[165]. The direct consequence of type II band alignment is to form an interlayer exciton, where the electrons and holes are located in the separated monolayer. To verify the theoretical prediction, femtosecond pump-probe spectroscopy has been used because it allows researchers to study charge and energy transfer. For example, the first experiment on ultrafast dynamics in heterostructure $MoS_2$-$WS_2$ was done by Hong et al. [518]. In their experiments, the pump photon energy is intentionally chosen as 1.86eV to excite the exciton of $MoS_2$ exclusively, and transient absorption spectra were measured and given in Fig.24a. Two resonant peaks are found at 2.06eV and 2.46eV, belonging to A exciton and B exciton of $WS_2$. The transient response at high photon energy represents the signature of the charge transfer of the hole from 1L-$MoS_2$ to 1L-$WS_2$. Moreover, they found that such a charge transfer process takes place within 50fs. Except for charge transfer, another prominent feature of the heterobilayer is the quenched PL compared to PL within the respective monolayer. The quenched effect of PL intensity was also reported in the as-grown $MoS_2$/$WS_2$ heterostructure (Fig.24c-d). Later, ultrafast dynamics has been studied on the other heterostructure with different conditions[519–522].

The existence of interlayer exciton was experimentally confirmed in heterostructure $MoSe_2$-$WSe_2$ by Rivera et al. [523]. The hetero-bilayer is manufactured by merely stacking mechanically exfoliated 1L-$MoSe_2$ on 1L-$WSe_2$, as shown in Fig. 24e. PL measurement at room temperature, as shown in Fig.24f, reveals that there are three main peaks in the spectrum. While the emission peaks at 1.65eV and 1.57eV are correlated to the exciton emission for $MoSe_2$ and $WSe_2$, respectively, the other one at 1.35eV is assigned as the emission of interlayer exciton due to type II band structure (Fig.24g). This was further confirmed by the PL mapping around interlayer exciton. Besides, it was also demonstrated that such an interlayer exciton can be dynamically controlled by external gating bias. Time-resolved PL measurement indicates that interlayer exciton's lifetime is 1.8ns (Fig.24h), that is almost two order longer than that of the intralayer exciton. The interlayer exciton was also observed by different groups in the other hetero-bilayers, such as $MoS_2$-$WS_2$[519,524,525], $MoS_2$-$MoSe_2$[526], $MoS_2$-$WSe_2$[527–530], $WS_2$-$MoSe_2$[531], $WS_2$-$WSe_2$[520], and $MoSe_2$-$WSe_2$[166,524,532–534].

Note that the interlayer exciton is not always bright (bright means observable in PL spectrum) in all hetero-bilayers. Its emission intensity strongly depends on the interlayer coupling strength[519,520,533,535]. The interlayer coupling strength can be enhanced by either thermal annealing or inserting hBN layers between 1L-TMDCs. Besides, Nayak et al found that twisted angle between two monolayers of heterostructure $MoSe_2$-$WSe_2$ also affect the PL intensity of interlayer exciton in a way that PL intensity reaches the maximum value at θ=0° and θ=60°, but suppresses to almost zero at 10°<θ<50°[533]. Additionally, interlayer coupling also gives rise to low-frequency Raman modes when the twisted angle is relatively small. Here, it is of great necessity to point out that the twisted angle of the constituent monolayers can be obtained from the polarization-resolved second harmonic generations.

Apart from heterostructure manufactured from manually stacking dissimilar 1L-TMDC, there are also tremendous efforts made on the growth of TMDC based vertical heterostructure[536–539]. For example, Gong et al reported a single-step vapor phase growth of both in-plane heterostructure and vertical heterostructure $WS_2$-$MoS_2$ by controlling the growth temperature[536]. Similar to manually stacking heterostructure, the interlayer exciton was observed at 875 nm due to Type II band alignment. At the same time, PL was quenched in the heterostructure area. Yu et al also reported the epitaxial growth of vertical heterostructure $MoS_2$-$WS_2$[537]. The efficient interlayer exciton relaxation was demonstrated in both epitaxial and non-epitaxial heterostructure, as evidenced by the two orders of PL magnitude quenching in the heterojunction area (Fig.24d) compared to 1L-$MoS_2$ and improved absorption that is particular prominent in the frequency below the bandgap of $MoS_2$.

Note that all above vertical heterostructure has fixed interlayer emission because of the fixed bandgap for each TMDC layer. More recently, Li et al realized a wavelength-tunable interlayer exciton emission in the near-infrared based on a vdW heterostructure that is composed of as-grown 1L-WSe$_2$ onto another as-grown monolayer alloy WS$_{2(1-x)}$Se$_{2x}$[540]. The experimental results show that the emission photon energy of interlayer exciton can be tuned from 1.52eV to 1.40eV, mainly arising from the tunable optical bandgap for alloy 1L-WS$_{2(1-x)}$Se$_{2x}$ covered a range of 1.97eV-1.40eV with modulated composition. In addition to the growth of vertical heterostructure, lateral heterojunction has been successfully synthesized by several different groups[536,538,541–543]. For example, Duan et al demonstrated the synthesis of lateral heterojunction MoS$_2$-MoSe$_2$ and WS$_2$-WSe$_2$ by in situ switchings of the vapor-phase reactants during the growth[541]. Li et al reported the two-step epitaxial growth of lateral WSe$_2$-MoS$_2$ heterojunction with atomically sharp transition in compositions at junction[542], where single-crystalline triangle 1L-WSe$_2$ is grown in one furnace and then the growth of 1L-MoS$_2$ is performed in a separate furnace. All these growth techniques may laydown the foundation of developing large-scale heterostructure-based electronic, photonic and optoelectronic devices.

Like the exciton of 1L-TMDC carrying information encoded in valleys, valley polarization is still maintained in the interlayer exciton in the heterostructure. Rivera et al studied the valley-polarization dynamics in heterostructure 1L-WSe$_2$-MoSe$_2$[419](Fig.24i). Polarization-resolved PL measurement inferred that the valley-specified interlayer exciton can be generated when the heterostructure is illuminated by the CW laser with circularised polarization. Moreover, the degree of valley polarization and lifetime for interlayer exciton can be dynamically tuned by the gating bias. Ultrafast dynamics measurement based on pulse pumping shows that the lifetime of spin-valley polarization for interlayer exciton 39±2ns at 60V (Fig.24j), several orders longer than that of intravalley excitons in monolayer, implying suppressed intervalley scattering for interlayer excitons. The long interlayer exciton lifetime enables visualization of lateral drift and diffusion. Similarly, Kim et al reports the observation of microsecond-long-lived valley polarization in heterostructure WSe$_2$-MoS$_2$ by ultra-fast pump-probe spectroscopy[544] (Fig.24k-l). Due to the ultrafast charge transfer process, the valley polarized hole can be generated in 1L-WSe$_2$ within 50fs (Fig.24l). The degree of valley polarization is close to unity for the valley polarized hole (Fig.24k). Except for electric tuning, the valley of interlayer exciton can also be manipulated by the magnetic field. It is found the lifetime of the valley in interlayer exciton can be changed.

It is well known that the p-n junction constitutes the fundamental element of modern semiconductors and has found numerous applications in LED, photodetector, and photovoltaics. The emergence of TMDC materials allows us to build a p-n junction with vdW vertical heterostructure. Three groups independently reported the fabrication of p-n heterojunction based on 1L-WSe$_2$-MoS$_2$[52,545,546]. Such a p-n junction enables the observation of gate tunable diode-like current rectification, photovoltaic response across the interface as well as electroluminescence. Except heterostructure based on exfoliated monolayers, the synthesized later heterojunction has also been demonstrated to be able to construct p-n junction[541,542]. A CMOS inverter is successfully fabricated based on WS$_2$-WSe$_2$ p-n heterojunction[541].

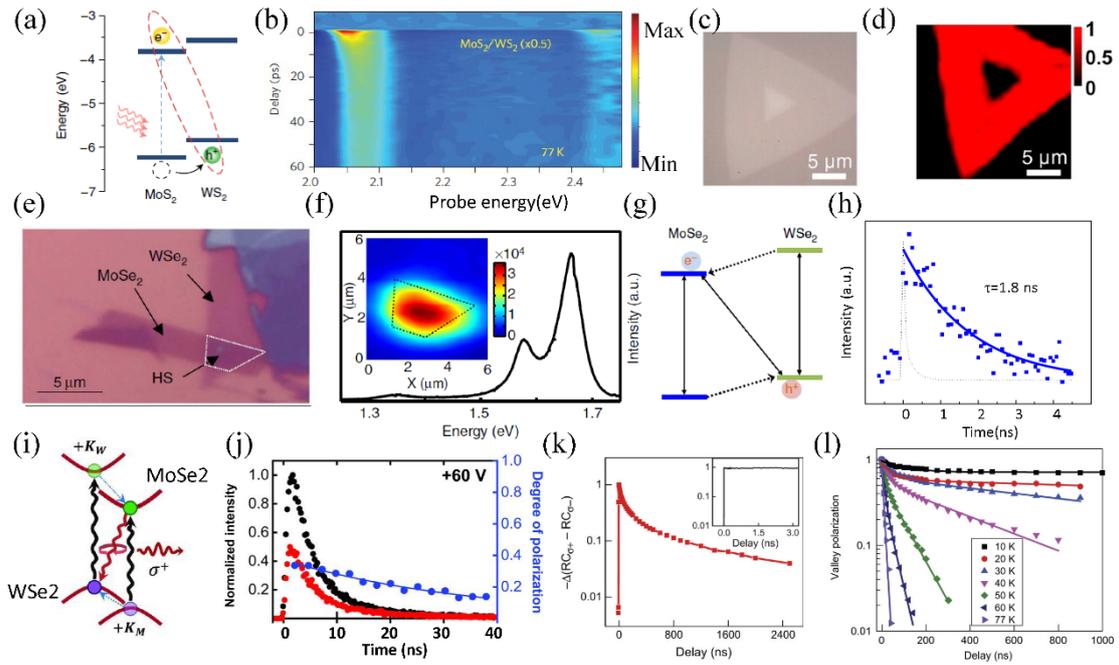

Figure 24. Exciton properties in hetero-bilayers (a) Schematic of the band alignment of type II heterojunction $MoS_2/WS_2$. (b) Transient absorption spectra mapping from a $MoS_2/WS_2$ heterostructure at 77K. (c) Optical image of the heterostructure of as-grown $MoS_2/WS_2$ heterostructure. (d) PL mapping of heterostructure that is shown in (c). (e) Optical image of a $MoSe_2$–$WSe_2$ heterostructure highlighted with the white dashed line. (f) PL spectrum of heterostructure under 20 mW laser excitation at 532 nm. Inset is spatial mapping of PL intensity integrated from the low-energy peak (1.273–1.400 eV). (g) Schematic drawing of Type-II band alignment diagram of a 2D $MoSe_2$-$WSe_2$ heterojunction. (h) Time-resolved PL of the interlayer exciton (1.35 eV). The life time is extracted as about 1.8 ns. (i) Schematic drawing of the interlayer exciton from a heterobilayer WSe2-MoSe2 in the +K valley. (j) Time-resolved interlayer exciton right (black) and left (red) circular polarization PL at 60V while the heterostructure is illuminated by laser with right circular polarization. (k) Decay dynamics of the photoinduced circular dichroism signal in a $WSe_2/MoS_2$ heterostructure at 10 K. (l) Decay dynamics of valley polarization within a wide range from 10 to 77 K. ((a)-(b) from Ref. [518], (c)-(d) from Ref. [537], (e)-(h) from Ref. [523], (i)-(j) from Ref. [419], (k)-(l) from Ref. [544])

## 9.2 Cavity-Enhanced Light-Matter Interactions in TMDC-Based Heterostructures

Given that TMDC-based heterostructures can emit light based on interlayer excitons with relatively long lifetime due to the spatial separation of electrons and holes in different layers, they may bring more opportunities in developing heterobilayer structure-based photonic devices. One of the most salient applications is lasing in TMDC based heterostructure. Liu et al. demonstrated room temperature interlayer exciton lasing with $MoS_2$-$WSe_2$ heterostructure, which was transferred on silicon photonic crystal nanocavity[168](Fig.25a-c). Heterobilayers $MoS_2$-$WSe_2$ was arranged as AA stacking such that it supports bright interlayer exciton in the near-infrared due to the nature of Type II band alignment. The PL spectra for such a hybrid heterobilayer-cavity system are shown in Fig.25b. Unlike exciton laser in 1L-TMDC, the requirement for the Q-factor of nanocavity can be relaxed due to such long life time of the interlayer exciton. In their design, the photonic crystal cavity's Q-factor is only 423, lower than that of the previous design in 1L-TMDC. The lasing behavior was confirmed by the distinct kink in the "L-L" curve and spectral linewidth narrowing (Fig.25c). The lasing threshold is low as 33μW at 5K and 54μW at room temperature.

Deng's group from the University of Michigan employed a different design to realized interlayer exciton laser in TMDC heterobilayer structure ($WSe_2$-$MoSe_2$), where heterostructure was integrated into a silicon nitride grating resonator[169] (Fig.25d-f). The cavity mode was confirmed by measuring angle-resolved reflectance spectroscopy, yielding Q factor between 500 and 680. The threshold of lasing is even as low as 0.18μW. Except for distinct kink in L-L curve and spectral linewidth narrowing, the spatial coherence of emission, as another important figure of merits of laser, was measured for the first time (Fig.25f). The coherence length increases abruptly from 2.38μm to around 5μm when the pump power passes through the threshold.

A more recent study by Khelifa et al indicates that the interlayer exciton of heterobilayer $MoSe_2$-$WSe_2$ can be coupled to a waveguide-coupled disk resonator made of hBN resonator[70]. The heterobilayer is sandwiched between two slabs of hBN, and then patterned as waveguide-coupled resonators. Compared to other designs, such a system allows it to position the TMDC heterostructure in the highest optical field intensity, thus maximizing light-matter interaction. Experimental results demonstrated that the emission of the interlayer exciton is effectively modulated by the optical modes in hBN nanodisks when the detection was put at the end of waveguide. When heterostructure $MoSe_2$-$WSe_2$ was embedded into a tunable microcavity system, Purcell enhancement was observed for the heterobilayer emission of interlayer exciton as evidenced by the simultaneous increase of PL and radiative decay rate[547]. The interlayer exciton in $WS_2$/$MoS_2$ can also be strongly coupled to Mie resonance in a single silicon NP[548].

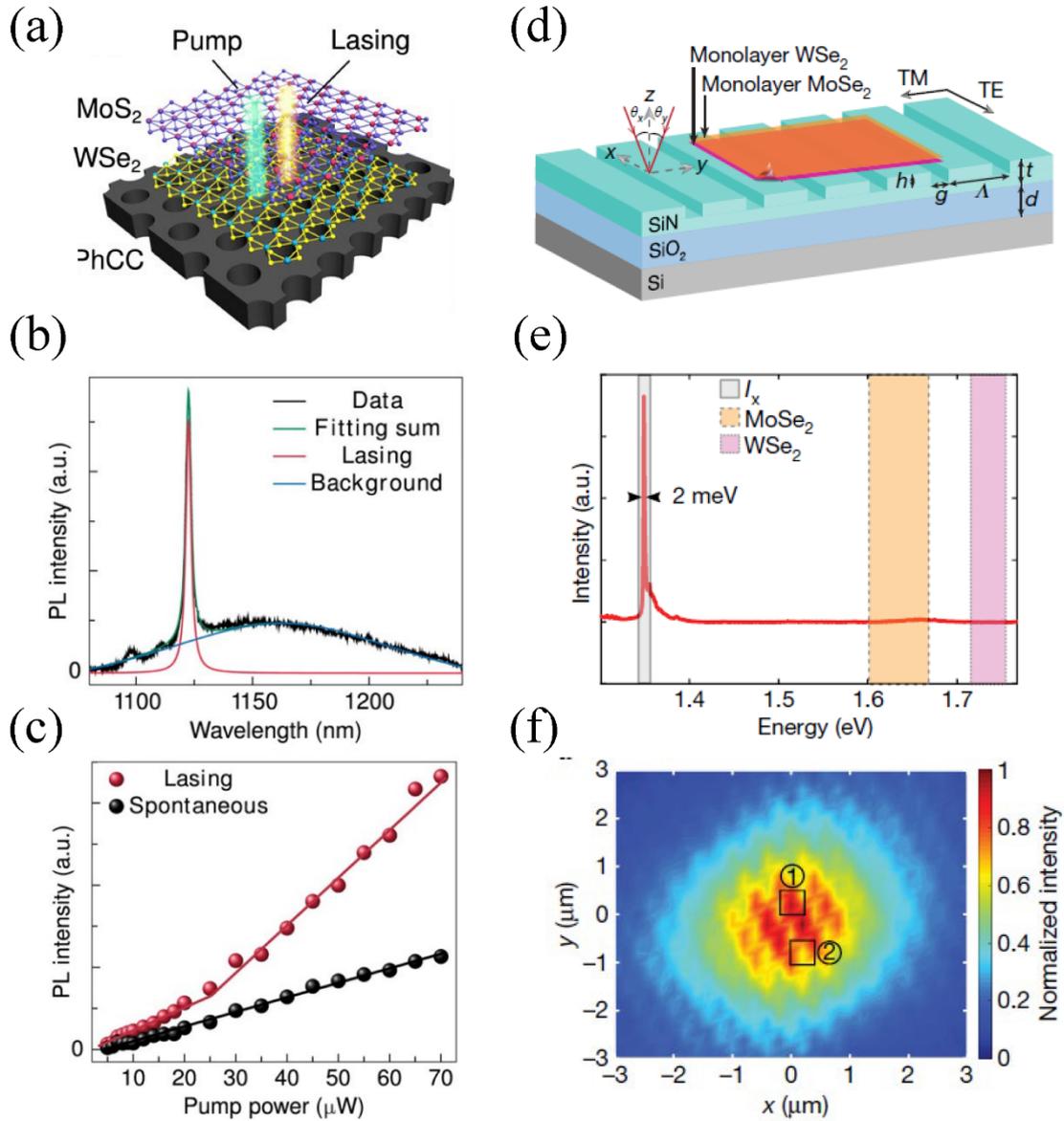

Figure 25. Lasing in van der Waals heterobilayer. (a) Schematics of the fabricated heterobilayer-photonic crystal cavity nanolaser. (b) lasing emission in heterobilayer-cavity region and PL emission of heterobilayer on Si slab. (c) Emission intensity at the laser wavelength versus the pump power. (d) 3D schematic illustration of the laser device which consists a heterobilayer 1L-WSe$_2$/1L-MoSe$_2$ on Si$_3$N$_4$ grating cavity. (e) PL spectrum from the heterobilayer under the pumping of the 633 nm laser with power of 20 μW. (f) Interference pattern above threshold power 20 μW. ((a)-(c) from Ref. [168], (d)-(f) from Ref. [169])

## *Summary and outlook*

2D TMDCs provide a versatile platform for engineering light-matter interactions and developing atomically thin electric, photonic and optoelectronic devices. This paper has reviewed the recent progress in enhanced light-matter interactions of 2D TMDC materials from the perspective of light emission, absorption, and scattering, which covers a broad range of topics from fundamental physics to practical applications. The performance of TMDC-based photonic and optoelectronic devices are optimized by either integrating them with resonant photonic cavities or exploiting themselves as resonant structures.

As a rapidly developing area and hot topic at the current stage, there are many newly published or ongoing works on TMDC materials and nanophotonics. Thus, it is impossible to cover all possible aspects related to TMDC photonics and optoelectronics. For instance, we did not discuss advances in photodetectors based on TMDC materials[549], especially on improving photodetectors' performance from different perspectives[550,551].

The marriage of TMDC materials and nanophotonics has witnessed great success in the past decade. Many interesting physics and applications have been explored by combining the advantages of both TMDC materials and resonant photonics. There are still many exciting opportunities and promising future directions.

(1) Growth of large scale TMDC materials with high quality. While significant progress has been made in growing wafer-scale 1L-TMDC with chemical vapor deposition or metal-organic chemical vapor deposition[30,552], high-quality 1L-TMDC with large mobility and high quantum yield is desired. Low mobility and quantum yield are two bottleneck factors that limit the performance of TMDC based photonic and optoelectronic devices. TMDC with large mobility and high quantum yield may significantly improve the performance of LED, ultralow threshold lasing, fast response photodetector. It might also be perfect if the controlled defect materials are grown because they are essential to realizing single quantum emitter. Also, as vdW heterostructure bears many exotic properties that monolayer does not have, it is desirable to grow a large scale of heterobilayer with a controlled rotation angle between them.

(2) High performance photonic and optoelectronic devices. Although tremendous progress has been made on photonics and optoelectronic devices, there are still many challenges and exciting opportunities. For example, all lasers reported so far are optically pumped. It might be more appealing to realize electrically pumped lasing. Gu et al. demonstrate efficient light emission by sandwiching the 1L-$WS_2$ into two DBR based cavities. Further increase of the cavity's Q-factor may help to realise electrically pumping laser[360]. Also, the unique chiral response of 1L-TMDC opens the door of realizing valley laser or circularly polarized laser. Although intervalley scattering reduces the valley coherent time, it is still possible to amplify the chiral response by combining the 1L-TMDC with plasmonic chiral structure[553]. A recently emerged topic called bound state in the continuum provides the solution of realizing large chirality by breaking the symmetry of resonant structure[554]. Thus, photonic crystal slab supporting BIC may be another nice platform of achieving circular polarization lasing in 1L-TMDC. In addition, due to the large binding energy of exciton, strong coupling has been reported in many hybrid cavity-1L-TMDC systems, accompanied by the formation of exciton-polariton. However, little attention has been paid to polariton lasing. A most recent study on coherent polariton lasing with an ultra-low threshold is reported by Zhao et al [555]. This work may stimulate further research on polariton superfluid and valley polariton lasing. For TMDC based photodetectors, there is also plenty of room to improve their performances. Massicotte et al demonstrated the ultrafast photodetector with several picosecond photoresponses based on the vertical vdW heterostructure graphene/WSe2/graphene[556]. The external quantum efficiency is only 7.3% due to the weak absorption in 1L-WSe2. It can be improved by integrating such vertical heterostructure with resonant photonic structures (optical waveguide, microring resonator, dielectric metasurface.

(3) Atomically thin metasurface[557]. 1L- and a few-layer TMDCs have been successfully used to construct metalenses. More functional metasurfaces should be explored to engineer the amplitude, phase, polarization and frequency of electric field based on TMDC materials. Also, given the excellent tunability compared to the conventional materials, tunable metasurface would be another interesting direction. Besides, because bulk TMDC has a large refractive index and supports exciton resonance, most of the applications developed in high-index semiconductors (i.e. Si, GaAs, AlGaAs) can be transplanted to TMDC materials.

(4) New opportunities of combining TMDC materials with topological photonics and non-Hermitian photonics. Topological photonics is one of the most active areas in recent photonic research[558]. It offers scientists the opportunity to fabricate tunable photonic devices that are robust against disorder and imperfections. Recently, two groups independently realized helical topological exciton-polaritons by strongly coupling of valley polarized exciton in 1L-TMDC with all-dielectric topological photonic metasurface[559,560]. The nature of topological polaritons makes the polariton with opposite helicity propagate in opposite directions. Also, a recently demonstrated topological nanocavity may be integrated with 1L-TMDC to enhance light-matter interactions[561]. In a related but distinct field of research, an interesting phenomenon related to non-Hermitian photonics is exceptional points[562], where both the eigenvalue and eigenstate of the system are degenerate. Exceptional points have been used to realize sensors with large sensitivity and chiral lasers. It will be interesting to see what may happen if we combine exceptional points with 1L-TMDC to further enhance light-matter interactions and nonlinearities, and leverage the valley response for unusual chiral emission.

(5) Light-matter interaction in heterostructures. There are few works on cavity-enhanced light-matter interactions in vdW heterostructures, focusing on improving the emission of interlayer excitons from type-II heterobilayers. On the one hand, other resonant photonic structures could be used further to reduce the lasing threshold at room temperature. On the other hand, it is also interesting to explore valleytronics in heterostructures by combining chiral metamaterials. Also, twistronics for photons gives us more freedom to engineer the transport of bright and dark excitons[563]. It is interesting to develop hybrid cavity-twistronic systems to realize strong coupling between them.

## Acknowledgments
L. H and A. E. M acknowledges support from the Australian Research Council Discovery Project (DP200101353). A. K and A. A acknowledge the Vannevar Bush Faculty Fellowship, the Air Force Office of Scientific Research, and the Simons Foundation. Y.Y acknowledges support from the U.S. Department of Energy, Office of Science, Basic Energy Sciences (BES), Materials Sciences and Engineering Division. D. N acknowledges support from the Australian Research Council Centre of Excellence for Transformative Meta-Optical Systems (CE20010001).

Reference

1. Novoselov, K. S. *et al.* Electric Field Effect in Atomically Thin Carbon Films. *Science (80-. ).* **306**, 666 LP-669 (2004).
2. Novoselov, K. S., Mishchenko, A., Carvalho, A. & Castro Neto, A. H. 2D materials and van der Waals heterostructures. *Science (80-. ).* **353**, (2016).
3. Butler, S. Z. *et al.* Progress, Challenges, and Opportunities in Two-Dimensional Materials Beyond Graphene. *ACS Nano* **7**, 2898–2926 (2013).
4. Bhimanapati, G. R. *et al.* Recent Advances in Two-Dimensional Materials beyond Graphene. *ACS Nano* **9**, 11509–11539 (2015).
5. Zhou, J. *et al.* A library of atomically thin metal chalcogenides. *Nature* **556**, 355–359 (2018).
6. Splendiani, A. *et al.* Emerging Photoluminescence in Monolayer MoS2. *Nano Lett.* **10**, 1271–1275 (2010).
7. Mak, K. F., Lee, C., Hone, J., Shan, J. & Heinz, T. F. Atomically Thin MoS2: A New Direct-Gap Semiconductor. *Phys. Rev. Lett.* **105**, 136805 (2010).
8. Radisavljevic, B., Radenovic, A., Brivio, J., Giacometti, V. & Kis, A. Single-layer MoS2 transistors. *Nat. Nanotechnol.* **6**, 147–150 (2011).
9. Wang, G. *et al.* Colloquium: Excitons in atomically thin transition metal dichalcogenides. *Rev. Mod. Phys.* **90**, 21001 (2018).
10. Schneider, C., Glazov, M. M., Korn, T., Höfling, S. & Urbaszek, B. Two-dimensional semiconductors in the regime of strong light-matter coupling. *Nat. Commun.* **9**, 2695 (2018).
11. Mak, K. F., Xiao, D. & Shan, J. Light–valley interactions in 2D semiconductors. *Nat. Photonics* **12**, 451–460 (2018).
12. Krasnok, A., Lepeshov, S. & Alú, A. Nanophotonics with 2D transition metal dichalcogenides [Invited]. *Opt. Express* **26**, 15972 (2018).
13. Mak, K. F. & Shan, J. Photonics and optoelectronics of 2D semiconductor transition metal dichalcogenides. *Nat. Photonics* **10**, 216 (2016).
14. Brar, V. W., Sherrott, M. C. & Jariwala, D. Emerging photonic architectures in two-dimensional opto-electronics. *Chem. Soc. Rev.* **47**, 6824–6844 (2018).
15. Xia, F., Wang, H., Xiao, D., Dubey, M. & Ramasubramaniam, A. Two-dimensional material nanophotonics. *Nat. Photonics* **8**, 899–907 (2014).
16. Sun, Z., Martinez, A. & Wang, F. Optical modulators with 2D layered materials. *Nat. Photonics* **10**, 227–238 (2016).
17. Xu, X., Yao, W., Xiao, D. & Heinz, T. F. Spin and pseudospins in layered transition metal dichalcogenides. *Nat. Phys.* **10**, 343–350 (2014).
18. Schaibley, J. R. *et al.* Valleytronics in 2D materials. *Nat. Rev. Mater.* **1**, 16055 (2016).
19. Cao, L. Two-dimensional transition-metal dichalcogenide materials: Toward an age of atomic-scale photonics. *MRS Bull.* **40**, 592–599 (2015).
20. Zheng, W. *et al.* Light Emission Properties of 2D Transition Metal Dichalcogenides: Fundamentals and Applications. *Adv. Opt. Mater.* **6**, 1800420 (2018).
21. Wilson, J. A. & Yoffe, A. D. The transition metal dichalcogenides discussion and interpretation of the observed optical, electrical and structural properties. *Adv. Phys.* **18**, 193–335 (1969).
22. Frindt, R. F. Single Crystals of MoS 2 Several Molecular Layers Thick. *J. Appl. Phys.* **37**, 1928–1929 (1966).
23. Bromley, R. A., Murray, R. B. & Yoffe, A. D. The band structures of some transition metal

dichalcogenides. III. Group VIA: trigonal prism materials. *J. Phys. C Solid State Phys.* **5**, 759–778 (1972).

24. Yun, W. S., Han, S. W., Hong, S. C., Kim, I. G. & Lee, J. D. Thickness and strain effects on electronic structures of transition metal dichalcogenides: 2H-MX2 semiconductors (M= Mo, W; X= S, Se, Te). *Phys. Rev. B* **85**, 033305 (2012).
25. Zhao, W. *et al.* Origin of Indirect Optical Transitions in Few-Layer MoS 2 , WS 2 , and WSe 2. *Nano Lett.* **13**, 5627–5634 (2013).
26. Novoselov, K. S. *et al.* Two-dimensional atomic crystals. *Proc. Natl. Acad. Sci.* **102**, 10451–10453 (2005).
27. Tan, C. *et al.* Recent Advances in Ultrathin Two-Dimensional Nanomaterials. *Chem. Rev.* **117**, 6225–6331 (2017).
28. Li, H. *et al.* Rapid and Reliable Thickness Identification of Two-Dimensional Nanosheets Using Optical Microscopy. *ACS Nano* **7**, 10344–10353 (2013).
29. Elías, A. L. *et al.* Controlled Synthesis and Transfer of Large-Area WS 2 Sheets: From Single Layer to Few Layers. *ACS Nano* **7**, 5235–5242 (2013).
30. Yu, Y. *et al.* Controlled Scalable Synthesis of Uniform, High-Quality Monolayer and Few-layer MoS2 Films. *Sci. Rep.* **3**, 1866 (2013).
31. van der Zande, A. M. *et al.* Grains and grain boundaries in highly crystalline monolayer molybdenum disulphide. *Nat. Mater.* **12**, 554–561 (2013).
32. Berkelbach, T. C. & Reichman, D. R. Optical and Excitonic Properties of Atomically Thin Transition-Metal Dichalcogenides. *Annu. Rev. Condens. Matter Phys.* **9**, 379–396 (2018).
33. Selig, M. *et al.* Excitonic linewidth and coherence lifetime in monolayer transition metal dichalcogenides. *Nat. Commun.* **7**, 13279 (2016).
34. Li, T. & Galli, G. Electronic Properties of MoS 2 Nanoparticles. *J. Phys. Chem. C* **111**, 16192–16196 (2007).
35. Zhu, Z. Y., Cheng, Y. C. & Schwingenschlögl, U. Giant spin-orbit-induced spin splitting in two-dimensional transition-metal dichalcogenide semiconductors. *Phys. Rev. B - Condens. Matter Mater. Phys.* **84**, 1–5 (2011).
36. Jin, W. *et al.* Direct Measurement of the Thickness-Dependent Electronic Band Structure of MoS2 Using Angle-Resolved Photoemission Spectroscopy. *Phys. Rev. Lett.* **111**, 106801 (2013).
37. Zhang, Y. *et al.* Direct observation of the transition from indirect to direct bandgap in atomically thin epitaxial MoSe2. *Nat. Nanotechnol.* **9**, 111–115 (2014).
38. Lee, C. *et al.* Anomalous Lattice Vibrations of Single- and Few-Layer MoS2. *ACS Nano* **4**, 2695–2700 (2010).
39. Zeng, H. *et al.* Optical signature of symmetry variations and spin-valley coupling in atomically thin tungsten dichalcogenides. *Sci. Rep.* **3**, 1608 (2013).
40. Li, Y. *et al.* Measurement of the optical dielectric function of monolayer transition-metal dichalcogenides: MoS2, Mo S e2, WS2, and WS e2. *Phys. Rev. B - Condens. Matter Mater. Phys.* **90**, 205422 (2014).
41. Bernardi, M., Palummo, M. & Grossman, J. C. Extraordinary Sunlight Absorption and One Nanometer Thick Photovoltaics Using Two-Dimensional Monolayer Materials. *Nano Lett.* **13**, 3664–3670 (2013).
42. Back, P., Zeytinoglu, S., Ijaz, A., Kroner, M. & Imamoğlu, A. Realization of an Electrically Tunable Narrow-Bandwidth Atomically Thin Mirror Using Monolayer ${\mathrm{MoSe}}_{2}$.

*Phys. Rev. Lett.* **120**, 37401 (2018).

43. Scuri, G. *et al.* Large Excitonic Reflectivity of Monolayer ${\mathrm{MoSe}}_{2}$ Encapsulated in Hexagonal Boron Nitride. *Phys. Rev. Lett.* **120**, 37402 (2018).
44. Krasnok, A. Metalenses go atomically thick and tunable. *Nat. Photonics* **14**, 409–410 (2020).
45. van de Groep, J. *et al.* Exciton resonance tuning of an atomically thin lens. *Nat. Photonics* **14**, 426–430 (2020).
46. Wang, Z. *et al.* Exciton-Enabled Meta-Optics in Two-Dimensional Transition Metal Dichalcogenides. *Nano Lett.* **20**, 7964–7972 (2020).
47. Qin, F. *et al.* π-phase modulated monolayer supercritical lens. *Nat. Commun.* **12**, 32 (2021).
48. Lopez-Sanchez, O., Lembke, D., Kayci, M., Radenovic, A. & Kis, A. Ultrasensitive photodetectors based on monolayer MoS 2. *Nat. Nanotechnol.* **8**, 497–501 (2013).
49. Perkins, F. K. *et al.* Chemical Vapor Sensing with Monolayer MoS 2. *Nano Lett.* **13**, 668–673 (2013).
50. Radisavljevic, B., Whitwick, M. B. & Kis, A. Integrated Circuits and Logic Operations Based on Single-Layer MoS 2. *ACS Nano* **5**, 9934–9938 (2011).
51. Wang, H. *et al.* Integrated Circuits Based on Bilayer MoS 2 Transistors. *Nano Lett.* **12**, 4674–4680 (2012).
52. Cheng, R. *et al.* Electroluminescence and Photocurrent Generation from Atomically Sharp WSe2/MoS2 Heterojunction p–n Diodes. *Nano Lett.* **14**, 5590–5597 (2014).
53. Lopez-Sanchez, O. *et al.* Light Generation and Harvesting in a van der Waals Heterostructure. *ACS Nano* **8**, 3042–3048 (2014).
54. Pospischil, A., Furchi, M. M. & Mueller, T. Solar-energy conversion and light emission in an atomic monolayer p–n diode. *Nat. Nanotechnol.* **9**, 257–261 (2014).
55. Chichibu, S., Azuhata, T., Sota, T. & Nakamura, S. Spontaneous emission of localized excitons in InGaN single and multiquantum well structures. *Appl. Phys. Lett.* **69**, 4188–4190 (1996).
56. Pelekanos, N. T. *et al.* Quasi-two-dimensional excitons in (Zn,Cd)Se/ZnSe quantum wells: Reduced exciton–LO-phonon coupling due to confinement effects. *Phys. Rev. B* **45**, 6037–6042 (1992).
57. Vinattieri, A. *et al.* Exciton dynamics in GaAs quantum wells under resonant excitation. *Phys. Rev. B* **50**, 10868–10879 (1994).
58. Zhang, C., Johnson, A., Hsu, C., Li, L. & Shih, C. Direct Imaging of Band Profile in Single Layer MoS 2 on Graphite: Quasiparticle Energy Gap, Metallic Edge States, and Edge Band Bending. *Nano Lett.* **14**, 2443–2447 (2014).
59. He, K. *et al.* Tightly bound excitons in monolayer WSe2. *Phys. Rev. Lett.* **113**, 1–5 (2014).
60. Chernikov, A. *et al.* Exciton binding energy and nonhydrogenic Rydberg series in monolayer WS2. *Phys. Rev. Lett.* **113**, 1–5 (2014).
61. Ye, Z. *et al.* Probing excitonic dark states in single-layer tungsten disulphide. *Nature* **513**, 214–218 (2014).
62. Ugeda, M. M. *et al.* Giant bandgap renormalization and excitonic effects in a monolayer transition metal dichalcogenide semiconductor. *Nat. Mater.* **13**, 1091–1095 (2014).
63. Mak, K. F. *et al.* Tightly bound trions in monolayer MoS2. *Nat. Mater.* **12**, 207–211 (2013).
64. Ross, J. S. *et al.* Electrical control of neutral and charged excitons in a monolayer semiconductor. *Nat. Commun.* **4**, 1474 (2013).
65. Ross, J. S. *et al.* Electrically tunable excitonic light-emitting diodes based on monolayer WSe2

p–n junctions. *Nat. Nanotechnol.* **9**, 268–272 (2014).

66. Cuadra, J. *et al.* Observation of Tunable Charged Exciton Polaritons in Hybrid Monolayer WS2−Plasmonic Nanoantenna System. *Nano Lett.* **18**, 1777–1785 (2018).
67. Courtade, E. *et al.* Charged excitons in monolayer WSe2: Experiment and theory. *Phys. Rev. B* **96**, 1–12 (2017).
68. Berkelbach, T. C., Hybertsen, M. S. & Reichman, D. R. Theory of neutral and charged excitons in monolayer transition metal dichalcogenides. *Phys. Rev. B* **88**, 045318 (2013).
69. You, Y. *et al.* Observation of biexcitons in monolayer WSe2. *Nat. Phys.* **11**, 477–481 (2015).
70. Khelifa, R. *et al.* Coupling Interlayer Excitons to Whispering Gallery Modes in van der Waals Heterostructures. *Nano Lett.* **20**, 6155–6161 (2020).
71. Nagler, P. *et al.* Interlayer exciton dynamics in a dichalcogenide monolayer heterostructure. *2D Mater.* **4**, 025112 (2017).
72. Ciarrocchi, A. *et al.* Polarization switching and electrical control of interlayer excitons in two-dimensional van der Waals heterostructures. *Nat. Photonics* **13**, 131–136 (2019).
73. Horng, J. *et al.* Observation of interlayer excitons in MoSe2 single crystals. *Phys. Rev. B* **97**, 241404 (2018).
74. Cotrufo, M., Sun, L., Choi, J., Alù, A. & Li, X. Enhancing functionalities of atomically thin semiconductors with plasmonic nanostructures. *Nanophotonics* **8**, 577–598 (2019).
75. Chernikov, A. *et al.* Electrical Tuning of Exciton Binding Energies in Monolayer ${\mathrm{WS}}_{2}$. *Phys. Rev. Lett.* **115**, 126802 (2015).
76. Shinada, M. & Sugano, S. Interband Optical Transitions in Extremely Anisotropic Semiconductors. I. Bound and Unbound Exciton Absorption. *J. Phys. Soc. Japan* **21**, 1936–1946 (1966).
77. Qiu, D. Y., da Jornada, F. H. & Louie, S. G. Optical Spectrum of MoS2: Many-Body Effects and Diversity of Exciton States. *Phys. Rev. Lett.* **111**, 216805 (2013).
78. Wang, G. *et al.* Exciton states in monolayer MoSe 2 : impact on interband transitions. *2D Mater.* **2**, 045005 (2015).
79. Stier, A. V., McCreary, K. M., Jonker, B. T., Kono, J. & Crooker, S. A. Exciton diamagnetic shifts and valley Zeeman effects in monolayer WS2 and MoS2 to 65 Tesla. *Nat. Commun.* **7**, 10643 (2016).
80. Raja, A. *et al.* Coulomb engineering of the bandgap and excitons in two-dimensional materials. *Nat. Commun.* **8**, 15251 (2017).
81. Stier, A. V., Wilson, N. P., Clark, G., Xu, X. & Crooker, S. A. Probing the Influence of Dielectric Environment on Excitons in Monolayer WSe2: Insight from High Magnetic Fields. *Nano Lett.* **16**, 7054–7060 (2016).
82. Wang, M. *et al.* Tunable Fano Resonance and Plasmon-Exciton Coupling in Single Au Nanotriangles on Monolayer WS 2 at Room Temperature. *Adv. Mater.* **30**, 1705779 (2018).
83. Lepeshov, S. *et al.* Tunable Resonance Coupling in Single Si Nanoparticle–Monolayer WS2 Structures. *ACS Appl. Mater. Interfaces* **10**, 16690–16697 (2018).
84. Cheiwchanchamnangij, T. & Lambrecht, W. R. L. Quasiparticle band structure calculation of monolayer, bilayer, and bulk MoS 2. *Phys. Rev. B* **85**, 205302 (2012).
85. Xiao, D., Liu, G.-B., Feng, W., Xu, X. & Yao, W. Coupled Spin and Valley Physics in Monolayers of MoS2 and Other Group-VI Dichalcogenides. *Phys. Rev. Lett.* **108**, 196802 (2012).
86. Zhang, X.-X., You, Y., Zhao, S. Y. F. & Heinz, T. F. Experimental Evidence for Dark Excitons in

Monolayer WSe2. *Phys. Rev. Lett.* **115**, 257403 (2015).

87. Feierabend, M., Berghäuser, G., Knorr, A. & Malic, E. Proposal for dark exciton based chemical sensors. *Nat. Commun.* **8**, 14776 (2017).
88. Malic, E. *et al.* Dark excitons in transition metal dichalcogenides. *Phys. Rev. Mater.* **2**, 014002 (2018).
89. Zhou, Y. *et al.* Probing dark excitons in atomically thin semiconductors via near-field coupling to surface plasmon polaritons. *Nat. Nanotechnol.* **12**, 856–860 (2017).
90. Baranowski, M. *et al.* Dark excitons and the elusive valley polarization in transition metal dichalcogenides. *2D Mater.* **4**, 025016 (2017).
91. Selig, M. *et al.* Dark and bright exciton formation, thermalization, and photoluminescence in monolayer transition metal dichalcogenides. *2D Mater.* **5**, 035017 (2018).
92. Lindlau, J. *et al.* Identifying optical signatures of momentum-dark excitons in transition metal dichalcogenide monolayers. 1–7
93. Wang, M. *et al.* Dark-Exciton-Mediated Fano Resonance from a Single Gold Nanostructure on Monolayer WS2 at Room Temperature. *Small* **15**, 1900982 (2019).
94. Shree, S., Paradisanos, I., Marie, X., Robert, C. & Urbaszek, B. Guide to optical spectroscopy of layered semiconductors. *Nat. Rev. Phys.* (2020). doi:10.1038/s42254-020-00259-1
95. Molas, M. R. *et al.* Brightening of dark excitons in monolayers of semiconducting transition metal dichalcogenides. *2D Mater.* **4**, 021003 (2017).
96. Zhang, X.-X. *et al.* Magnetic brightening and control of dark excitons in monolayer WSe2. *Nat. Nanotechnol.* **12**, 883–888 (2017).
97. Koitzsch, A. *et al.* Nonlocal dielectric function and nested dark excitons in MoS2. *npj 2D Mater. Appl.* **3**, 41 (2019).
98. Zhang, C. *et al.* Probing Critical Point Energies of Transition Metal Dichalcogenides: Surprising Indirect Gap of Single Layer WSe 2. *Nano Lett.* **15**, 6494–6500 (2015).
99. Koperski, M. *et al.* Optical properties of atomically thin transition metal dichalcogenides: Observations and puzzles. *Nanophotonics* **6**, 1289–1308 (2017).
100. Kośmider, K., González, J. W. & Fernández-Rossier, J. Large spin splitting in the conduction band of transition metal dichalcogenide monolayers. *Phys. Rev. B* **88**, 245436 (2013).
101. Yuan, L., Wang, T., Zhu, T., Zhou, M. & Huang, L. Exciton Dynamics, Transport, and Annihilation in Atomically Thin Two-Dimensional Semiconductors. *J. Phys. Chem. Lett.* **8**, 3371–3379 (2017).
102. Arora, A. *et al.* Excitonic resonances in thin films of WSe 2 : from monolayer to bulk material. *Nanoscale* **7**, 10421–10429 (2015).
103. Wang, G. *et al.* Spin-orbit engineering in transition metal dichalcogenide alloy monolayers. *Nat. Commun.* **6**, 10110 (2015).
104. Withers, F. *et al.* WSe2 Light-Emitting Tunneling Transistors with Enhanced Brightness at Room Temperature. *Nano Lett.* **15**, 8223–8228 (2015).
105. Kormányos, A. *et al.* k · p theory for two-dimensional transition metal dichalcogenide semiconductors. *2D Mater.* **2**, 022001 (2015).
106. Echeverry, J. P., Urbaszek, B., Amand, T., Marie, X. & Gerber, I. C. Splitting between bright and dark excitons in transition metal dichalcogenide monolayers. *Phys. Rev. B* **93**, 121107 (2016).
107. Carozo, V. *et al.* Optical identification of sulfur vacancies: Bound excitons at the edges of monolayer tungsten disulfide. *Sci. Adv.* **3**, e1602813 (2017).

108. Liu, X. *et al.* High Performance Field-Effect Transistor Based on Multilayer Tungsten Disulfide. *ACS Nano* **8**, 10396–10402 (2014).
109. Cadiz, F. *et al.* Excitonic Linewidth Approaching the Homogeneous Limit in ${\mathrm{MoS}}_{2}$-Based van der Waals Heterostructures. *Phys. Rev. X* **7**, 21026 (2017).
110. Tonndorf, P. *et al.* Single-photon emission from localized excitons in an atomically thin semiconductor. *Optica* **2**, 347 (2015).
111. Srivastava, A. *et al.* Optically active quantum dots in monolayer WSe2. *Nat. Nanotechnol.* **10**, 491–496 (2015).
112. He, Y. M. *et al.* Single quantum emitters in monolayer semiconductors. *Nat. Nanotechnol.* **10**, 497–502 (2015).
113. Koperski, M. *et al.* Single photon emitters in exfoliated WSe2 structures. *Nat. Nanotechnol.* **10**, 503–506 (2015).
114. Chakraborty, C., Kinnischtzke, L., Goodfellow, K. M., Beams, R. & Vamivakas, A. N. Voltage-controlled quantum light from an atomically thin semiconductor. *Nat. Nanotechnol.* **10**, 507–511 (2015).
115. Iff, O. *et al.* Substrate engineering for high quality emission of free and localized excitons from atomic monolayers in hybrid architectures. **4**, 2–6 (2017).
116. Palacios-Berraquero, C. *et al.* Large-scale quantum-emitter arrays in atomically thin semiconductors. *Nat. Commun.* **8**, 1–6 (2017).
117. Branny, A., Kumar, S., Proux, R. & Gerardot, B. D. Deterministic strain-induced arrays of quantum emitters in a two-dimensional semiconductor. *Nat. Commun.* **8**, 1–7 (2017).
118. Dhakal, K. P. *et al.* Local Strain Induced Band Gap Modulation and Photoluminescence Enhancement of Multilayer Transition Metal Dichalcogenides. *Chem. Mater.* **29**, 5124–5133 (2017).
119. Cai, T. *et al.* Coupling Emission from Single Localized Defects in Two-Dimensional Semiconductor to Surface Plasmon Polaritons. *Nano Lett.* **17**, 6564–6568 (2017).
120. Cai, T. *et al.* Radiative Enhancement of Single Quantum Emitters in WSe2 Monolayers Using Site-Controlled Metallic Nanopillars. *ACS Photonics* **5**, 3466–3471 (2018).
121. Blauth, M. *et al.* Coupling Single Photons from Discrete Quantum Emitters in WSe2 to Lithographically Defined Plasmonic Slot Waveguides. *Nano Lett.* **18**, 6812–6819 (2018).
122. Shang, J. *et al.* Observation of Excitonic Fine Structure in a 2D Transition-Metal Dichalcogenide Semiconductor. *ACS Nano* **9**, 647–655 (2015).
123. Kim, M. S. *et al.* Biexciton Emission from Edges and Grain Boundaries of Triangular WS2 Monolayers. *ACS Nano* **10**, 2399–2405 (2016).
124. Vitale, S. A. *et al.* Valleytronics: Opportunities, Challenges, and Paths Forward. *Small* **14**, 1801483 (2018).
125. Mak, K. F., He, K., Shan, J. & Heinz, T. F. Control of valley polarization in monolayer MoS2 by optical helicity. *Nat. Nanotechnol.* **7**, 494–498 (2012).
126. Zeng, H., Dai, J., Yao, W., Xiao, D. & Cui, X. Valley polarization in MoS 2 monolayers by optical pumping. *Nat. Nanotechnol.* **7**, 490–493 (2012).
127. Sun, L. *et al.* Separation of valley excitons in a MoS$_2$ monolayer using a subwavelength asymmetric groove array. *Nat. Photonics* **13**, (2019).
128. Chen, P. *et al.* Chiral Coupling of Valley Excitons and Light through Photonic Spin–Orbit Interactions. *Adv. Opt. Mater.* **8**, 1901233 (2020).

129. Onga, M., Zhang, Y., Ideue, T. & Iwasa, Y. Exciton Hall effect in monolayer MoS2. *Nat. Mater.* **216**, 155449 (2017).
130. Lee, J., Mak, K. F. & Shan, J. Electrical control of the valley Hall effect in bilayer MoS2 transistors. *Nat. Nanotechnol.* **11**, 421–425 (2016).
131. Wu, S. *et al.* Electrical tuning of valley magnetic moment through symmetry control in bilayer MoS2. *Nat. Phys.* **9**, 149–153 (2013).
132. Sie, E. J. *et al.* Valley-selective optical Stark effect in monolayer WS2. *Nat. Mater.* **14**, 290–294 (2015).
133. Kim, J. *et al.* Ultrafast generation of pseudo-magnetic field for valley excitons in WSe 2 monolayers. *Science (80-. ).* **346**, 1205–1208 (2014).
134. Hsu, C. *et al.* Thickness-Dependent Refractive Index of 1L, 2L, and 3L MoS 2 , MoSe 2 , WS 2 , and WSe 2. *Adv. Opt. Mater.* **7**, 1900239 (2019).
135. Yu, Y. *et al.* Exciton-dominated Dielectric Function of Atomically Thin MoS2 Films. *Sci. Rep.* **5**, 16996 (2015).
136. Brem, S., Selig, M., Berghaeuser, G. & Malic, E. Exciton Relaxation Cascade in two-dimensional Transition Metal Dichalcogenides. *Sci. Rep.* **8**, 8238 (2018).
137. Korn, T., Heydrich, S., Hirmer, M., Schmutzler, J. & Schüller, C. Low-temperature photocarrier dynamics in monolayer MoS 2. *Appl. Phys. Lett.* **99**, 102109 (2011).
138. Lagarde, D. *et al.* Carrier and Polarization Dynamics in Monolayer MoS 2. **047401**, 1–5 (2014).
139. Wang, G. *et al.* Valley dynamics probed through charged and neutral exciton emission in monolayer WSe2. *Phys. Rev. B* **90**, 075413 (2014).
140. Shi, H. *et al.* Exciton Dynamics in Suspended Monolayer and Few-Layer MoS 2 2D Crystals. *ACS Nano* **7**, 1072–1080 (2013).
141. Mouri, S. *et al.* Nonlinear photoluminescence in atomically thin layered WSe2 arising from diffusion-assisted exciton-exciton annihilation. *Phys. Rev. B* **90**, 155449 (2014).
142. Palummo, M., Bernardi, M. & Grossman, J. C. Exciton radiative lifetimes in two-dimensional transition metal dichalcogenides. *Nano Lett.* **15**, 2794–2800 (2015).
143. Amani, M. *et al.* Near-unity photoluminescence quantum yield in MoS2. *Science (80-. ).* **350**, 1065–1068 (2015).
144. Amani, M. *et al.* Recombination Kinetics and Effects of Superacid Treatment in Sulfur- and Selenium-Based Transition Metal Dichalcogenides. *Nano Lett.* **16**, 2786–2791 (2016).
145. Johnson, A. D., Cheng, F., Tsai, Y. & Shih, C.-K. Giant Enhancement of Defect-Bound Exciton Luminescence and Suppression of Band-Edge Luminescence in Monolayer WSe2–Ag Plasmonic Hybrid Structures. *Nano Lett.* **17**, 4317–4322 (2017).
146. Zhang, X. *et al.* Unidirectional Doubly Enhanced MoS2 Emission via Photonic Fano Resonances. *Nano Lett.* **17**, 6715–6720 (2017).
147. Huang, J., Akselrod, G. M., Ming, T., Kong, J. & Mikkelsen, M. H. Tailored Emission Spectrum of 2D Semiconductors Using Plasmonic Nanocavities. *ACS Photonics* **5**, 552–558 (2018).
148. Park, K. D., Jiang, T., Clark, G., Xu, X. & Raschke, M. B. Radiative control of dark excitons at room temperature by nano-optical antenna-tip Purcell effect. *Nat. Nanotechnol.* **13**, 59–64 (2018).
149. Eda, G. & Maier, S. A. Two-dimensional crystals: Managing light for optoelectronics. *ACS Nano* **7**, 5660–5665 (2013).
150. Wang, Z. *et al.* Giant photoluminescence enhancement in tungsten-diselenide–gold plasmonic

hybrid structures. *Nat. Commun.* **7**, 11283 (2016).

151. Akselrod, G. M. *et al.* Leveraging Nanocavity Harmonics for Control of Optical Processes in 2D Semiconductors. *Nano Lett.* **15**, 3578–3584 (2015).
152. Galfsky, T. *et al.* Broadband Enhancement of Spontaneous Emission in Two-Dimensional Semiconductors Using Photonic Hypercrystals. *Nano Lett.* **16**, 4940–4945 (2016).
153. Sortino, L. *et al.* Enhanced light-matter interaction in an atomically thin semiconductor coupled with dielectric nano-antennas. *Nat. Commun.* **10**, 5119 (2019).
154. Moody, G. *et al.* Intrinsic homogeneous linewidth and broadening mechanisms of excitons in monolayer transition metal dichalcogenides. *Nat. Commun.* **6**, 8315 (2015).
155. Yuan, L. & Huang, L. Exciton dynamics and annihilation in $WS_2$ 2D semiconductors. *Nanoscale* **7**, 7402–7408 (2015).
156. Ryder, C. R., Wood, J. D., Wells, S. A. & Hersam, M. C. Chemically Tailoring Semiconducting Two-Dimensional Transition Metal Dichalcogenides and Black Phosphorus. *ACS Nano* **10**, 3900–3917 (2016).
157. Cunningham, P. D., Hanbicki, A. T., McCreary, K. M. & Jonker, B. T. Photoinduced Bandgap Renormalization and Exciton Binding Energy Reduction in WS2. *ACS Nano* **11**, 12601–12608 (2017).
158. Fan, X. *et al.* Nonlinear photoluminescence in monolayer $WS_2$: parabolic emission and excitation fluence-dependent recombination dynamics. *Nanoscale* **9**, 7235–7241 (2017).
159. Peng, Z., Chen, X., Fan, Y., Srolovitz, D. J. & Lei, D. Strain engineering of 2D semiconductors and graphene: from strain fields to band-structure tuning and photonic applications. *Light Sci. Appl.* **9**, 190 (2020).
160. Kobayashi, Y. *et al.* Growth and Optical Properties of High-Quality Monolayer $WS_2$ on Graphite. *ACS Nano* **9**, 4056–4063 (2015).
161. Liu, X. *et al.* Control of Coherently Coupled Exciton Polaritons in Monolayer Tungsten Disulphide. *Phys. Rev. Lett.* **119**, 27403 (2017).
162. Moilanen, A. J., Hakala, T. K. & Törmä, P. Active Control of Surface Plasmon–Emitter Strong Coupling. *ACS Photonics* acsphotonics.7b00655 (2017). doi:10.1021/acsphotonics.7b00655
163. Rivera, P. *et al.* Interlayer valley excitons in heterobilayers of transition metal dichalcogenides. *Nat. Nanotechnol.* **13**, 1004–1015 (2018).
164. Jin, C. *et al.* Ultrafast dynamics in van der Waals heterostructures. *Nat. Nanotechnol.* **13**, 994–1003 (2018).
165. Komsa, H.-P. & Krasheninnikov, A. V. Electronic structures and optical properties of realistic transition metal dichalcogenide heterostructures from first principles. *Phys. Rev. B* **88**, 85318 (2013).
166. Wilson, N. R. *et al.* Determination of band offsets, hybridization, and exciton binding in 2D semiconductor heterostructures. *Sci. Adv.* **3**, e1601832 (2017).
167. Chiu, M.-H. *et al.* Determination of band alignment in the single-layer MoS2/WSe2 heterojunction. *Nat. Commun.* **6**, 7666 (2015).
168. Liu, Y. *et al.* Room temperature nanocavity laser with interlayer excitons in 2D heterostructures. *Sci. Adv.* **5**, eaav4506 (2019).
169. Paik, E. Y. *et al.* Interlayer exciton laser of extended spatial coherence in atomically thin heterostructures. *Nature* **576**, 80–84 (2019).
170. Tran, K. *et al.* Evidence for moiré excitons in van der Waals heterostructures. *Nature* **567**,

71–75 (2019).

171. Seyler, K. L. *et al.* Signatures of moiré-trapped valley excitons in MoSe2/WSe2 heterobilayers. *Nature* **567**, 66–70 (2019).
172. Jin, C. *et al.* Observation of moiré excitons in WSe2/WS2 heterostructure superlattices. *Nature* **567**, 76–80 (2019).
173. Alexeev, E. M. *et al.* Resonantly hybridized excitons in moiré superlattices in van der Waals heterostructures. *Nature* **567**, 81–86 (2019).
174. Ra'di, Y., Krasnok, A. & Alù, A. Virtual Critical Coupling. *ACS Photonics* **7**, 1468–1475 (2020).
175. Maier, S. A. Plasmonic field enhancement and SERS in the effective mode volume picture. *Opt. Express* **14**, 1957 (2006).
176. Choi, H., Heuck, M. & Englund, D. Self-Similar Nanocavity Design with Ultrasmall Mode Volume for Single-Photon Nonlinearities. *Phys. Rev. Lett.* **118**, 223605 (2017).
177. Marquier, F., Sauvan, C. & Greffet, J. J. Revisiting quantum optics with surface plasmons and plasmonic resonators. *ACS Photonics* **4**, 2091–2101 (2017).
178. Kristensen, P. T. & Hughes, S. Modes and Mode Volumes of Leaky Optical Cavities and Plasmonic Nanoresonators. *ACS Photonics* **1**, 2–10 (2014).
179. Sauvan, C., Hugonin, J. P., Maksymov, I. S. & Lalanne, P. Theory of the Spontaneous Optical Emission of Nanosize Photonic and Plasmon Resonators. *Phys. Rev. Lett.* **110**, 237401 (2013).
180. Lalanne, P., Yan, W., Vynck, K., Sauvan, C. & Hugonin, J.-P. Light Interaction with Photonic and Plasmonic Resonances. *Laser Photon. Rev.* **12**, 1700113 (2018).
181. Koenderink, A. F. On the use of Purcell factors for plasmon antennas. *Opt. Lett.* **35**, 4208 (2010).
182. Muljarov, E. A. & Langbein, W. Exact mode volume and Purcell factor of open optical systems. *Phys. Rev. B* **94**, 235438 (2016).
183. Anger, P., Bharadwaj, P. & Novotny, L. Enhancement and quenching of single-molecule fluorescence. *Phys. Rev. Lett.* **96**, 3–6 (2006).
184. Yu, X. *et al.* Strong Coupling in Microcavity Structures: Principle, Design, and Practical Application. *Laser Photon. Rev.* **13**, 1800219 (2019).
185. Baranov, D. G., Wersäll, M., Cuadra, J., Antosiewicz, T. J. & Shegai, T. Novel Nanostructures and Materials for Strong Light–Matter Interactions. *ACS Photonics* **5**, 24–42 (2018).
186. Liu, W. *et al.* Understanding the Different Exciton-Plasmon Coupling Regimes in Two-Dimensional Semiconductors Coupled with Plasmonic Lattices: A Combined Experimental and Unified Equation of Motion Approach. *ACS Photonics* **5**, 192–204 (2018).
187. Hertzog, M., Wang, M., Mony, J. & Börjesson, K. Strong light-matter interactions: A new direction within chemistry. *Chem. Soc. Rev.* **48**, 937–961 (2019).
188. Krasnok, A., Caldarola, M., Bonod, N. & Alú, A. Spectroscopy and Biosensing with Optically Resonant Dielectric Nanostructures. *Adv. Opt. Mater.* **6**, 1701094 (2018).
189. Krasnok, A. E. *et al.* An antenna model for the Purcell effect. *Sci. Rep.* **5**, 12956 (2015).
190. Hensen, M., Heilpern, T., Gray, S. K. & Pfeiffer, W. Strong Coupling and Entanglement of Quantum Emitters Embedded in a Nanoantenna-Enhanced Plasmonic Cavity. *ACS Photonics* **5**, 240–248 (2018).
191. Kavokin, A., Baumberg, J. J., Malpuech, G. & Laussy, F. P. *Microcavities*. *Microcavities* (Oxford University Press, 2007). doi:10.1093/acprof:oso/9780199228942.001.0001
192. Novotny, L. & Hecht, B. *Principles of Nano-Optics*. *Principles of Nano-Optics* **9781107005**,

(Cambridge University Press, 2012).

193. Fang, H. H. *et al.* Control of the Exciton Radiative Lifetime in van der Waals Heterostructures. *Phys. Rev. Lett.* **123**, 067401 (2019).

194. Luo, Y. *et al.* Purcell-enhanced quantum yield from carbon nanotube excitons coupled to plasmonic nanocavities. *Nat. Commun.* **8**, 1413 (2017).

195. Luo, Y. *et al.* Deterministic coupling of site-controlled quantum emitters in monolayer WSe 2 to plasmonic nanocavities. *Nat. Nanotechnol.* **13**, 1137–1142 (2018).

196. Flatten, L. C. *et al.* Microcavity enhanced single photon emission from two-dimensional WSe 2. *Appl. Phys. Lett.* **112**, 191105 (2018).

197. Lepeshov, S., Krasnok, A. & Alù, A. Enhanced excitation and emission from 2D transition metal dichalcogenides with all-dielectric nanoantennas. *Nanotechnology* **30**, 254004 (2019).

198. Noori, Y. J. *et al.* Photonic Crystals for Enhanced Light Extraction from 2D Materials. *ACS Photonics* **3**, 2515–2520 (2016).

199. Ao, X., Xu, X., Dong, J. & He, S. Unidirectional Enhanced Emission from 2D Monolayer Suspended by Dielectric Pillar Array. *ACS Appl. Mater. Interfaces* **10**, 34817–34821 (2018).

200. Wang, M. *et al.* Dark-Exciton-Mediated Fano Resonance from a Single Gold Nanostructure on Monolayer WS 2 at Room Temperature. *Small* **15**, 1900982 (2019).

201. Li, Z. *et al.* Tailoring MoS2 Exciton-Plasmon Interaction by Optical Spin-Orbit Coupling. *ACS Nano* **11**, 1165–1171 (2017).

202. Wu, S. *et al.* Monolayer semiconductor nanocavity lasers with ultralow thresholds. *Nature* **520**, 69 (2015).

203. Krasnok, A. & Alu, A. Active Nanophotonics. *Proc. IEEE* **108**, 628–654 (2020).

204. Gong, S.-H. H., Alpeggiani, F., Sciacca, B., Garnett, E. C. & Kuipers, L. Nanoscale chiral valley-photon interface through optical spin-orbit coupling. *Science (80-. ).* **359**, 443–447 (2018).

205. Krasnok, A. & Alù, A. Valley-Selective Response of Nanostructures Coupled to 2D Transition-Metal Dichalcogenides. *Appl. Sci.* **8**, 1157 (2018).

206. Chen, H., Liu, M., Xu, L. & Neshev, D. N. Valley-selective directional emission from a transition-metal dichalcogenide monolayer mediated by a plasmonic nanoantenna. *Beilstein J. Nanotechnol.* **9**, 780–788 (2018).

207. Zheng, L. *et al.* Deep subwavelength control of valley polarized cathodoluminescence in h-BN/WSe2/h-BN heterostructure. *Nat. Commun.* **12**, 291 (2021).

208. Chervy, T. *et al.* Room Temperature Chiral Coupling of Valley Excitons with Spin-Momentum Locked Surface Plasmons. *ACS Photonics* **5**, 1281–1287 (2018).

209. Wang, J. *et al.* Routing valley exciton emission of a WS2 monolayer via delocalized Bloch modes of in-plane inversion-symmetry-broken photonic crystal slabs. *Light Sci. Appl.* **9**, 148 (2020).

210. Krasnok, A. Photonic Rashba effect. *Nat. Nanotechnol.* **15**, 893–894 (2020).

211. Rong, K. *et al.* Photonic Rashba effect from quantum emitters mediated by a Berry-phase defective photonic crystal. *Nat. Nanotechnol.* **15**, 927–933 (2020).

212. Maier, S. A. & Atwater, H. A. Plasmonics: Localization and guiding of electromagnetic energy in metal/dielectric structures. *J. Appl. Phys.* **98**, 11101 (2005).

213. Wang, H., Brandl, D. W., Le, F., Nordlander, P. & Halas, N. J. Nanorice: A hybrid plasmonic nanostructure. *Nano Lett.* **6**, 827–832 (2006).

214. Chen, J. *et al.* Gold nanocages: Engineering their structure for biomedical applications. *Adv. Mater.* **17**, 2255–2261 (2005).
215. Nehl, C. L., Liao, H. & Hafner, J. H. Optical Properties of Star-Shaped Gold Nanoparticles. (2006).
216. Shumaker-Parry, J. S., Rochholz, H. & Kreiter, M. Fabrication of crescent-shaped optical antennas. *Adv. Mater.* **17**, 2131–2134 (2005).
217. Sherry, L. J. *et al.* Localized surface plasmon resonance spectroscopy of single silver nanocubes. *Nano Lett.* **5**, 2034–2038 (2005).
218. Oldenburg, S. J., Averitt, R. D., Westcott, S. L. & Halas, N. J. Nanoengineering of optical resonances. *Chem. Phys. Lett.* **288**, 243–247 (1998).
219. Boltasseva, A. & Atwater, H. A. Low-Loss Plasmonic Metamaterials. *Science (80-. ).* **331**, 290–291 (2011).
220. West, P. R. *et al.* Searching for better plasmonic materials. *Laser Photonics Rev.* **4**, 795–808 (2010).
221. Papasimakis, N. *et al.* Graphene in a photonic metamaterial. *Opt. Express* **18**, 8353 (2010).
222. Naik, G. V., Shalaev, V. M. & Boltasseva, A. Alternative plasmonic materials: Beyond gold and silver. *Adv. Mater.* **25**, 3264–3294 (2013).
223. Vakil, A. & Engheta, N. Transformation optics using graphene. *Science (80-. ).* **332**, 1291–1294 (2011).
224. Wen, J. *et al.* Room-Temperature Strong Light–Matter Interaction with Active Control in Single Plasmonic Nanorod Coupled with Two-Dimensional Atomic Crystals. *Nano Lett.* **17**, 4689–4697 (2017).
225. Khurgin, J. B. How to deal with the loss in plasmonics and metamaterials. *Nat. Nanotechnol.* **10**, 2–6 (2015).
226. Mie, G. Beiträge zur optik trüber medien speziell kolloidaler metallösungen. *Ann. Phys.* **330**, 377–445 (1908).
227. Fenollosa, B. R., Meseguer, F. & Tymczenko, M. Silicon Colloids : From Microcavities to Photonic Sponges **. 95–98 (2008). doi:10.1002/adma.200701589
228. Xifré-Pérez, E., Fenollosa, R. & Meseguer, F. Low order modes in microcavities based on silicon colloids. *Opt. Express* **19**, 3455 (2011).
229. Evlyukhin, A. B. *et al.* Demonstration of magnetic dipole resonances of dielectric nanospheres in the visible region. *Nano Lett.* **12**, 3749–3755 (2012).
230. Kuznetsov, A. I., Miroshnichenko, A. E., Fu, Y. H., Zhang, J. & Luk'yanchuk, B. Magnetic light. *Sci. Rep.* **2**, 492 (2012).
231. Zywietz, U., Evlyukhin, A. B., Reinhardt, C. & Chichkov, B. N. Laser printing of silicon nanoparticles with resonant optical electric and magnetic responses. *Nat. Commun.* **5**, 3402 (2014).
232. Baranov, D. G. *et al.* All-dielectric nanophotonics: the quest for better materials and fabrication techniques. *Optica* **4**, 814 (2017).
233. Savelev, R. S., Makarov, S. V., Krasnok, A. E. & Belov, P. A. From optical magnetic resonance to dielectric nanophotonics (A review). *Opt. Spectrosc.* **119**, 551–568 (2015).
234. Bohren, Craig F, Huffman, D. R. *Absorption and Scattering of Light by Small Particles*. *Wiley Science* (Wiley-VCH Verlag GmbH, 1998). doi:10.1002/9783527618156
235. Mulholland, G. W., Bohren, C. F. & Fuller, K. A. Light Scattering by Agglomerates: Coupled

Electric and Magnetic Dipole Method. *Langmuir* **10**, 2533–2546 (1994).
236. Evlyukhin, A. B., Reinhardt, C., Seidel, A., Luk'yanchuk, B. S. & Chichkov, B. N. Optical response features of Si-nanoparticle arrays. *Phys. Rev. B* **82**, 045404 (2010).
237. Wang, M. *et al.* Suppressing material loss in the visible and near-infrared range for functional nanophotonics using bandgap engineering. *Nat. Commun.* **11**, 5055 (2020).
238. Feng, T., Xu, Y., Zhang, W. & Miroshnichenko, A. E. Ideal Magnetic Dipole Scattering. *Phys. Rev. Lett.* **118**, 173901 (2017).
239. Kapitanova, P. *et al.* Seeing the Unseen: Experimental Observation of Magnetic Anapole State Inside a High-Index Dielectric Particle. *Ann. Phys.* **2000293**, 1–7 (2020).
240. Miroshnichenko, A. E. *et al.* Nonradiating anapole modes in dielectric nanoparticles. *Nat. Commun.* **6**, 1–8 (2015).
241. Lepeshov, S., Krasnok, A. & Alù, A. Nonscattering-to-Superscattering Switch with Phase-Change Materials. *ACS Photonics* **6**, 2126–2132 (2019).
242. Monticone, F., Sounas, D., Krasnok, A. & Alù, A. Can a Nonradiating Mode Be Externally Excited? Nonscattering States versus Embedded Eigenstates. *ACS Photonics* **6**, 3108–3114 (2019).
243. Krasnok, A. *et al.* Anomalies in light scattering. *Adv. Opt. Photonics* **11**, 892 (2019).
244. Rybin, M. V *et al.* High-$Q$ Supercavity Modes in Subwavelength Dielectric Resonators. *Phys. Rev. Lett.* **119**, 243901 (2017).
245. Koshelev, K. *et al.* Subwavelength dielectric resonators for nonlinear nanophotonics. *Science (80-. ).* **367**, 288–292 (2020).
246. Liu, S.-D., Wang, Z.-X., Wang, W.-J., Chen, J.-D. & Chen, Z.-H. High Q-factor with the excitation of anapole modes in dielectric split nanodisk arrays. *Opt. Express* **25**, 22375 (2017).
247. Huang, L., Xu, L., Rahmani, M., Neshev, D. N. & Miroshnichenko, A. E. Pushing the limit of high-Q mode of a single dielectric nanocavity. *Adv. Photonics* **3**, 1–9 (2021).
248. Krasnok, A. E., Miroshnichenko, A. E., Belov, P. A. & Kivshar, Y. S. All-dielectric optical nanoantennas. *Opt. Express* **20**, 20599 (2012).
249. Zambrana-Puyalto, X. & Bonod, N. Purcell factor of spherical Mie resonators. *Phys. Rev. B - Condens. Matter Mater. Phys.* **91**, 1–11 (2015).
250. Krasnok, A. *et al.* Demonstration of the enhanced Purcell factor in all-dielectric structures. *Appl. Phys. Lett.* **108**, 211105 (2016).
251. Dmitriev, P. A. *et al.* Resonant Raman scattering from silicon nanoparticles enhanced by magnetic response. *Nanoscale* **8**, 9721–9726 (2016).
252. Rolly, B., Bebey, B., Bidault, S., Stout, B. & Bonod, N. Promoting magnetic dipolar transition in trivalent lanthanide ions with lossless Mie resonances. *Phys. Rev. B - Condens. Matter Mater. Phys.* **85**, 2–7 (2012).
253. Li, S. V., Baranov, D. G., Krasnok, A. E. & Belov, P. A. All-dielectric nanoantennas for unidirectional excitation of electromagnetic guided modes. *Appl. Phys. Lett.* **107**, 171101 (2015).
254. Liu, S. *et al.* Light-Emitting Metasurfaces: Simultaneous Control of Spontaneous Emission and Far-Field Radiation. *Nano Lett.* **18**, 6906–6914 (2018).
255. Geffrin, J. M. *et al.* Magnetic and electric coherence in forward-and back-scattered electromagnetic waves by a single dielectric subwavelength sphere. *Nat. Commun.* **3**, (2012).
256. Fu, Y. H., Kuznetsov, A. I., Miroshnichenko, A. E., Yu, Y. F. & Luk'yanchuk, B. Directional visible

light scattering by silicon nanoparticles. *Nat. Commun.* **4**, 1527 (2013).

257. Kruk, S. S. & Kivshar, Y. S. Functional Meta-Optics and Nanophotonics Govern by Mie Resonances. *ACS Photonics* **4**, 2638–2649 (2017).
258. Kryzhanovskaya, N. *et al.* Enhanced light outcoupling in microdisk lasers via Si spherical nanoantennas. *J. Appl. Phys.* **124**, 163102 (2018).
259. Albella, P., Shibanuma, T. & Maier, S. A. Switchable directional scattering of electromagnetic radiation with subwavelength asymmetric silicon dimers. *Sci. Rep.* **5**, 1–8 (2015).
260. Staude, I. *et al.* Tailoring directional scattering through magnetic and electric resonances in subwavelength silicon nanodisks. *ACS Nano* **7**, 7824–7832 (2013).
261. Shcherbakov, M. R. *et al.* Enhanced third-harmonic generation in silicon nanoparticles driven by magnetic response. *Nano Lett.* **14**, 6488–6492 (2014).
262. Makarov, S. *et al.* Tuning of Magnetic Optical Response in a Dielectric Nanoparticle by Ultrafast Photoexcitation of Dense Electron-Hole Plasma. *Nano Lett.* **15**, (2015).
263. Shcherbakov, M. R. *et al.* Ultrafast All-Optical Switching with Magnetic Resonances in Nonlinear Dielectric Nanostructures. *Nano Lett.* **15**, 6985–6990 (2015).
264. Baranov, D. G. D. G. *et al.* Nonlinear Transient Dynamics of Photoexcited Resonant Silicon Nanostructures. *ACS Photonics* **3**, 1546–1551 (2016).
265. Staude, I. & Schilling, J. Metamaterial-inspired silicon nanophotonics. *Nat. Photonics* **11**, 274–284 (2017).
266. Smirnova, D. & Kivshar, Y. S. Multipolar nonlinear nanophotonics. *Optica* **3**, 1241 (2016).
267. Kruk, S. S. *et al.* Nonlinear optical magnetism revealed by second-harmonic generation in nanoantennas. *Nano Lett.* **17**, 3914–3918 (2017).
268. Baranov, D. G. D. G., Makarov, S. V. S. V. S. V., Krasnok, A. E., Belov, P. A. P. A. P. A. & Alù, A. Tuning of near- and far-field properties of all-dielectric dimer nanoantennas via ultrafast electron-hole plasma photoexcitation. *Laser Photonics Rev.* **10**, 1009–1015 (2016).
269. Wang, H. *et al.* Resonance Coupling in Heterostructures Composed of Silicon Nanosphere and Monolayer WS2: A Magnetic-Dipole-Mediated Energy Transfer Process. *ACS Nano* **13**, 1739–1750 (2019).
270. Jahani, S. & Jacob, Z. All-dielectric metamaterials. *Nat. Nanotechnol.* **11**, 23–36 (2016).
271. Kuznetsov, A. I., Miroshnichenko, A. E., Brongersma, M. L., Kivshar, Y. S. & Luk'yanchuk, B. Optically resonant dielectric nanostructures. *Science (80-. ).* **354**, aag2472 (2016).
272. Schuller, J. A., Zia, R., Taubner, T. & Brongersma, M. L. Dielectric metamaterials based on electric and magnetic resonances of silicon carbide particles. *Phys. Rev. Lett.* **99**, 1–4 (2007).
273. Ginn, J. C. *et al.* Realizing optical magnetism from dielectric metamaterials. *Phys. Rev. Lett.* **108**, 1–5 (2012).
274. Moitra, P. *et al.* Large-Scale All-Dielectric Metamaterial Perfect Reflectors. *ACS Photonics* **2**, 692–698 (2015).
275. Arbabi, A., Horie, Y., Bagheri, M. & Faraon, A. Dielectric metasurfaces for complete control of phase and polarization with subwavelength spatial resolution and high transmission. *Nat. Nanotechnol.* **10**, 937–943 (2015).
276. Shalaev, M. I. *et al.* High-Efficiency All-Dielectric Metasurfaces for Ultracompact Beam Manipulation in Transmission Mode. *Nano Lett.* **15**, 6261–6266 (2015).
277. Khorasaninejad, M. *et al.* Metalenses at visible wavelengths: Diffraction-limited focusing and subwavelength resolution imaging. *Science (80-. ).* **352**, 1190–1194 (2016).

278. Krasnok, A., Tymchenko, M. & Alù, A. Nonlinear metasurfaces: a paradigm shift in nonlinear optics. *Mater. Today* **21**, 8–21 (2018).

279. Zhou, Z. *et al.* Efficient Silicon Metasurfaces for Visible Light. *ACS Photonics* **4**, 544–551 (2017).

280. Yu, Y. F. *et al.* High-transmission dielectric metasurface with 2π phase control at visible wavelengths. *Laser Photonics Rev.* **9**, 412–418 (2015).

281. Lodahl, P., Mahmoodian, S. & Stobbe, S. Interfacing single photons and single quantum dots with photonic nanostructures. *Rev. Mod. Phys.* **87**, 347–400 (2015).

282. Latini, S., Ronca, E., De Giovannini, U., Hübener, H. & Rubio, A. Cavity Control of Excitons in Two-Dimensional Materials. *Nano Lett.* **19**, 3473–3479 (2019).

283. Grynberg, G., Aspect, A., Fabre, C. & Cohen-Tannoudji, C. *Introduction to Quantum Optics*. *Introduction to Quantum Optics* (Cambridge University Press, 2010). doi:10.1017/CBO9780511778261

284. Scully, M. O. & Zubairy, M. S. *Quantum Optics*. *Cambridge University Press* (Cambridge University Press, 1997). doi:10.1017/CBO9780511813993

285. Mandel, L. & Wolf, E. *Optical Coherence and Quantum Optics*. (Cambridge University Press, 1995). doi:10.1017/CBO9781139644105

286. Albash, T., Boixo, S., Lidar, D. A. & Zanardi, P. Quantum adiabatic Markovian master equations. *New J. Phys.* **14**, 123016 (2012).

287. Cattaneo, M., Giorgi, G. L., Maniscalco, S. & Zambrini, R. Local versus global master equation with common and separate baths: Superiority of the global approach in partial secular approximation. *New J. Phys.* **21**, 113045 (2019).

288. Törmä, P. & Barnes, W. L. Strong coupling between surface plasmon polaritons and emitters. *Reports Prog. Phys.* **78**, 013901 (2014).

289. Wu, X., Gray, S. K. & Pelton, M. Quantum-dot-induced transparency in a nanoscale plasmonic resonator. *Opt. Express* **18**, 23633 (2010).

290. Novotny, L. Strong coupling, energy splitting, and level crossings: A classical perspective. *Am. J. Phys.* **78**, 1199–1202 (2010).

291. Stete, F., Koopman, W. & Bargheer, M. Signatures of Strong Coupling on Nanoparticles: Revealing Absorption Anticrossing by Tuning the Dielectric Environment. *ACS Photonics* **4**, 1669–1676 (2017).

292. Lien, D.-H. *et al.* Electrical suppression of all nonradiative recombination pathways in monolayer semiconductors. *Science (80-. ).* **364**, 468 LP-471 (2019).

293. Chakraborty, B. *et al.* Symmetry-dependent phonon renormalization in monolayer $MoS_{2}$ transistor. *Phys. Rev. B* **85**, 161403 (2012).

294. Yu, Y. *et al.* Enhancing Multifunctionalities of Transition-Metal Dichalcogenide Monolayers via Cation Intercalation. *ACS Nano* **11**, 9390–9396 (2017).

295. Amani, M. *et al.* High Luminescence Efficiency in MoS2 Grown by Chemical Vapor Deposition. *ACS Nano* **10**, 6535–6541 (2016).

296. Han, H.-V. *et al.* Photoluminescence Enhancement and Structure Repairing of Monolayer MoSe2 by Hydrohalic Acid Treatment. *ACS Nano* **10**, 1454–1461 (2016).

297. Tanoh, A. O. A. *et al.* Enhancing Photoluminescence and Mobilities in WS2 Monolayers with Oleic Acid Ligands. *Nano Lett.* **19**, 6299–6307 (2019).

298. Mouri, S., Miyauchi, Y. & Matsuda, K. Tunable Photoluminescence of Monolayer MoS2 via Chemical Doping. *Nano Lett.* **13**, 5944–5948 (2013).

299. Peimyoo, N. *et al.* Chemically Driven Tunable Light Emission of Charged and Neutral Excitons in Monolayer WS2. *ACS Nano* **8**, 11320–11329 (2014).
300. Tongay, S. *et al.* Broad-Range Modulation of Light Emission in Two-Dimensional Semiconductors by Molecular Physisorption Gating. *Nano Lett.* **13**, 2831–2836 (2013).
301. Yang, L. *et al.* Chloride Molecular Doping Technique on 2D Materials: WS2 and MoS2. *Nano Lett.* **14**, 6275–6280 (2014).
302. Li, Y., Xu, C.-Y., Hu, P. & Zhen, L. Carrier Control of MoS2 Nanoflakes by Functional Self-Assembled Monolayers. *ACS Nano* **7**, 7795–7804 (2013).
303. Kang, D.-H. *et al.* High-Performance Transition Metal Dichalcogenide Photodetectors Enhanced by Self-Assembled Monolayer Doping. *Adv. Funct. Mater.* **25**, 4219–4227 (2015).
304. Huang Lujun. Engineering Optical and Thermal Properties of Two dimensional Transition Metal Dichalcogenide Monolayer. (North Carolina State University, 2017).
305. Nan, H. *et al.* Strong Photoluminescence Enhancement of MoS2 through Defect Engineering and Oxygen Bonding. *ACS Nano* **8**, 5738–5745 (2014).
306. Ajayi, O. A. *et al.* Approaching the intrinsic photoluminescence linewidth in transition metal dichalcogenide monolayers. *2D Mater.* **4**, 31011 (2017).
307. Courtade, E. *et al.* Spectrally narrow exciton luminescence from monolayer MoS2 and MoSe2 exfoliated onto epitaxially grown hexagonal BN. *Appl. Phys. Lett.* **113**, 32106 (2018).
308. Cong, C. *et al.* Intrinsic excitonic emission and valley Zeeman splitting in epitaxial MS2 (M = Mo and W) monolayers on hexagonal boron nitride. *Nano Res.* **11**, 6227–6236 (2018).
309. Wurdack, M. *et al.* Ultrathin Ga2O3 Glass: A Large-Scale Passivation and Protection Material for Monolayer WS2. *Adv. Mater.* **33**, 2005732 (2021).
310. Lin, Y. *et al.* Dielectric Screening of Excitons and Trions in Single-Layer MoS2. *Nano Lett.* **14**, 5569–5576 (2014).
311. Kim, H., Lien, D.-H., Amani, M., Ager, J. W. & Javey, A. Highly Stable Near-Unity Photoluminescence Yield in Monolayer MoS2 by Fluoropolymer Encapsulation and Superacid Treatment. *ACS Nano* **11**, 5179–5185 (2017).
312. Buscema, M., Steele, G. A., van der Zant, H. S. J. & Castellanos-Gomez, A. The effect of the substrate on the Raman and photoluminescence emission of single-layer MoS2. *Nano Res.* **7**, 561–571 (2014).
313. Varghese, J. O. *et al.* The Influence of Water on the Optical Properties of Single-Layer Molybdenum Disulfide. *Adv. Mater.* **27**, 2734–2740 (2015).
314. Yu, Y. *et al.* Engineering Substrate Interactions for High Luminescence Efficiency of Transition-Metal Dichalcogenide Monolayers. *Adv. Funct. Mater.* **26**, 4733–4739 (2016).
315. Roddaro, S., Pingue, P., Piazza, V., Pellegrini, V. & Beltram, F. The Optical Visibility of Graphene:  Interference Colors of Ultrathin Graphite on SiO2. *Nano Lett.* **7**, 2707–2710 (2007).
316. Lien, D.-H. *et al.* Engineering Light Outcoupling in 2D Materials. *Nano Lett.* **15**, 1356–1361 (2015).
317. Eizagirre Barker, S. *et al.* Preserving the Emission Lifetime and Efficiency of a Monolayer Semiconductor upon Transfer. *Adv. Opt. Mater.* **7**, 1900351 (2019).
318. Zhang, H. *et al.* Interference effect on optical signals of monolayer MoS2. *Appl. Phys. Lett.* **107**, 101904 (2015).
319. Li, S.-L. *et al.* Quantitative Raman Spectrum and Reliable Thickness Identification for Atomic Layers on Insulating Substrates. *ACS Nano* **6**, 7381–7388 (2012).

320. Jeong, H. Y. *et al.* Optical Gain in MoS2 via Coupling with Nanostructured Substrate: Fabry–Perot Interference and Plasmonic Excitation. *ACS Nano* **10**, 8192–8198 (2016).
321. Najmaei, S. *et al.* Plasmonic Pumping of Excitonic Photoluminescence in Hybrid MoS2–Au Nanostructures. *ACS Nano* **8**, 12682–12689 (2014).
322. Kern, J. *et al.* Nanoantenna-Enhanced Light–Matter Interaction in Atomically Thin WS2. *ACS Photonics* **2**, 1260–1265 (2015).
323. Butun, S., Tongay, S. & Aydin, K. Enhanced Light Emission from Large-Area Monolayer MoS2 Using Plasmonic Nanodisc Arrays. *Nano Lett.* **15**, 2700–2704 (2015).
324. Lee, B. *et al.* Fano Resonance and Spectrally Modified Photoluminescence Enhancement in Monolayer MoS2 Integrated with Plasmonic Nanoantenna Array. *Nano Lett.* **15**, 3646–3653 (2015).
325. Hao, Q. *et al.* Boosting the Photoluminescence of Monolayer MoS2 on High-Density Nanodimer Arrays with Sub-10 nm Gap. *Adv. Opt. Mater.* **6**, 1700984 (2018).
326. Chen, H. *et al.* Manipulation of photoluminescence of two-dimensional MoSe2 by gold nanoantennas. *Sci. Rep.* **6**, 22296 (2016).
327. Cheng, F. *et al.* Enhanced Photoluminescence of Monolayer WS2 on Ag Films and Nanowire–WS2–Film Composites. *ACS Photonics* **4**, 1421–1430 (2017).
328. Han, C. & Ye, J. Polarized resonant emission of monolayer WS2 coupled with plasmonic sawtooth nanoslit array. *Nat. Commun.* **11**, 713 (2020).
329. Kivshar, Y. & Miroshnichenko, A. Meta-Optics with Mie Resonances. *Opt. Photonics News* **28**, 24–31 (2017).
330. Chen, H. *et al.* Enhanced Directional Emission from Monolayer WSe2 Integrated onto a Multiresonant Silicon-Based Photonic Structure. *ACS Photonics* **4**, 3031–3038 (2017).
331. Bucher, T. *et al.* Tailoring Photoluminescence from MoS2 Monolayers by Mie-Resonant Metasurfaces. *ACS Photonics* **6**, 1002–1009 (2019).
332. Ma, C., Yan, J., Huang, Y. & Yang, G. Photoluminescence manipulation of WS2 flakes by an individual Si nanoparticle. *Mater. Horizons* **6**, 97–106 (2019).
333. Cihan, A. F., Curto, A. G., Raza, S., Kik, P. G. & Brongersma, M. L. Silicon Mie resonators for highly directional light emission from monolayer MoS2. *Nat. Photonics* **12**, 284–290 (2018).
334. Akahane, Y., Asano, T., Song, B.-S. & Noda, S. High-Q photonic nanocavity in a two-dimensional photonic crystal. *Nature* **425**, 944–947 (2003).
335. Song, B.-S., Noda, S., Asano, T. & Akahane, Y. Ultra-high-Q photonic double-heterostructure nanocavity. *Nat. Mater.* **4**, 207–210 (2005).
336. Fine-tuned high-Q photonic-crystal nanocavity. *Opt. Express* **13**, 1202–1214 (2005).
337. Li, Y. *et al.* Room-temperature continuous-wave lasing from monolayer molybdenum ditelluride integrated with a silicon nanobeam cavity. *Nat. Nanotechnol.* **12**, 987 (2017).
338. Deotare, P. B., McCutcheon, M. W., Frank, I. W., Khan, M. & Lončar, M. High quality factor photonic crystal nanobeam cavities. *Appl. Phys. Lett.* **94**, 121106 (2009).
339. Fang, H. *et al.* 1305 nm Few-Layer MoTe2-on-Silicon Laser-Like Emission. *Laser Photon. Rev.* **12**, 1800015 (2018).
340. Fang, H. *et al.* Laser-Like Emission from a Sandwiched MoTe2 Heterostructure on a Silicon Single-Mode Resonator. *Adv. Opt. Mater.* **7**, 1900538 (2019).
341. Vahala, K. J. Optical microcavities. *Nature* **424**, 839 (2003).
342. Ye, Y. *et al.* Monolayer excitonic laser. *Nat. Photonics* **9**, 733 (2015).

343. Salehzadeh, O., Djavid, M., Tran, N. H., Shih, I. & Mi, Z. Optically Pumped Two-Dimensional MoS2 Lasers Operating at Room-Temperature. *Nano Lett.* **15**, 5302–5306 (2015).
344. Zhao, L. *et al.* High-Temperature Continuous-Wave Pumped Lasing from Large-Area Monolayer Semiconductors Grown by Chemical Vapor Deposition. *ACS Nano* **12**, 9390–9396 (2018).
345. Javerzac-Galy, C. *et al.* Excitonic Emission of Monolayer Semiconductors Near-Field Coupled to High-Q Microresonators. *Nano Lett.* **18**, 3138–3146 (2018).
346. Shang, J. *et al.* Room-temperature 2D semiconductor activated vertical-cavity surface-emitting lasers. *Nat. Commun.* **8**, 543 (2017).
347. Ge, X., Minkov, M., Fan, S., Li, X. & Zhou, W. Laterally confined photonic crystal surface emitting laser incorporating monolayer tungsten disulfide. *npj 2D Mater. Appl.* **3**, 16 (2019).
348. Marinica, D. C., Borisov, A. G. & Shabanov, S. V. Bound states in the continuum in photonics. *Phys. Rev. Lett.* **100**, 1–4 (2008).
349. Hsu, C. W., Zhen, B., Stone, A. D., Joannopoulos, J. D. & Soljačić, M. Bound states in the continuum. *Nat. Rev. Mater.* **1**, 16048 (2016).
350. Liao, F. *et al.* Enhancing monolayer photoluminescence on optical micro/nanofibers for low-threshold lasing. *Sci. Adv.* **5**, eaax7398 (2019).
351. Reeves, L., Wang, Y. & Krauss, T. F. 2D Material Microcavity Light Emitters: To Lase or Not to Lase? *Adv. Opt. Mater.* **6**, 1800272 (2018).
352. Samuel, I. D. W., Namdas, E. B. & Turnbull, G. A. How to recognize lasing. *Nat. Photonics* **3**, 546–549 (2009).
353. Sundaram, R. S. *et al.* Electroluminescence in Single Layer MoS2. *Nano Lett.* **13**, 1416–1421 (2013).
354. Jo, S., Ubrig, N., Berger, H., Kuzmenko, A. B. & Morpurgo, A. F. Mono- and Bilayer WS2 Light-Emitting Transistors. *Nano Lett.* **14**, 2019–2025 (2014).
355. Baugher, B. W. H., Churchill, H. O. H., Yang, Y. & Jarillo-Herrero, P. Optoelectronic devices based on electrically tunable p–n diodes in a monolayer dichalcogenide. *Nat. Nanotechnol.* **9**, 262–267 (2014).
356. Withers, F. *et al.* Light-emitting diodes by band-structure engineering in van der Waals heterostructures. *Nat. Mater.* **14**, 301 (2015).
357. Lien, D.-H. *et al.* Large-area and bright pulsed electroluminescence in monolayer semiconductors. *Nat. Commun.* **9**, 1229 (2018).
358. Li, D. *et al.* Electric-field-induced strong enhancement of electroluminescence in multilayer molybdenum disulfide. *Nat. Commun.* **6**, 7509 (2015).
359. Liu, C.-H. *et al.* Nanocavity Integrated van der Waals Heterostructure Light-Emitting Tunneling Diode. *Nano Lett.* **17**, 200–205 (2017).
360. Gu, J., Chakraborty, B., Khatoniar, M. & Menon, V. M. A room-temperature polariton light-emitting diode based on monolayer WS2. *Nat. Nanotechnol.* **14**, 1024–1028 (2019).
361. Del Pozo-Zamudio, O. *et al.* Electrically pumped WSe2-based light-emitting van der Waals heterostructures embedded in monolithic dielectric microcavities. *2D Mater.* **7**, 31006 (2020).
362. Autere, A. *et al.* Nonlinear Optics with 2D Layered Materials. *Adv. Mater.* **30**, 1705963 (2018).
363. You, J. W., Bongu, S. R., Bao, Q. & Panoiu, N. C. Nonlinear optical properties and applications of 2D materials: theoretical and experimental aspects. *Nanophotonics* **8**, 63–97
364. Fryett, T., Zhan, A. & Majumdar, A. Cavity nonlinear optics with layered materials.

*Nanophotonics* **7**, 355–370

365. Li, Y. *et al.* Probing Symmetry Properties of Few-Layer MoS2 and h-BN by Optical Second-Harmonic Generation. *Nano Lett.* **13**, 3329–3333 (2013).

366. Malard, L. M., Alencar, T. V, Barboza, A. P. M., Mak, K. F. & de Paula, A. M. Observation of intense second harmonic generation from $MoS_{2}$ atomic crystals. *Phys. Rev. B* **87**, 201401 (2013).

367. Kumar, N. *et al.* Second harmonic microscopy of monolayer $MoS_{2}$. *Phys. Rev. B* **87**, 161403 (2013).

368. Wang, G. *et al.* Giant Enhancement of the Optical Second-Harmonic Emission of $\mathrm{WSe}_{2}$ Monolayers by Laser Excitation at Exciton Resonances. *Phys. Rev. Lett.* **114**, 97403 (2015).

369. Seyler, K. L. *et al.* Electrical control of second-harmonic generation in a WSe2 monolayer transistor. *Nat. Nanotechnol.* **10**, 407–411 (2015).

370. Yin, X. *et al.* Edge Nonlinear Optics on a $MoS_{2}$ Atomic Monolayer. *Science (80-. ).* **344**, 488 LP-490 (2014).

371. Säynätjoki, A. *et al.* Ultra-strong nonlinear optical processes and trigonal warping in MoS2 layers. *Nat. Commun.* **8**, 893 (2017).

372. Day, J. K., Chung, M.-H., Lee, Y.-H. & Menon, V. M. Microcavity enhanced second harmonic generation in 2D MoS2. *Opt. Mater. Express* **6**, 2360–2365 (2016).

373. Yi, F. *et al.* Optomechanical Enhancement of Doubly Resonant 2D Optical Nonlinearity. *Nano Lett.* **16**, 1631–1636 (2016).

374. Shi, J. *et al.* Plasmonic Enhancement and Manipulation of Optical Nonlinearity in Monolayer Tungsten Disulfide. *Laser Photon. Rev.* **12**, 1800188 (2018).

375. Wang, Z. *et al.* Selectively Plasmon-Enhanced Second-Harmonic Generation from Monolayer Tungsten Diselenide on Flexible Substrates. *ACS Nano* **12**, 1859–1867 (2018).

376. Han, X. *et al.* Harmonic Resonance Enhanced Second-Harmonic Generation in the Monolayer WS2–Ag Nanocavity. *ACS Photonics* **7**, 562–568 (2020).

377. Li, X. *et al.* Enhancement of the Second Harmonic Generation from WS2 Monolayers by Cooperating with Dielectric Microspheres. *Adv. Opt. Mater.* **7**, 1801270 (2019).

378. Chen, H. *et al.* Enhanced second-harmonic generation from two-dimensional MoSe2 on a silicon waveguide. *Light Sci. Appl.* **6**, e17060–e17060 (2017).

379. Bernhardt, N. *et al.* Quasi-BIC Resonant Enhancement of Second-Harmonic Generation in WS2 Monolayers. *Nano Lett.* **20**, 5309–5314 (2020).

380. Guo, Q. *et al.* Efficient Frequency Mixing of Guided Surface Waves by Atomically Thin Nonlinear Crystals. *Nano Lett.* **20**, 7956–7963 (2020).

381. Yu, H., Talukdar, D., Xu, W., Khurgin, J. B. & Xiong, Q. Charge-Induced Second-Harmonic Generation in Bilayer WSe2. *Nano Lett.* **15**, 5653–5657 (2015).

382. Wen, X., Xu, W., Zhao, W., Khurgin, J. B. & Xiong, Q. Plasmonic Hot Carriers-Controlled Second Harmonic Generation in WSe2 Bilayers. *Nano Lett.* **18**, 1686–1692 (2018).

383. Xiao, J. *et al.* Nonlinear optical selection rule based on valley-exciton locking in monolayer ws2. *Light Sci. Appl.* **4**, e366–e366 (2015).

384. Hu, G. *et al.* Coherent steering of nonlinear chiral valley photons with a synthetic Au–WS2 metasurface. *Nat. Photonics* **13**, 467–472 (2019).

385. Dasgupta, A., Gao, J. & Yang, X. Atomically Thin Nonlinear Transition Metal Dichalcogenide

Holograms. *Nano Lett.* **19**, 6511–6516 (2019).

386. Dasgupta, A., Yang, X. & Gao, J. Nonlinear Beam Shaping with Binary Phase Modulation on Patterned WS2 Monolayer. *ACS Photonics* **7**, 2506–2514 (2020).
387. Forn-Díaz, P., Lamata, L., Rico, E., Kono, J. & Solano, E. Ultrastrong coupling regimes of light-matter interaction. *Rev. Mod. Phys.* **91**, 25005 (2019).
388. Deng, H., Weihs, G., Santori, C., Bloch, J. & Yamamoto, Y. Condensation of Semiconductor Microcavity Exciton Polaritons. *Science (80-. ).* **298**, 199 LP-202 (2002).
389. Huang, L., Xu, L., Woolley, M. & Miroshnichenko, A. E. Trends in Quantum Nanophotonics. *Adv. Quantum Technol.* **3**, 1900126 (2020).
390. Liu, X. *et al.* Strong light–matter coupling in two-dimensional atomic crystals. *Nat. Photonics* **9**, 30–34 (2015).
391. Dufferwiel, S. *et al.* Exciton–polaritons in van der Waals heterostructures embedded in tunable microcavities. *Nat. Commun.* **6**, 8579 (2015).
392. Dhara, S. *et al.* Anomalous dispersion of microcavity trion-polaritons. *Nat. Phys.* **14**, 130–133 (2018).
393. Sidler, M. *et al.* Fermi polaron-polaritons in charge-tunable atomically thin semiconductors. *Nat. Phys.* **13**, 255–261 (2017).
394. Zheng, D. *et al.* Manipulating Coherent Plasmon–Exciton Interaction in a Single Silver Nanorod on Monolayer WSe2. *Nano Lett.* **17**, 3809–3814 (2017).
395. Wang, S. *et al.* Coherent Coupling of WS2 Monolayers with Metallic Photonic Nanostructures at Room Temperature. *Nano Lett.* **16**, 4368–4374 (2016).
396. Liu, W. *et al.* Strong Exciton–Plasmon Coupling in MoS2 Coupled with Plasmonic Lattice. *Nano Lett.* **16**, 1262–1269 (2016).
397. Deng, F., Liu, H., Xu, L., Lan, S. & Miroshnichenko, A. E. Strong Exciton–Plasmon Coupling in a WS2 Monolayer on Au Film Hybrid Structures Mediated by Liquid Ga Nanoparticles. *Laser Photon. Rev.* **14**, 1900420 (2020).
398. Kleemann, M.-E. *et al.* Strong-coupling of WSe2 in ultra-compact plasmonic nanocavities at room temperature. *Nat. Commun.* **8**, 1296 (2017).
399. Liu, W. *et al.* Observation and Active Control of a Collective Polariton Mode and Polaritonic Band Gap in Few-Layer WS2 Strongly Coupled with Plasmonic Lattices. *Nano Lett.* **20**, 790–798 (2020).
400. Lee, B. *et al.* Electrical Tuning of Exciton–Plasmon Polariton Coupling in Monolayer MoS2 Integrated with Plasmonic Nanoantenna Lattice. *Nano Lett.* **17**, 4541–4547 (2017).
401. Zhang, H. *et al.* Hybrid exciton-plasmon-polaritons in van der Waals semiconductor gratings. *Nat. Commun.* **11**, 3552 (2020).
402. Zhang, L., Gogna, R., Burg, W., Tutuc, E. & Deng, H. Photonic-crystal exciton-polaritons in monolayer semiconductors. *Nat. Commun.* **9**, 713 (2018).
403. Chen, Y. *et al.* Metasurface Integrated Monolayer Exciton Polariton. *Nano Lett.* **20**, 5292–5300 (2020).
404. Wang, S. *et al.* Collective Mie Exciton-Polaritons in an Atomically Thin Semiconductor. *J. Phys. Chem. C* **124**, 19196–19203 (2020).
405. Photonic crystals for controlling strong coupling in van der Waals materials. *Opt. Express* **27**, 22700–22707 (2019).
406. Strong coupling between excitons and guided modes in WS2-based nanostructures. *J. Opt. Soc.*

*Am. B* **37**, 1447–1452 (2020).

407. Kravtsov, V. *et al.* Nonlinear polaritons in a monolayer semiconductor coupled to optical bound states in the continuum. *Light Sci. Appl.* **9**, 56 (2020).
408. Cao, S. *et al.* Normal-Incidence-Excited Strong Coupling between Excitons and Symmetry-Protected Quasi-Bound States in the Continuum in Silicon Nitride–WS2 Heterostructures at Room Temperature. *J. Phys. Chem. Lett.* **11**, 4631–4638 (2020).
409. Xiao, D., Chang, M.-C. & Niu, Q. Berry phase effects on electronic properties. *Rev. Mod. Phys.* **82**, 1959–2007 (2010).
410. Rycerz, A., Tworzydło, J. & Beenakker, C. W. J. Valley filter and valley valve in graphene. *Nat. Phys.* **3**, 172–175 (2007).
411. Gunlycke, D. & White, C. T. Graphene Valley Filter Using a Line Defect. *Phys. Rev. Lett.* **106**, 136806 (2011).
412. Jiang, Y., Low, T., Chang, K., Katsnelson, M. I. & Guinea, F. Generation of Pure Bulk Valley Current in Graphene. *Phys. Rev. Lett.* **110**, 046601 (2013).
413. Offidani, M. & Ferreira, A. Anomalous Hall Effect in 2D Dirac Materials. *Phys. Rev. Lett.* **121**, 126802 (2018).
414. Cao, T. *et al.* Valley-selective circular dichroism of monolayer molybdenum disulphide. *Nat. Commun.* **3**, 885–887 (2012).
415. Plechinger, G. *et al.* Trion fine structure and coupled spin–valley dynamics in monolayer tungsten disulfide. *Nat. Commun.* **7**, 12715 (2016).
416. Zhu, C. R. *et al.* Exciton valley dynamics probed by Kerr rotation in WSe2 monolayers. *Phys. Rev. B - Condens. Matter Mater. Phys.* **90**, 1–5 (2014).
417. Molina-Sánchez, A., Sangalli, D., Wirtz, L. & Marini, A. Ab Initio Calculations of Ultrashort Carrier Dynamics in Two-Dimensional Materials: Valley Depolarization in Single-Layer WSe2. *Nano Lett.* **17**, 4549–4555 (2017).
418. Volmer, F. *et al.* Intervalley dark trion states with spin lifetimes of 150 ns in WSe2. *Phys. Rev. B* **95**, 235408 (2017).
419. Rivera, P. *et al.* Valley-polarized exciton dynamics in a 2D semiconductor heterostructure. *Science (80-. ).* **351**, 688 LP-691 (2016).
420. Yang, L. *et al.* Long-lived nanosecond spin relaxation and spin coherence of electrons in monolayer MoS2 and WS2. *Nat. Phys.* **11**, 830–834 (2015).
421. Dey, P. *et al.* Gate-Controlled Spin-Valley Locking of Resident Carriers in WSe2 Monolayers. *Phys. Rev. Lett.* **119**, 137401 (2017).
422. Mak, K. F., McGill, K. L., Park, J. & McEuen, P. L. The valley Hall effect in MoS2 transistors. *Science (80-. ).* **344**, 1489–1492 (2014).
423. Li, L. *et al.* Room-temperature valleytronic transistor. *Nat. Nanotechnol.* (2020). doi:10.1038/s41565-020-0727-0
424. Srivastava, A. *et al.* Valley Zeeman effect in elementary optical excitations of monolayer WSe2. *Nat. Phys.* **11**, 141–147 (2015).
425. Wang, G. *et al.* Polarization and time-resolved photoluminescence spectroscopy of excitons in MoSe 2 monolayers. *Appl. Phys. Lett.* **106**, 112101 (2015).
426. Yang, W. *et al.* Electrically Tunable Valley-Light Emitting Diode (vLED) Based on CVD-Grown Monolayer WS 2. *Nano Lett.* **16**, 1560–1567 (2016).
427. Jones, A. M. *et al.* Optical generation of excitonic valley coherence in monolayer WSe 2. *Nat.*

*Nanotechnol.* **8**, 634–638 (2013).

428. Gunawan, O. *et al.* Valley Susceptibility of an Interacting Two-Dimensional Electron System. *Phys. Rev. Lett.* **97**, 186404 (2006).
429. Dufferwiel, S. *et al.* Valley coherent exciton-polaritons in a monolayer semiconductor. *Nat. Commun.* **9**, 4797 (2018).
430. Qiu, L., Chakraborty, C., Dhara, S. & Vamivakas, A. N. Room-temperature valley coherence in a polaritonic system. *Nat. Commun.* **10**, 1–5 (2019).
431. Sun, Z. *et al.* Optical control of roomerature valley polaritons. *Nat. Photonics* **11**, 491–496 (2017).
432. Dufferwiel, S. *et al.* Valley-addressable polaritons in atomically thin semiconductors. *Nat. Photonics* **11**, 497–501 (2017).
433. Jha, P. K., Shitrit, N., Ren, X., Wang, Y. & Zhang, X. Spontaneous Exciton Valley Coherence in Transition Metal Dichalcogenide Monolayers Interfaced with an Anisotropic Metasurface. *Phys. Rev. Lett.* **121**, 116102 (2018).
434. Chen, Y. *et al.* Tunable Band Gap Photoluminescence from Atomically Thin Transition-Metal Dichalcogenide Alloys. *ACS Nano* **7**, 4610–4616 (2013).
435. Gong, Y. *et al.* Band Gap Engineering and Layer-by-Layer Mapping of Selenium-Doped Molybdenum Disulfide. *Nano Lett.* **14**, 442–449 (2014).
436. Feng, Q. *et al.* Growth of Large-Area 2D $MoS_{2(1-x)}Se_{2x}$ Semiconductor Alloys. *Adv. Mater.* **26**, 2648–2653 (2014).
437. Mann, J. *et al.* 2-Dimensional Transition Metal Dichalcogenides with Tunable Direct Band Gaps: $MoS_{2(1-x)}Se_{2x}$ Monolayers. *Adv. Mater.* **26**, 1399–1404 (2014).
438. Li, H. *et al.* Growth of Alloy $MoS_{2x}Se_{2(1-x)}$ Nanosheets with Fully Tunable Chemical Compositions and Optical Properties. *J. Am. Chem. Soc.* **136**, 3756–3759 (2014).
439. Duan, X. *et al.* Synthesis of $WS_{2x}Se_{2-2x}$ Alloy Nanosheets with Composition-Tunable Electronic Properties. *Nano Lett.* **16**, 264–269 (2016).
440. Li, H. *et al.* Lateral Growth of Composition Graded Atomic Layer $MoS_{2(1-x)}Se_{2x}$ Nanosheets. *J. Am. Chem. Soc.* **137**, 5284–5287 (2015).
441. Wu, X. *et al.* Spatially composition-modulated two-dimensional $WS_{2x}Se_{2(1-x)}$ nanosheets. *Nanoscale* **9**, 4707–4712 (2017).
442. Kobayashi, Y., Mori, S., Maniwa, Y. & Miyata, Y. Bandgap-tunable lateral and vertical heterostructures based on monolayer $Mo_{1-x}W_xS_2$ alloys. *Nano Res.* **8**, 3261–3271 (2015).
443. Zhang, W. *et al.* CVD synthesis of $Mo_{(1-x)}W_xS_2$ and $MoS_{2(1-x)}Se_{2x}$ alloy monolayers aimed at tuning the bandgap of molybdenum disulfide. *Nanoscale* **7**, 13554–13560 (2015).
444. Li, H. *et al.* Composition-Modulated Two-Dimensional Semiconductor Lateral Heterostructures via Layer-Selected Atomic Substitution. *ACS Nano* **11**, 961–967 (2017).
445. Conley, H. J. *et al.* Bandgap Engineering of Strained Monolayer and Bilayer MoS2. *Nano Lett.* **13**, 3626–3630 (2013).
446. He, K., Poole, C., Mak, K. F. & Shan, J. Experimental Demonstration of Continuous Electronic Structure Tuning via Strain in Atomically Thin MoS2. *Nano Lett.* **13**, 2931–2936 (2013).
447. Zhu, C. R. *et al.* Strain tuning of optical emission energy and polarization in monolayer and bilayer $MoS_{2}$. *Phys. Rev. B* **88**, 121301 (2013).
448. Wang, Y. *et al.* Strain-induced direct–indirect bandgap transition and phonon modulation in monolayer WS2. *Nano Res.* **8**, 2562–2572 (2015).

449. Desai, S. B. *et al.* Strain-Induced Indirect to Direct Bandgap Transition in Multilayer WSe2. *Nano Lett.* **14**, 4592–4597 (2014).
450. Niehues, I. *et al.* Strain Control of Exciton–Phonon Coupling in Atomically Thin Semiconductors. *Nano Lett.* **18**, 1751–1757 (2018).
451. Hui, Y. Y. *et al.* Exceptional Tunability of Band Energy in a Compressively Strained Trilayer MoS2 Sheet. *ACS Nano* **7**, 7126–7131 (2013).
452. Lloyd, D. *et al.* Band Gap Engineering with Ultralarge Biaxial Strains in Suspended Monolayer MoS2. *Nano Lett.* **16**, 5836–5841 (2016).
453. Liu, Z. *et al.* Strain and structure heterogeneity in MoS2 atomic layers grown by chemical vapour deposition. *Nat. Commun.* **5**, 5246 (2014).
454. Li, Z. *et al.* Efficient strain modulation of 2D materials via polymer encapsulation. *Nat. Commun.* **11**, 1151 (2020).
455. Zeng, M. *et al.* Bandgap tuning of two-dimensional materials by sphere diameter engineering. *Nat. Mater.* **19**, 528–533 (2020).
456. Pak, S. *et al.* Strain-Mediated Interlayer Coupling Effects on the Excitonic Behaviors in an Epitaxially Grown MoS2/WS2 van der Waals Heterobilayer. *Nano Lett.* **17**, 5634–5640 (2017).
457. He, X. *et al.* Strain engineering in monolayer WS2, MoS2, and the WS2/MoS2 heterostructure. *Appl. Phys. Lett.* **109**, 173105 (2016).
458. Maiti, R. *et al.* Strain-engineered high-responsivity MoTe2 photodetector for silicon photonic integrated circuits. *Nat. Photonics* **14**, 578–584 (2020).
459. Yu, Y. *et al.* Giant Gating Tunability of Optical Refractive Index in Transition Metal Dichalcogenide Monolayers. *Nano Lett.* **17**, 3613–3618 (2017).
460. Chakraborty, B. *et al.* Control of Strong Light–Matter Interaction in Monolayer WS2 through Electric Field Gating. *Nano Lett.* **18**, 6455–6460 (2018).
461. Dibos, A. M. *et al.* Electrically Tunable Exciton–Plasmon Coupling in a WSe2 Monolayer Embedded in a Plasmonic Crystal Cavity. *Nano Lett.* **19**, 3543–3547 (2019).
462. Munkhbat, B. *et al.* Electrical Control of Hybrid Monolayer Tungsten Disulfide–Plasmonic Nanoantenna Light–Matter States at Cryogenic and Room Temperatures. *ACS Nano* **14**, 1196–1206 (2020).
463. Yan, J., Ma, C., Huang, Y. & Yang, G. Single silicon nanostripe gated suspended monolayer and bilayer WS2 to realize abnormal electro-optical modulation. *Mater. Horizons* **6**, 334–342 (2019).
464. Kravets, V. G. *et al.* Measurements of electrically tunable refractive index of MoS2 monolayer and its usage in optical modulators. *npj 2D Mater. Appl.* **3**, 36 (2019).
465. Zhou, Y. *et al.* Controlling Excitons in an Atomically Thin Membrane with a Mirror. *Phys. Rev. Lett.* **124**, 27401 (2020).
466. Datta, I. *et al.* Low-loss composite photonic platform based on 2D semiconductor monolayers. *Nat. Photonics* **14**, 256–262 (2020).
467. Mitioglu, A. A. *et al.* Optical manipulation of the exciton charge state in single-layer tungsten disulfide. *Phys. Rev. B* **88**, 245403 (2013).
468. Mai, C. *et al.* Many-Body Effects in Valleytronics: Direct Measurement of Valley Lifetimes in Single-Layer MoS2. *Nano Lett.* **14**, 202–206 (2014).
469. Plechinger, G. *et al.* Identification of excitons, trions and biexcitons in single-layer WS2. *Phys. status solidi – Rapid Res. Lett.* **9**, 457–461 (2015).

470. Pei, J. *et al.* Excited State Biexcitons in Atomically Thin MoSe2. *ACS Nano* **11**, 7468–7475 (2017).
471. Lee, H. S., Kim, M. S., Kim, H. & Lee, Y. H. Identifying multiexcitons in $\mathrm{Mo}{\mathrm{S}}_{2}$ monolayers at room temperature. *Phys. Rev. B* **93**, 140409 (2016).
472. Barbone, M. *et al.* Charge-tuneable biexciton complexes in monolayer WSe2. *Nat. Commun.* **9**, 3721 (2018).
473. Yu, Y. *et al.* Room-Temperature Electron–Hole Liquid in Monolayer MoS2. *ACS Nano* **13**, 10351–10358 (2019).
474. Bataller, A. W. *et al.* Dense Electron–Hole Plasma Formation and Ultralong Charge Lifetime in Monolayer MoS2 via Material Tuning. *Nano Lett.* **19**, 1104–1111 (2019).
475. Arp, T. B., Pleskot, D., Aji, V. & Gabor, N. M. Electron–hole liquid in a van der Waals heterostructure photocell at room temperature. *Nat. Photonics* **13**, 245–250 (2019).
476. Li, Z. *et al.* Active Light Control of the MoS2 Monolayer Exciton Binding Energy. *ACS Nano* **9**, 10158–10164 (2015).
477. Yu, Y. *et al.* Ultrafast Plasmonic Hot Electron Transfer in Au Nanoantenna/MoS2 Heterostructures. *Adv. Funct. Mater.* **26**, 6394–6401 (2016).
478. Zu, S. *et al.* Active Control of Plasmon–Exciton Coupling in MoS2–Ag Hybrid Nanostructures. *Adv. Opt. Mater.* **4**, 1463–1469 (2016).
479. Deng, M. *et al.* Light-Controlled Near-Field Energy Transfer in Plasmonic Metasurface Coupled MoS2 Monolayer. *Small* **16**, 2003539 (2020).
480. Zhang, X. *et al.* Dynamic Photochemical and Optoelectronic Control of Photonic Fano Resonances via Monolayer MoS2 Trions. *Nano Lett.* **18**, 957–963 (2018).
481. Taghinejad, M. *et al.* Photocarrier-Induced Active Control of Second-Order Optical Nonlinearity in Monolayer MoS2. *Small* **16**, 1906347 (2020).
482. Liu, J.-T., Wang, T.-B., Li, X.-J. & Liu, N.-H. Enhanced absorption of monolayer MoS2 with resonant back reflector. *J. Appl. Phys.* **115**, 193511 (2014).
483. Nearly perfect absorption of light in monolayer molybdenum disulfide supported by multilayer structures. *Opt. Express* **25**, 21630–21636 (2017).
484. Epstein, I. *et al.* Near-Unity Light Absorption in a Monolayer WS2 Van der Waals Heterostructure Cavity. *Nano Lett.* **20**, 3545–3552 (2020).
485. Horng, J. *et al.* Perfect Absorption by an Atomically Thin Crystal. *Phys. Rev. Appl.* **14**, 24009 (2020).
486. Kats, M. A., Blanchard, R., Genevet, P. & Capasso, F. Nanometre optical coatings based on strong interference effects in highly absorbing media. *Nat. Mater.* **12**, 20–24
487. Jariwala, D. *et al.* Near-Unity Absorption in van der Waals Semiconductors for Ultrathin Optoelectronics. *Nano Lett.* **16**, 5482–5487 (2016).
488. Wong, J. *et al.* High Photovoltaic Quantum Efficiency in Ultrathin van der Waals Heterostructures. *ACS Nano* **11**, 7230–7240 (2017).
489. Wang, Q. *et al.* Fabry–Perot Cavity-Enhanced Optical Absorption in Ultrasensitive Tunable Photodiodes Based on Hybrid 2D Materials. *Nano Lett.* **17**, 7593–7598 (2017).
490. Piper, J. R. & Fan, S. Total Absorption in a Graphene Monolayer in the Optical Regime by Critical Coupling with a Photonic Crystal Guided Resonance. *ACS Photonics* **1**, 347–353 (2014).
491. Jiang, X. *et al.* Approaching perfect absorption of monolayer molybdenum disulfide at visible

wavelengths using critical coupling. *Nanotechnology* **29**, 335205 (2018).
492. Total absorption of light in monolayer transition-metal dichalcogenides by critical coupling. *Opt. Express* **25**, 31612–31621 (2017).
493. Li, H., Ren, Y., Hu, J., Qin, M. & Wang, L. Wavelength-Selective Wide-Angle Light Absorption Enhancement in Monolayers of Transition-Metal Dichalcogenides. *J. Light. Technol.* **36**, 3236–3241 (2018).
494. Nong, J., Da, H., Fang, Q., Yu, Y. & Yan, X. Perfect absorption in transition metal dichalcogenides-based dielectric grating. *J. Phys. D. Appl. Phys.* **51**, 375105 (2018).
495. Huang, L. *et al.* Atomically Thin MoS2 Narrowband and Broadband Light Superabsorbers. *ACS Nano* **10**, 7493–7499 (2016).
496. Wang, W. *et al.* Enhanced absorption in two-dimensional materials via Fano-resonant photonic crystals. *Appl. Phys. Lett.* **106**, 181104 (2015).
497. Ultra-narrowband visible light absorption in a monolayer MoS2 based resonant nanostructure. *Opt. Express* **28**, 27608–27614 (2020).
498. Piper, J. R. & Fan, S. Broadband Absorption Enhancement in Solar Cells with an Atomically Thin Active Layer. *ACS Photonics* **3**, 571–577 (2016).
499. Bahauddin, S. M., Robatjazi, H. & Thomann, I. Broadband Absorption Engineering to Enhance Light Absorption in Monolayer MoS2. *ACS Photonics* **3**, 853–862 (2016).
500. Plasmon-enhanced broadband absorption of MoS2-based structure using Au nanoparticles. *Opt. Express* **27**, 2305–2316 (2019).
501. Luk'yanchuk, B. *et al.* The Fano resonance in plasmonic nanostructures and metamaterials. *Nat. Mater.* **9**, 707–715 (2010).
502. Miroshnichenko, A. E., Flach, S. & Kivshar, Y. S. Fano resonances in nanoscale structures. *Rev. Mod. Phys.* **82**, 2257–2298 (2010).
503. Zhao, W. *et al.* Exciton–Plasmon Coupling and Electromagnetically Induced Transparency in Monolayer Semiconductors Hybridized with Ag Nanoparticles. *Adv. Mater.* **28**, 2709–2715 (2016).
504. Khurgin, J. B. Two-dimensional exciton–polariton—light guiding by transition metal dichalcogenide monolayers. *Optica* **2**, 740–742 (2015).
505. Zhang, X. *et al.* Guiding of visible photons at the ångström thickness limit. *Nat. Nanotechnol.* **14**, 844–850 (2019).
506. Zeytinoğlu, S., Roth, C., Huber, S. & İmamoğlu, A. Atomically thin semiconductors as nonlinear mirrors. *Phys. Rev. A* **96**, 31801 (2017).
507. Huang, L., Yu, Y. & Cao, L. General Modal Properties of Optical Resonances in Subwavelength Nonspherical Dielectric Structures. *Nano Lett.* **13**, 3559–3565 (2013).
508. Cao, L. *et al.* Engineering light absorption in semiconductor nanowire devices. *Nat. Mater.* **8**, 643 (2009).
509. Brongersma, M. L., Cui, Y. & Fan, S. Light management for photovoltaics using high-index nanostructures. *Nat. Mater.* **13**, 451–460 (2014).
510. Verre, R. *et al.* Transition metal dichalcogenide nanodisks as high-index dielectric Mie nanoresonators. *Nat. Nanotechnol.* **14**, 679–683 (2019).
511. Busschaert, S. *et al.* Transition Metal Dichalcogenide Resonators for Second Harmonic Signal Enhancement. *ACS Photonics* **7**, 2482–2488 (2020).
512. Munkhbat, B. *et al.* Transition metal dichalcogenide metamaterials with atomic precision. *Nat.*

*Commun.* **11**, 4604 (2020).
513. Yang, J. *et al.* Atomically thin optical lenses and gratings. *Light Sci. Appl.* **5**, e16046–e16046 (2016).
514. Lin, H. *et al.* Diffraction-limited imaging with monolayer 2D material-based ultrathin flat lenses. *Light Sci. Appl.* **9**, 137 (2020).
515. Liu, C.-H. *et al.* Ultrathin van der Waals Metalenses. *Nano Lett.* **18**, 6961–6966 (2018).
516. Wang, Y. *et al.* Atomically Thin Noble Metal Dichalcogenides for Phase-Regulated Meta-optics. *Nano Lett.* **20**, 7811–7818 (2020).
517. Yang, H. *et al.* Optical Waveplates Based on Birefringence of Anisotropic Two-Dimensional Layered Materials. *ACS Photonics* **4**, 3023–3030 (2017).
518. Hong, X. *et al.* Ultrafast charge transfer in atomically thin MoS2/WS2 heterostructures. *Nat. Nanotechnol.* **9**, 682–686 (2014).
519. Heo, H. *et al.* Interlayer orientation-dependent light absorption and emission in monolayer semiconductor stacks. *Nat. Commun.* **6**, 7372 (2015).
520. Wang, K. *et al.* Interlayer Coupling in Twisted WSe2/WS2 Bilayer Heterostructures Revealed by Optical Spectroscopy. *ACS Nano* **10**, 6612–6622 (2016).
521. Ji, Z. *et al.* Robust Stacking-Independent Ultrafast Charge Transfer in MoS2/WS2 Bilayers. *ACS Nano* **11**, 12020–12026 (2017).
522. Zhu, H. *et al.* Interfacial Charge Transfer Circumventing Momentum Mismatch at Two-Dimensional van der Waals Heterojunctions. *Nano Lett.* **17**, 3591–3598 (2017).
523. Rivera, P. *et al.* Observation of long-lived interlayer excitons in monolayer MoSe2–WSe2 heterostructures. *Nat. Commun.* **6**, 6242 (2015).
524. Rigosi, A. F., Hill, H. M., Li, Y., Chernikov, A. & Heinz, T. F. Probing Interlayer Interactions in Transition Metal Dichalcogenide Heterostructures by Optical Spectroscopy: MoS2/WS2 and MoSe2/WSe2. *Nano Lett.* **15**, 5033–5038 (2015).
525. Tongay, S. *et al.* Tuning Interlayer Coupling in Large-Area Heterostructures with CVD-Grown MoS2 and WS2 Monolayers. *Nano Lett.* **14**, 3185–3190 (2014).
526. Ceballos, F., Bellus, M. Z., Chiu, H.-Y. & Zhao, H. Ultrafast Charge Separation and Indirect Exciton Formation in a MoS2–MoSe2 van der Waals Heterostructure. *ACS Nano* **8**, 12717–12724 (2014).
527. Chiu, M.-H. *et al.* Spectroscopic Signatures for Interlayer Coupling in MoS2–WSe2 van der Waals Stacking. *ACS Nano* **8**, 9649–9656 (2014).
528. Zhang, C. *et al.* Interlayer couplings, Moiré patterns, and 2D electronic superlattices in $MoS_2/WSe_2$ hetero-bilayers. *Sci. Adv.* **3**, e1601459 (2017).
529. Karni, O. *et al.* Infrared Interlayer Exciton Emission in ${\mathrm{MoS}}_{2}/{\mathrm{WSe}}_{2}$ Heterostructures. *Phys. Rev. Lett.* **123**, 247402 (2019).
530. Fang, H. *et al.* Strong interlayer coupling in van der Waals heterostructures built from single-layer chalcogenides. *Proc. Natl. Acad. Sci.* **111**, 6198 LP-6202 (2014).
531. Bellus, M. Z., Ceballos, F., Chiu, H.-Y. & Zhao, H. Tightly Bound Trions in Transition Metal Dichalcogenide Heterostructures. *ACS Nano* **9**, 6459–6464 (2015).
532. Schaibley, J. R. *et al.* Directional interlayer spin-valley transfer in two-dimensional heterostructures. *Nat. Commun.* **7**, 13747 (2016).

533. Nayak, P. K. *et al.* Probing Evolution of Twist-Angle-Dependent Interlayer Excitons in MoSe2/WSe2 van der Waals Heterostructures. *ACS Nano* **11**, 4041–4050 (2017).
534. Miller, B. *et al.* Long-Lived Direct and Indirect Interlayer Excitons in van der Waals Heterostructures. *Nano Lett.* **17**, 5229–5237 (2017).
535. Liu, K. *et al.* Evolution of interlayer coupling in twisted molybdenum disulfide bilayers. *Nat. Commun.* **5**, 4966 (2014).
536. Gong, Y. *et al.* Vertical and in-plane heterostructures from WS2/MoS2 monolayers. *Nat. Mater.* **13**, 1135–1142 (2014).
537. Yu, Y. *et al.* Equally Efficient Interlayer Exciton Relaxation and Improved Absorption in Epitaxial and Nonepitaxial MoS2/WS2 Heterostructures. *Nano Lett.* **15**, 486–491 (2015).
538. Heo, H. *et al.* Rotation-Misfit-Free Heteroepitaxial Stacking and Stitching Growth of Hexagonal Transition-Metal Dichalcogenide Monolayers by Nucleation Kinetics Controls. *Adv. Mater.* **27**, 3803–3810 (2015).
539. Gong, Y. *et al.* Two-Step Growth of Two-Dimensional WSe2/MoSe2 Heterostructures. *Nano Lett.* **15**, 6135–6141 (2015).
540. Li, L. *et al.* Wavelength-Tunable Interlayer Exciton Emission at the Near-Infrared Region in van der Waals Semiconductor Heterostructures. *Nano Lett.* **20**, 3361–3368 (2020).
541. Duan, X. *et al.* Lateral epitaxial growth of two-dimensional layered semiconductor heterojunctions. *Nat. Nanotechnol.* **9**, 1024–1030 (2014).
542. Li, M.-Y. *et al.* Epitaxial growth of a monolayer $WSe_2$-$MoS_2$ lateral p-n junction with an atomically sharp interface. *Science (80-. ).* **349**, 524 LP-528 (2015).
543. Huang, C. *et al.* Lateral heterojunctions within monolayer MoSe2–WSe2 semiconductors. *Nat. Mater.* **13**, 1096–1101 (2014).
544. Kim, J. *et al.* Observation of ultralong valley lifetime in $WSe_2$/$MoS_2$ heterostructures. *Sci. Adv.* **3**, e1700518 (2017).
545. Lee, C.-H. *et al.* Atomically thin p–n junctions with van der Waals heterointerfaces. *Nat. Nanotechnol.* **9**, 676–681 (2014).
546. Furchi, M. M., Pospischil, A., Libisch, F., Burgdörfer, J. & Mueller, T. Photovoltaic Effect in an Electrically Tunable van der Waals Heterojunction. *Nano Lett.* **14**, 4785–4791 (2014).
547. Förg, M. *et al.* Cavity-control of interlayer excitons in van der Waals heterostructures. *Nat. Commun.* **10**, 3697 (2019).
548. Yan, J., Ma, C., Huang, Y. & Yang, G. Tunable Control of Interlayer Excitons in WS2/MoS2 Heterostructures via Strong Coupling with Enhanced Mie Resonances. *Adv. Sci.* **6**, 1802092 (2019).
549. Long, M., Wang, P., Fang, H. & Hu, W. Progress, Challenges, and Opportunities for 2D Material Based Photodetectors. *Adv. Funct. Mater.* **29**, 1803807 (2019).
550. Guo, N. *et al.* Light-Driven WSe2-ZnO Junction Field-Effect Transistors for High-Performance Photodetection. *Adv. Sci.* **7**, 1901637 (2020).
551. Xie, C., Mak, C., Tao, X. & Yan, F. Photodetectors Based on Two-Dimensional Layered Materials Beyond Graphene. *Adv. Funct. Mater.* **27**, 1603886 (2017).
552. Kang, K. *et al.* High-mobility three-atom-thick semiconducting films with wafer-scale homogeneity. *Nature* **520**, 656–660 (2015).

553. Hentschel, M., Schäferling, M., Duan, X., Giessen, H. & Liu, N. Chiral plasmonics. *Sci. Adv.* **3**, (2017).

554. Liu, W. *et al.* Circularly Polarized States Spawning from Bound States in the Continuum. *Phys. Rev. Lett.* **123**, 116104 (2019).

555. Zhao, J. *et al.* Ultralow threshold polariton condensate in a monolayer semiconductor microcavity at room temperature. *arXiv:2010.04381* (2020).

556. Massicotte, M. *et al.* Picosecond photoresponse in van der Waals heterostructures. *Nat. Nanotechnol.* **11**, 42–46 (2016).

557. Brongersma, M. L. The road to atomically thin metasurface optics. *Nanophotonics* **10**, 643–654

558. Ozawa, T. *et al.* Topological photonics. *Rev. Mod. Phys.* **91**, 15006 (2019).

559. Liu, W. *et al.* Generation of helical topological exciton-polaritons. *Science (80-. ).* **370**, 600–604 (2020).

560. Li, M. *et al.* Experimental observation of topological exciton-polaritons in transition metal dichalcogenide monolayers. *arXiv* (2020).

561. Gao, X. *et al.* Dirac-vortex topological cavities. *Nat. Nanotechnol.* **15**, 1012–1018 (2020).

562. Miri, M.-A. & Alù, A. Exceptional points in optics and photonics. *Science (80-. ).* **363**, (2019).

563. Hu, G. *et al.* Topological polaritons and photonic magic angles in twisted α-MoO3 bilayers. *Nature* **582**, 209–213 (2020).